\documentclass[10pt,letterpaper]{article}

\usepackage{amsmath,amssymb}

\usepackage{changepage}

\usepackage[utf8x]{inputenc}

\usepackage{textcomp,marvosym}

\usepackage{cite}

\usepackage{nameref,hyperref}

\usepackage[right]{lineno}

\usepackage{microtype}
\DisableLigatures[f]{encoding = *, family = * }

\usepackage[table]{xcolor}

\usepackage{array}

\newcolumntype{+}{!{\vrule width 2pt}}

\newlength\savedwidth

\newcommand\thickhline{\noalign{\global\savedwidth\arrayrulewidth\global\arrayrulewidth 2pt}%
\hline
\noalign{\global\arrayrulewidth\savedwidth}}

\usepackage[aboveskip=1pt,labelfont=bf,labelsep=period,justification=raggedright,singlelinecheck=off]{caption}

\makeatletter
\renewcommand{\@biblabel}[1]{\quad#1.}
\makeatother

\date{}

\usepackage{lastpage,fancyhdr,graphicx}
\usepackage{epstopdf}
\usepackage{fullpage}

\usepackage{multirow}

\usepackage{color}

\begin{document}
\vspace*{0.2in}

\begin{flushleft}
{\Large
\textbf\newline{A Theoretical Framework for Analyzing Coupled Neuronal Networks: Application to the Olfactory System} 
}
\newline
\\
Andrea K. Barreiro \textsuperscript{1},
Shree Hari Gautam \textsuperscript{2},
Woodrow L. Shew \textsuperscript{2},
Cheng Ly \textsuperscript{3*},
\\
\bigskip
\textbf{1} Department of Mathematics, Southern Methodist University, Dallas, TX 75275  U.S.A.
\\
\textbf{2} Department of Physics, University of Arkansas, Fayetteville, AR  72701  U.S.A.
\\
\textbf{3} Department of Statistical Sciences and Operations Research, Virginia Commonwealth University, Richmond, VA 23284  U.S.A.
\\
\bigskip

%
%






* CLy@vcu.edu

\end{flushleft}
\section*{Abstract}
Determining how synaptic coupling within and between regions is modulated during sensory processing is an important topic in neuroscience.  Electrophysiological recordings provide detailed information about neural spiking but have traditionally been confined to a particular region or layer of cortex.  
Here we develop new theoretical methods to study interactions between and within two brain regions, based on experimental measurements of spiking activity simultaneously recorded from the two regions.  By 
systematically comparing experimentally-obtained spiking statistics to (efficiently computed) model spike rate statistics, we identify regions in model parameter space that are consistent with the experimental 
data. We apply our new technique to dual micro-electrode array \textit{in vivo} recordings from two distinct regions: olfactory bulb ({\bf OB}) and anterior piriform cortex ({\bf PC}).  
Our analysis predicts that: i) inhibition within the afferent region (OB) has to be weaker than the inhibition within PC, ii) excitation from PC to OB is generally stronger than excitation from OB to PC, 
iii) excitation from PC to OB and inhibition 
within PC have to both be relatively strong compared to presynaptic inputs from OB.  
These predictions are validated in a spiking 
neural network model of the OB--PC pathway that satisfies the many constraints from our experimental data.  
We find when the derived relationships are violated, the spiking statistics no longer satisfy the constraints from the data.  In principle this modeling framework can be adapted to other systems and be used to investigate 
relationships between other neural attributes besides network connection strengths.  
Thus, this work can serve as a guide to further investigations into the relationships of various neural attributes within and across different regions during sensory processing.

\section*{Author Summary}

Sensory processing is known to span multiple regions of the nervous system. 
However, electrophysiological recordings during sensory processing have traditionally been limited to a single region or brain layer.  
With recent advances in experimental techniques, recorded spiking activity from multiple regions simultaneously is feasible. However, other important quantities--- such as inter-region connection strengths --- 
cannot yet be measured. 
Here, we develop new theoretical tools to leverage data obtained by recording from two different brain regions simultaneously.   We address the following questions: 
what are the crucial neural network attributes that enable sensory processing across different regions, and how are these attributes related to one another?  

With a novel theoretical framework to efficiently calculate spiking statistics, we can characterize a high dimensional parameter space that satisfies 
data constraints. We apply our results to the olfactory system to make specific predictions about effective network connectivity.  
Our framework relies on incorporating relatively easy-to-measure quantities to predict hard-to-measure interactions across multiple brain regions. 
Because this work is adaptable to other systems, we anticipate it will be a valuable tool for analysis of other larger scale brain recordings.  


\section*{Introduction}

As experimental tools advance, measuring whole-brain dynamics with single-neuron resolution becomes closer to reality~\cite{prevedel14,ahrens13,lemon15,brainInit_13}.  
However, a task that remains technically elusive is to measure the interactions within and across brain regions that govern such system-wide dynamics.  
Here we develop a theoretical approach to elucidate such interactions based on easily-recorded properties such as mean and (co-)variance of firing rates, when they can be measured in multiple regions and in multiple activity states.  
Although previous theoretical studies have addressed how spiking statistics depend on various mechanisms~\cite{brunel,brunelhakim,doiron16,BarreiroLy_RecrCorr_17}, these studies have typically been limited to a single region, 
leaving open the challenge of how inter-regional interactions impact the system dynamics, and ultimately the coding of sensory signals ~\cite{zohary94,bair01,ecker11,moreno14,kohn16}.

As a test case for our new theoretical tools, we studied interactions in the olfactory system.  We used two micro-electrode arrays to simultaneously record from olfactory bulb ({\bf OB}) and anterior piriform cortex ({\bf PC}).  
Constrained by these experimental data, we developed computational models and theory to investigate interactions within and between OB and PC.  
The modeling framework includes two distinct regions: a 
network that receives direct sensory stimuli (here, the OB), and a second neural network (PC) that is reciprocally coupled to the afferent region. Each region contains multiple individual populations, each of which is modeled with a firing rate model~\cite{wilsoncowan1}; thus even this minimal model involves several coupled stochastic differential equations (here, six) and has a large-dimensional parameter space.
Analysis of this system would be unwieldy in general; we address this by developing a novel method to compute firing statistics that is computationally efficient, captures the results of Monte Carlo simulations, and can provide analytic insight.  

Thorough analysis of experimental data in both the spontaneous and stimulus-evoked states leads to a number of constraints on first- and second-order spiking statistics--- many of which could not be observed using data 
from just one micro-electrode array.  In particular, we find twelve (12) constraints that are consistent across different odorant stimuli.  We use our theory and modeling to study an important subset of neural attributes (synaptic strengths) 
and investigate what relationships, if any, must be satisfied in order to robustly capture the many constraints observed in the data.  We find that: i) inhibition within OB has to be weaker than the inhibition in PC, 
ii) excitation from PC to OB is generally stronger than excitation from OB to PC, iii) excitation from PC to OB and inhibition within PC have to both be relatively strong compared to inputs originating in OB (inhibition within OB and excitation from OB to PC).  
We validate these guiding principles in a large spiking neural network (leaky integrate-and-fire, or {\bf LIF}) model, 
by showing that the many constraints from the experimental data are {\it all} satisfied.  
Finally, we demonstrate that violating these relationships in the LIF model results in spiking statistics that do not satisfy all of the data constraints.


Our predictions provide insights into interactions in the olfactory system that are difficult to directly measure experimentally. Importantly, these predictions were inferred from spike rates and variability, which are relatively easy to measure.  
We believe that the general approach we have developed -- using easy-to-measure quantities to predict hard-to-measure interactions -- will be valuable in diverse future investigations of how whole-brain function 
emerges from interactions among its constituent components.

\section*{Results}

Our main result is the development of a theoretical framework to infer hard-to-measure connection strengths in a minimal firing rate model, constrained by spike count statistics from  
simultaneous array recordings. 

We performed simultaneous dual micro-electrode recordings in the olfactory bulb ({\bf OB}) and the anterior piriform cortex ({\bf PC})
(see {\bf Materials and Methods}). First, we use the experimental data to 
compute population-averaged (across cells or cell pairs) first and second order spike count statistics, 
comparing across regions (OB or PC) and activity states (spontaneous or stimulus-evoked).  
We use these statistics to 
constrain a minimal firing rate model of the coupled OB-PC system, aided by a quick and efficient method for calculating firing statistics without Monte Carlo simulations. 

As a test case for our methods, we investigate the structure of four important parameters: within-region inhibitory connection strengths and between-region excitatory connection strengths. We find several relationships that must hold, in
order to satisfy all constraints from the experimental data.   These results are then validated with a large 
spiking network of leaky integrate-and-fire ({\bf LIF}) model neurons.  

\subsection*{Consistent Trends in the Experimental Data}

\begin{figure}[!h]
\centering
	\includegraphics[width=5.25in]{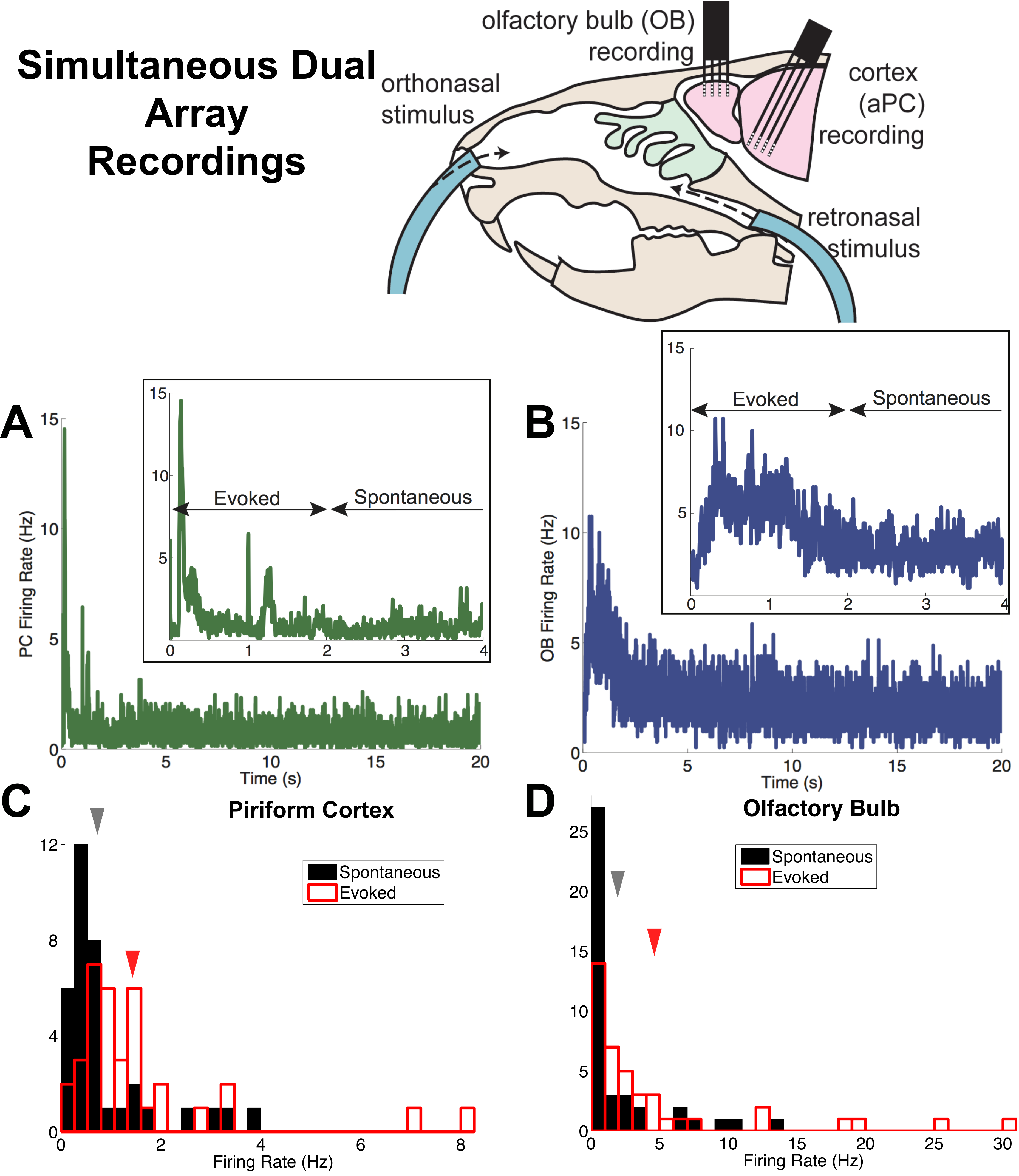}
\caption{{\bf Population firing rates in anterior piriform cortex (PC) and olfactory bulb (OB) from simultaneous dual array recordings.}
(A) Trial-averaged population firing rate in time from 73 PC cells (38 and 35 cells from two recordings).  The inset shows a closeup view, to highlight the distinction between spontaneous and evoked states.  
(B) Trial-averaged population firing rate in time from 41 OB cells (23 and 18 cells from two recordings).  Inset as in (A); both (A) and (B) use 5\,ms time bins.  
(C) The PC firing rate (averaged in time and over trials) of individual cells in the spontaneous (black) and 
evoked states (red).  The arrows indicate the mean across 73 cells; the mean$\pm$std. dev. in the spontaneous state is: $0.75\pm 0.93$\,Hz, in the evoked state is: $1.5\pm1.6$\,Hz.  
(D) Similar to (C), but for the OB cells described in (B).  The mean$\pm$std. dev. in the spontaneous state is: $2\pm3.3$\,Hz, in the evoked state is: $4.7\pm7.1$\,Hz.}
\label{fig1}
\end{figure}

We first present our data from simultaneous dual micro-electrode array recordings in anesthetized rats. 
During each 30-second trial an odor was presented for roughly one second; recordings continued for a total of 30 seconds. 
This sequence was repeated for 10 trials with 2-3 minutes in between trials; the protocol was repeated for another odor. 
Recordings were processed to extract single-unit activity; the number of units identified was: 23 in OB and 38 in PC (first recording, two odors), 18 in OB and 35 in PC (second recording, another two odors). In total, there were four
different odors presented.

In this paper, we focus on the spike count statistics rather than the detailed temporal structure of the neural activity (Fig~\ref{fig1}A--B). We divided each 30\,s trial into two segments, representing the odor-{\bf evoked} state (first 2 seconds) and the {\bf spontaneous} state (remaining 28 seconds).  We computed first- and second-order statistics for identified units; i.e., firing rate 
$\nu_k$, spike count variance, and spike count covariance (we also computed two derived statistics, Fano Factor and Pearson's correlation coefficient, for each cell or cell pair). Spike count variances, covariances and correlations were computed 
using time windows $T_{win}$ ranging between 5\,ms and 2\,s.  In computing population statistics we distinguished between different odors (four total), different regions (OB vs. PC), and different activity states (spontaneous vs. evoked); 
otherwise, we assumed statistics were stationary over time.

\begin{figure}[!h]
\centering
	\includegraphics[width=5.25in]{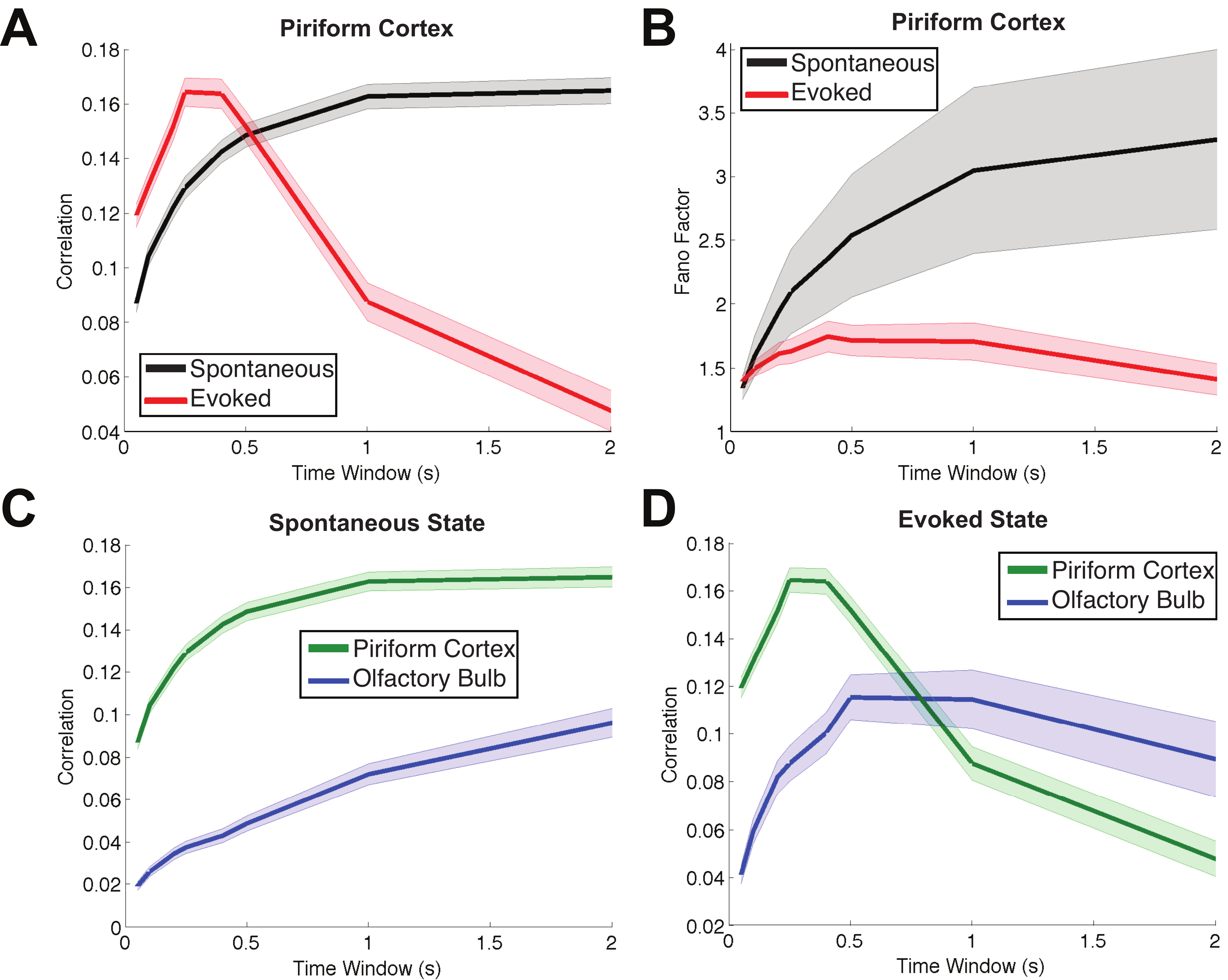}
\caption{{\bf A subset of the important relationships between the spiking statistics in spontaneous and evoked states.}
Consistent trends that hold for {\it all} 4 odorant stimuli in the experimental data. Each panel shows two spike count statistics, as a function of the time window.  
The shaded error bars show the \textit{standard error of the mean} above and below the mean statistic.
(A) Stimulus-induced decorrelation of PC cell pairs (red) compared to the spontaneous state (black).
(B) The variability in PC (measured by Fano Factor) is lower in the evoked state (red) than in the spontaneous state (black). 
(C) In the spontaneous state, the average correlation of PC pairs (green) is \textit{higher} than that of OB pairs (blue).
(D) In the evoked state, the average correlation of PC pairs (green) is \textit{lower} than that of OB pairs (blue), for long time windows. There were 406 total OB pairs and 1298 total PC pairs.  (Although the trends reverse in (A) and (D) for smaller time windows, our focus is on the larger time windows because stimuli were held for 1\,s; smaller time windows are shown for completeness.)}
\label{fig2}
\end{figure}

\begin{figure}[!h]
\centering
 \includegraphics[width=5.25in]{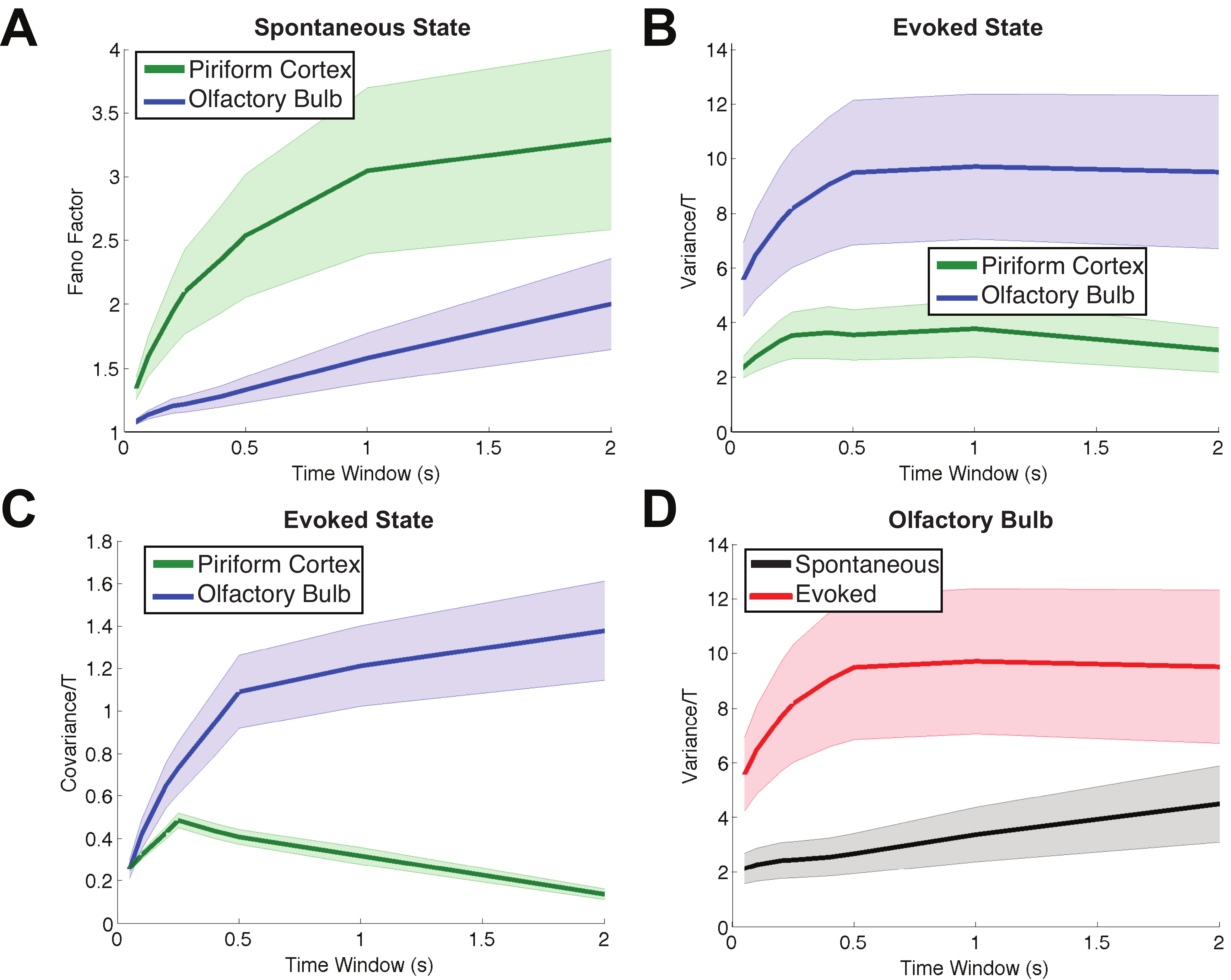}
\caption{\label{fig3} Showing the other trends from the experimental data that are consistent with all odors and for all time windows.  The shaded error bars show the \textit{standard error of the mean} above and below the mean statistic.
(A) Fano Factor of spontaneous activity is larger in PC (green) than in OB (blue).  (B) The spike count variance in the evoked state is smaller in PC (green) than in OB (blue).  
(C) Spike count covariance in the evoked state is smaller in PC (green) than in OB (blue).  (D) In OB cells, the evoked spike count variance (red) is larger than the spontaneous (black).  
The number of cells and number of pairs are the same as in Fig~\ref{fig2}.  Throughout we scale spike count variance and covariance by time window $T$ for aesthetic reasons. 
} 
\end{figure}

We then sought to identify relationships among these standard measures of spiking activity.  For example, we found that mean firing rate of OB cells in the evoked state was higher than the mean firing rate in the spontaneous state, or $\nu_{OB}^{Ev} > \nu_{OB}^{Sp}$ (although there is significant variability across the population, we focus on population-averaged statistics here).
We found twelve (12) robust relationships 
that held across all odors.  
Table~\ref{table:constr} summarizes the consistent 
relationships we found in our data, and Fig~\ref{fig1}C--D, Fig~\ref{fig2}, Fig~\ref{fig3} show the data exhibiting these relationships when combining all odorant stimuli (see \nameref{S1_file} for statistics plotted by distinct odors).  
Throughout the paper, when comparing activity states the spontaneous state is in black and the evoked state in red; when comparing regions the OB cells are in blue and PC cells in green.

\begin{table}[!ht]
\centering
\caption{ {\bf The 12 relationships (constraints) that hold in the experimental data across all odors.}}
\label{table:constr}
\begin{tabular}{l|l|l|l|}
\cline{2-4}
                                                                & \textbf{Spont.}                       & \textbf{Evoked}                       & \textbf{Spon. to Evoked}        \\ \thickhline
\multicolumn{1}{|l|}{}                                          &                                       &                                       & $\nu_{PC}^{Sp}<\nu_{PC}^{Ev}$   \\ \cline{4-4} 
\multicolumn{1}{|l|}{\multirow{-2}{*}{\textbf{Firing Rate}}}    & \multirow{-2}{*}{$\nu_{PC}<\nu_{OB}$} & \multirow{-2}{*}{$\nu_{PC}<\nu_{OB}$} & $\nu_{OB}^{Sp}<\nu_{OB}^{Ev}$   \\ \hline
\multicolumn{1}{|l|}{}                                          & \cellcolor[HTML]{C0C0C0}              & $\text{Var}_{PC}<\text{Var}_{OB}$                   & $\text{Var}_{OB}^{Sp}<\text{Var}_{OB}^{Ev} $        \\ \cline{2-4} 
\multicolumn{1}{|l|}{\multirow{-2}{*}{\textbf{Variability}}}  & $FF_{PC}>FF_{OB}$                     & \cellcolor[HTML]{C0C0C0}              & $FF_{PC}^{Sp}>FF_{PC}^{Ev}$     \\ \hline
\multicolumn{1}{|l|}{}                                          & \cellcolor[HTML]{C0C0C0}              & $\text{Cov}_{PC}<\text{Cov}_{OB}$                   & \cellcolor[HTML]{C0C0C0}        \\ \cline{2-4} 
\multicolumn{1}{|l|}{\multirow{-2}{*}{\textbf{Co-variability}}} & $\rho_{PC}>\rho_{OB}$                 & $\rho_{PC}<\rho_{OB}$                 & $\rho_{PC}^{Sp}>\rho_{PC}^{Ev}$ \\ \hline
\end{tabular}
\begin{flushleft} Relationships between population-averaged statistics (averages are across all cells or cell pairs)  that were consistent across all odors.  Other possible relationships were left out because they were 
ambiguous and/or odor dependent.
\end{flushleft}
\end{table}

A common observation across different animals and sensory systems, is that firing rates increase in the evoked state (see, for example, Figure 3 in~\cite{churchland10}). 
Indeed, we observed that 
average firing rates in both the OB and PC were  higher in the evoked state than in the spontaneous state (Fig~\ref{fig1}C--D).  
Furthermore, the firing rate in the OB was larger than the firing rate in the PC, in both spontaneous and evoked states (see mean values in Fig~\ref{fig1}C--D).  

\textit{Stimulus-induced decorrelation} appears to be a widespread phenomena in 
many sensory systems and in many animals~\cite{doiron16}; stimulus-induced decorrelation was previously reported in PC cells under different experimental conditions~\cite{miura12}.
Here, we found that in the PC, the average spike count correlation is lower in the evoked state (red) than in the spontaneous state (black), at least for time windows of 0.5\,s and above (Fig~\ref{fig2}A).  
Although we show a range of time windows for completeness, 
we focus on the larger time windows because in our experiments the odors are held for 1\,s; furthermore, our theoretical methods only address long time-averaged spiking statistics.  
Note that stimulus-induced decorrelation in the OB cells was not consistently observed across odors. 

Another common observation in cortex, is for variability to decrease at the onset of stimulus~\cite{churchland10}: 
in Fig~\ref{fig2}B we see that the Fano Factor of spike counts in PC cells decreases in the evoked state (red) compared to the spontaneous state (black); note that other experimental labs have also
observed this decrease in the Fano factor of PC cells (see supplemental figure S6D in~\cite{miura12}). 
Fig~\ref{fig2}C--D shows a comparison of PC and OB spike count correlation in the spontaneous state and evoked state, respectively.  Spike count correlation in PC (green) 
is larger than correlation in OB (blue) in the spontaneous state, but in the evoked state the relationship switches, at least for time windows larger than 0.5\,sec.  

Fig~\ref{fig3} shows the four remaining constraints that are consistent for all odors and for all time windows.  The Fano Factor in PC (green) is larger than in OB (blue), in the spontaneous state (Fig~\ref{fig3}A); spike count variance in PC (green) 
is smaller than in OB (blue) in the evoked state (Fig~\ref{fig3}B); spike count covariance in PC (green) is smaller than in OB (blue) in the evoked state (Fig~\ref{fig3}C); and in OB the spike count variance in the evoked state (red) is larger than spontaneous (black, Fig~\ref{fig3}D).  
Throughout the paper, we scale the spike count variance and covariance by time window for aesthetic reasons; 
this does not affect the relative relationships.

\subsection*{A Minimal Firing Rate Model to Capture Data Constraints}

\begin{figure}[!h]
\centering
 \includegraphics[width=4.25in]{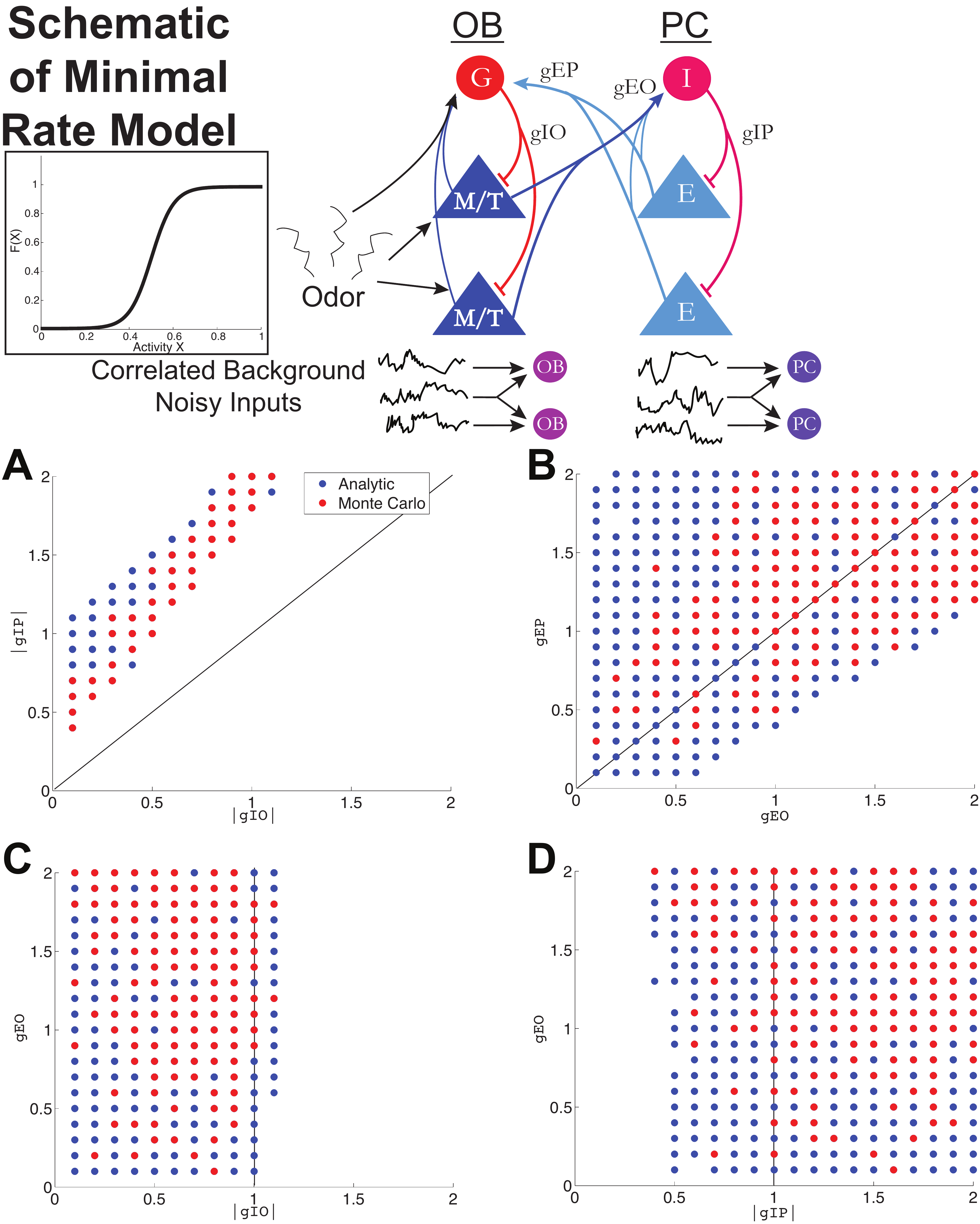}
\caption{{\bf Minimal firing rate model to analyze important synaptic conductance strengths.}
A firing rate model (Wilson-Cowan) with background correlated noisy inputs is analyzed to derive principles relating these network attributes (see Eq~\ref{eqn:gen_WC_pop} and {\bf Materials and Methods} section).  
This model only incorporates some of the anatomical connections that are known to exist {\it and} are important for modulation of statistics of firing (see main text for further discussion).  
Each neuron within a region (OB or PC) receives correlated background noisy input with $c_{OB}<c_{PC}$. 
Each plot shows parameter sets (4-tuples) that satisfy all 12 data constraints in Table~\ref{table:constr}, projected into a two-dimensional plane in parameter space. 
The blue dots show the result of the fast analytic method that satisfy all constraints; the red dots show the Monte Carlo simulations that satisfy 
all of the constraints.  For computational purposes, we only tested the Monte Carlo on parameter sets that first satisfied the constraints in the fast analytic method.  
(A) The magnitude of the inhibition within PC ($|gIP|$) is greater than the magnitude of the inhibition within OB ($|gIO|$);  all dots are above the diagonal line.  
(B) The excitation from PC to OB ($gEP$) is generally (but not always) larger than the excitation from OB to PC ($gEO$).  
(C) The inhibition within OB is generally weak; dots are to the left of the vertical line.  
(D) The inhibition within PC is generally strong; dots are to the right of the vertical line.  Table~\ref{table:rateMod_parms} states the parameter values.  
}
\label{fig4}
\end{figure}

We model two distinct regions (OB and PC) with a system of six (6) stochastic differential equations, each representing the averaged activity of a neural 
population~\cite{wilsoncowan1} or representative cell (see Fig~\ref{fig4} for a schematic of the network).  
For simplicity, in this section we use the word ``cell"  to refer to one of these populations.  Each region has two excitatory ({\bf E}) and one inhibitory ({\bf I}) cell to account for a variety of spiking correlations.  

We chose to include two E cells for two reasons: first, excitatory cells are the dominant source of projections between regions; we need at least two E cells to compute an E-to-E correlation.  Moreover, in our experimental data, we are most likely 
recording from excitatory mitral and tufted cells (we do not distinguish between mitral vs tufted here, and therefore refer to them as M/T cells); therefore, the experimental measurements of correlations are likely to have many E-to-E correlations.  
The arrays likely record from I cell spiking activity as well, and the inclusion of the I cell is also important for capturing the stimulus-induced decreases in correlation and Fano factor~\cite{churchland10,doiron16} 
(also see~\cite{LMD_whisker_12} who similarly used these same cell types to analyze spiking correlations in larger spiking network models).

We use $j\in\{1,2,3\}$ to denote three OB ``cells" and $j\in\{4,5,6\}$ for three PC cells, with $j=1$ as the inhibitory OB granule cell and $j=4$ as the inhibitory PC cell.
The equations are:
\begin{eqnarray}\label{eqn:gen_WC_pop}
	\tau \frac{d x_j}{dt} & = & -x_j + \mu_j + \sigma_j \eta_j + \sum_k g_{jk} F(x_k)   \label{eqn:dxdt_SDE}
\end{eqnarray}
where $F(x_k)$ is a transfer function mapping activity to firing rate.  Thus, the firing rate is:
\begin{equation}
	\nu_j = F(x_j).  
\end{equation}
We set the transfer function to $F(X)=\frac{1}{2}\left(1+\tanh((X-0.5)/0.1) \right)$, a commonly used sigmoidal function~\cite{wilsoncowan1} for all cells; experimental recordings of this function demonstrate it can be sigmoidal~\cite{fellous03,prescott03,cardin08}.  
All cells receive noise $\eta_j$, the increment of a Weiner process, uncorrelated in time but correlated within a region: i.e. $\langle \eta_j(t) \rangle = 0$, $\langle \eta_j(t) \eta_j(t+s) \rangle  = \delta(s)$, 
and $\langle \eta_j(t) \eta_k(t+s) \rangle  = c_{jk} \delta(s)$.  We set $c_{jk}$ to:
\begin{eqnarray} c_{jk} = \left\{
\begin{array}{ll}
      0, & \hbox{if }j\in\{1,2,3\}; k\in\{4,5,6\} \\
      1, & \hbox{if } j=k \\
      c_{OB} & \hbox{if }j\neq k; j,k\in\{1,2,3\} \\
      c_{PC} & \hbox{if }j\neq k; j,k\in\{4,5,6\} \\
\end{array} \right.
\end{eqnarray}
The parameters $\mu_j$ and $\sigma_j$ are constants that give the input mean and input standard deviation, respectively.
Within a particular region (OB or PC), all three cells receive correlated background noisy input, but there is {\bf no} correlated background input provided to both PC and OB cells.  This is justified 
by the experimental data (see Fig S9 in \nameref{S2_file}); average pairwise OB-to-PC correlations are all relatively small, and in particular, less than pairwise correlations \textit{within} the OB and PC.  Furthermore, 
anatomically there are no known common inputs to both regions that are active at the same time.  

We also set the background correlations to be higher in PC than in OB: i.e., 
$$ c_{PC} > c_{OB} . $$
This is justified in part by our array recordings, where correlated local field potential fluctuations are larger in PC than in OB.  
Furthermore, one source of background correlation is global synchronous activity; Murakami et al.~\cite{murakami05} has demonstrated that state changes 
(i.e., slow or fast waves as measured by EEG) strongly affect odorant responses in piriform cortex but only minimally effect olfactory bulb cells.  
Finally, PC has more recurrent activity than the olfactory bulb; this
could lead to more recurrent common input, if not cancelled by inhibition~\cite{renart10}.

We constructed our model to have two distinct activity states, spontaneous and evoked. 
We modeled the evoked state by increasing the three parameters $\mu_1,\mu_2,\mu_3$, representing mean input to the olfactory bulb (values given in Table~\ref{table:rateMod_parms}).  All other parameters were the same for both states.  
While increasing the input to the I cells in OB in the evoked state ($\mu_1$) is not anatomically accurate because granule cells do not receive direct sensory input~\cite{oswald12}, overall this captures the net effect of stimulus input to 
granule cells (see section {\bf Generality of Firing Rate Model Predictions} for how we apply this method to a specific olfactory system).

The model we have described is less realistic than a large network of spiking models (such as Hodgkin-Huxley or leaky integrate-and-fire neurons).  However, its simplicity permits fast and efficient evaluation of firing rate 
statistics, a necessity in exploring a large parameter space. 
Specifically, we calculate the statistics of the coupled network by solving a system of transcendental equations Eq~\ref{mn_x1}--\ref{cov_56}, 
rather than using Monte 
Carlo simulations. These equations were derived using an approximation based on asymptotic expansions  (see {\bf Materials and Methods: Approximation of Firing Statistics in the Firing Rate Model} for details).

This fast method allowed us to evaluate many parameter combinations, and therefore constrain the unknown coupling parameters, $g_{jk}$, which would otherwise be an intractable problem. 
Comparisons of the firing statistics computed from our method and 
Monte Carlo simulations show that the mean activity and firing rates are very accurate; variance and covariance (and thus correlation) are not as accurate, for larger 
coupling strengths (see Fig S10 in \nameref{S2_file} comparing 
100 random parameter sets).  
Nonetheless, we will find that these reduced model results are replicated by more realistic and larger spiking network models. 

In principle, there can be up to 36 coupling strengths, which is intractable to explore in detail.  We make the following assumptions:
\begin{itemize}
\item No cross-region inhibitory projections: $g_{41} = g_{51} = g_{61} = g_{14} = g_{24} = g_{34} = 0$.
\item Excitatory $PC \rightarrow OB$ output will synapse only onto the inhibitory population: $g_{25} = g_{26} = g_{35} = g_{36} = 0$ (see~\cite{oswald12}).  
This reflects experimental evidence that the feedback projections from PC to OB are dominated by inhibition~\cite{boyd12,markopoulos12}.  
\item Excitatory $OB \rightarrow PC$ output will synapse only onto the inhibitory population: $g_{52} = g_{62} = g_{53} = g_{63} = 0$.  Although this is not anatomically accurate because the mitral/tufted cells also project to the E cells in PC, our goal is to 
(heuristically) model the prominent role of I cells in PC.  Recent work has shown that within PC the recurrent activity is dominated by inhibition~\cite{large16}. Previous work has also shown that inhibitory synaptic events are much more common in PC and are much easier to elicit~\cite{poo09}.  
Thus, the connections from excitatory OB to inhibitory PC (Fig~\ref{fig4}) should be thought of as the \textit{net} effect of OB-to-PC connections.  
\end{itemize}
Within OB, there is also excitatory (M/T) 
input to the inhibitory (granule) cells: $g_{12}=g_{13}=0.1$ -- these values are small because 
feedforward inhibition is known to be a significant component in this circuit~\cite{burton15}.  Within PC, we also include similar connections from E to I cells: $g_{45}=g_{46}=0.1$.  
Recurrent E to E connections in PC are omitted; such connections can cause problems for our reduction method, resulting in oscillatory firing rates that cannot be efficiently captured.

We also make the following simplifying assumptions to limit the dimension of the parameter space of interest: 
\begin{itemize}
\item Feedforward inhibitory connections within a population were identical: $gIO \equiv g_{21} = g_{31}$ and $gIP \equiv g_{54} = g_{64}$.
\item Excitatory connections projecting outward from each region to the other region were identical: $gEO \equiv g_{42} = g_{43}$ and $gEP \equiv g_{15} = g_{16}$.
\item No within-region excitatory connections; $g_{23}=g_{32}=g_{56}=g_{65}=0$. 
\end{itemize}
The resulting network model is illustrated in Fig~\ref{fig4}.  Here we use non-standard notation  for the 4 main connections of interest; instead of subscripts, we use two indicative capital letters 
(e.g., $gIP$) so that readers can easily distinguish the connections we explore, vs. unexplored connections.

Thus, we were left with four undetermined coupling strengths: $gIO$, $gIP$, $gEO$ and $gEP$. We comprehensively surveyed a four-dimensional parameter space in which each coupling strength $|gIO|$, $|gIP|$, $gEO$, $gEP$ was chosen between 0.1 and 2, with a interval of 0.1, giving us $20^4 = 1.6 \times 10^5$ total models. 
Given each choice of 4-tuple $\{ gIO, gIP, gEO, gEP \}$, we computed first- and second-order statistics of both activity $x_k$ and firing rates $F(x_k)$ using the formulas given in Eq~\ref{mn_x1}--\ref{cov_56}, and 
checked whether the results satisfied the constraints listed in Table~\ref{table:constr} -- comparing the mean statistic across all 3 cells or all 3 possible pairs in various states and regions.  
We found that approximately 1.1\% of all 4-tuples satisfied the constraints; we display them in Fig~\ref{fig4}, by projecting all constraint-satisfying 
4-tuples onto a two-dimensional plane where the axes are two of the four coupling parameters. We show four out of six possible pairs (the other two show qualitatively similar patterns, see Fig S11 in \nameref{S2_file}): 
$|gIO|$ vs. $|gIP|$ (Fig~\ref{fig4}A), $gEO$ vs. $gEP$ (Fig~\ref{fig4}B), $|gIO|$ vs. $gEP$ (Fig~\ref{fig4}C), and $|gIP|$ vs. $gEO$ (Fig~\ref{fig4}D).

The results from the minimal firing rate model are:
\begin{itemize}
\item The magnitude of the inhibition within PC, $|gIP|$, is greater than the magnitude of the inhibition within OB, $|gIO|$ (Fig~\ref{fig4}A: all dots are above the diagonal line).  
\item The excitation from PC to OB, $gEP$, is generally larger than the excitation from OB to PC, $gEO$ (Fig~\ref{fig4}B).  
\item The inhibition within OB is generally weak (Fig~\ref{fig4}C: dots are to the left of the vertical line).  
\item  The inhibition within PC is generally strong (Fig~\ref{fig4}D: dots are to the right of the vertical line).  
\end{itemize}

The statistics computed in Eq~\ref{mn_x1}--\ref{cov_56} rely on the assumption that the activity distributions $x_k$ are only weakly perturbed from a normal distribution; this may be violated for larger coupling strengths.  
Thus, we used Monte Carlo simulations of Eq~\ref{eqn:dxdt_SDE} to check the accuracy of this approximation; specifically we performed Monte Carlo simulations \underline{only} 
on each 4-tuple of parameters for which the analytic approximation met our constraints.  
The resulting parameter sets that satisfied all 12 constraints are included as red dots in Fig~\ref{fig4}A--D (therefore a red dot indicates that all 12 constraints were satisfied both for the analytic approximation \textit{and} for the Monte Carlo simulations).  
The result was a smaller set of parameters, but 
it is evident that the qualitative results derived from the fast analytic solver hold for the Monte Carlo simulations.  
Moreover, these results were robust to the choice of transfer function: in Fig. S12 of \nameref{S2_file}, we show that the same constraints are obtained when using a ``square root" transfer function, rather than a sigmoid.


\subsection*{Admissible firing rate model parameters}

\begin{figure}[!h]
\centering
    \includegraphics[width=4.25in]{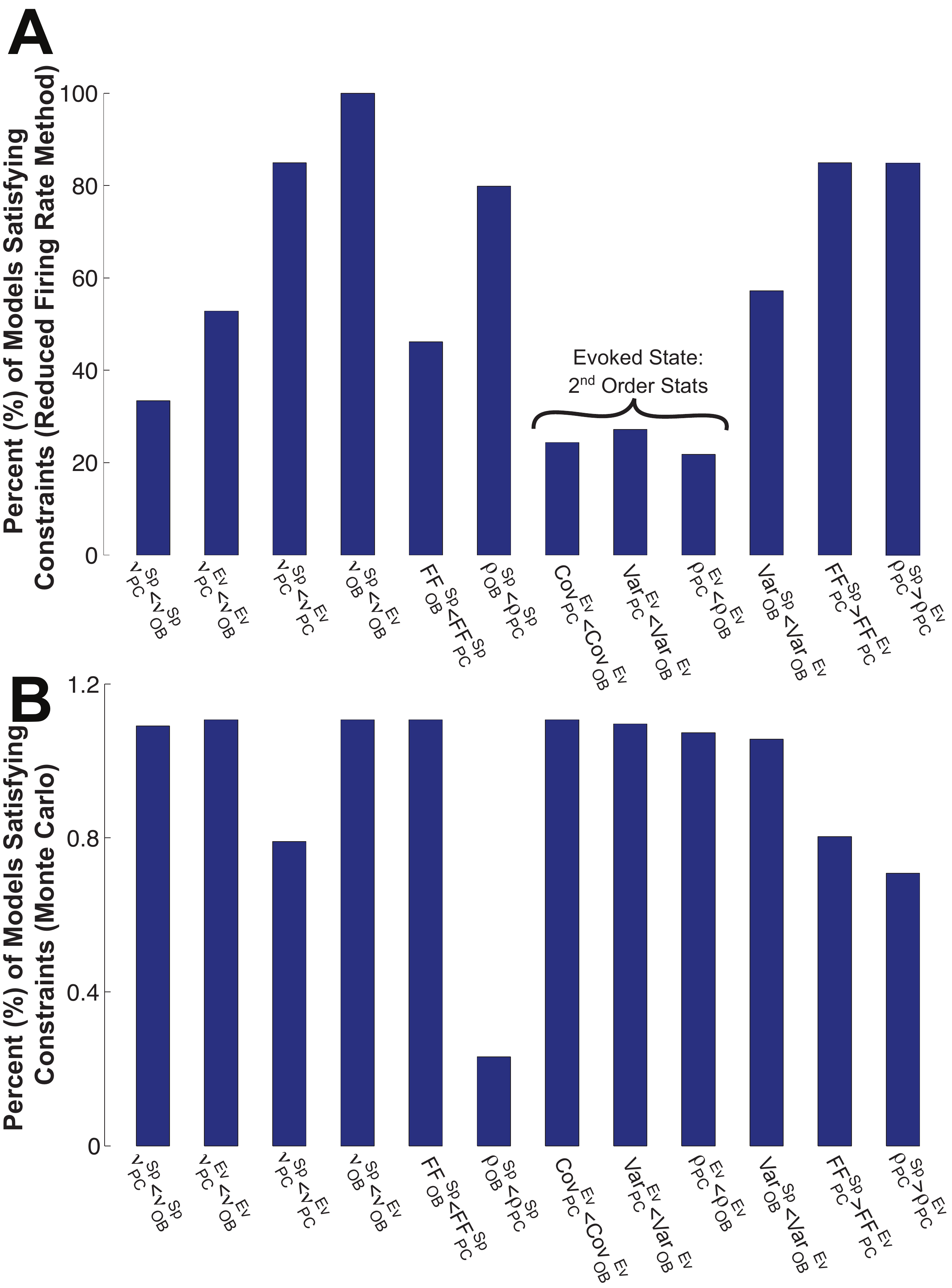}
\caption{{\bf Each constraint limits the set of admissible models.} 
(A) The percentage of all $20^4$ parameter sets that satisfy each particular constraint (for each of the 12 constraints in Table~\ref{table:constr}) in the minimal firing rate model.  See Eq~\ref{eqn:gen_WC_pop} and {\bf Materials and Methods}.  
We see that some constraints are more restrictive than others (e.g., second-order firing statistics comparing OB and PC in the evoked state --- flagged in (A) above --- are particularly restrictive).  Only 1.1\% of all parameter sets satisfy all 12 constraints.  
(B)  The percentage of models that satisfy each particular constraint, in both the reduction method \textit{and} Monte Carlo simulations. Recall that we take the relatively conservative approach of only testing the Monte Carlo simulations on the admissible set (1.1\%), thus resulting in a small fraction 
(0.13\%) of the total sets that satisfy all constraints.  
}
\label{fig5}
\end{figure}

How do each of the 12 data constraints (Table~\ref{table:constr}) restrict the set of possible model parameters? Figure~\ref{fig5} addresses this question in two ways.  In Fig~\ref{fig5}A, we show, for each constraint, the fraction (as a percent) of all $20^4$ parameter
sets for which that constraint is satisfied, when statistics are computed via the reduction method (see {\bf Materials and Methods, Approximation of Firing Statistics in the Firing Rate Model}). 
Constraints have varying levels of restrictions, but the second order firing statistics in the evoked state appear more 
restrictive than the others.  Together, only 1.1\% of the values in parameter space satisfy all 12 constraints.  

In Fig~\ref{fig5}B, we show, for each constraint, the fraction 
of all $20^4$ parameter sets for which that constraint is satisfied, in both the reduction method \textit{and} in Monte Carlo simulations (recall that we took the relatively conservative approach of only testing the Monte Carlo simulations on the admissible set 
from the reduction method (1.1\%);  this yielded only 0.13\% of parameter space. 
The constraint that $\rho_{OB}^{Sp}<\rho_{PC}^{Sp}$ has the smallest percent by far in Fig~\ref{fig5}B.  
We attribute this ``mismatch" to inaccuracies in our method
with stronger coupling (note that $gIP$ and $gEP$ are both relatively strong in the admissible set); the smaller percentages in Fig~\ref{fig5}B compared to Fig~\ref{fig5}A are likely due to errors in the Cov and Var calculations (see Fig S10 in \nameref{S2_file}), 
as well as possible amplification of these errors when dividing by Var in the $\rho$ calculation.

Another way to succinctly examine the structure of the four neural attributes: $gIO, gEO, gIP, gEP$ is to consider a matrix:
\begin{equation}\label{mat_gs}
	A(j,:) = [gIO(j),\hspace{.05in} gEO(j), \hspace{.05in}gIP(j), \hspace{.05in}gEP(j)]
\end{equation}
where the $j^{th}$ row of $A$ corresponds to a parameter set where all 12 constraints are satisfied.
We first subtract the mean, finding that $[gIO, gEO, gIP, gEP]^T=[-0.62, 1.11, -1.38, 1.29]^T$, which is consistent with the results described in Fig~\ref{fig4}.
A standard singular value composition (SVD) of the mean-corrected matrix,
$$A=U\Sigma V^T,$$ 
shows that two dimensions in the parameter space accounts for 82\,\% of the remaining variance (as quantified by the singular values) and thus provide an approximation to the structure of the valid $gIO, gEO, gIP, gEP$ values.  
The eigenvectors corresponding to the largest singular values are:  
$[gIO, gEO, gIP, gEP]^T=[-0.05, 0.60, -0.07, 0.79]^T$
and $[gIO, gEO, gIP, gEP]^T=[0.56, 0.05, 0.82, 0.08]^T$; that is, they reflect high positive correlations between the two inhibitory strengths $gIP$ and $gIO$, and between the two excitatory strengths $gEP$ and $gEO$.  Therefore, with the 
minimal firing rate model we predict the connectivity strengths generally satisfy:
$$ |gIO| < gEO < gEP <  |gIP|. $$

We next asked whether the full set of data constraints were necessary; would we have seen a similar relationship between connectivity strengths, while using only a subset of the constraints
outlined in Table~\ref{table:constr}?
Because the admissible set is defined as the intersection over all constraints, removing any constraint would likely result in a different and (if different) larger parameter space.  
We considered i) keeping 
8 of the 12 constraints in Table~\ref{table:constr}, neglecting the constraints on the {\bf Co-variability row}, and ii) keeping only 4 of the 12 constraints in Table~\ref{table:constr}, 
neglecting both the {\bf Variability} and {\bf Co-variability} rows (i.e., only with the firing rate).  Briefly, the result is that i) 21.5\% of parameters in the
analytic method satisfy the constraints; 
ii) 33.4\% of parameters in the analytic method satisfy the constraints; 
compare this to 1.1\% (and 0.13\% Monte Carlo) with all 12 constraints. The relationships of the connection strengths are different than when all 12 constraints are included: for example, it is no longer true that $gEP > gEO$, once the covariance constraints are omitted.
%

\subsection*{Generality of Firing Rate Model Predictions}

In general, we should expect that if we change the wiring diagram of our simple firing rate model (Fig~\ref{fig4}), then the same experimental constraints might result in different predictions.  
This could be a concern since our simple firing rate model is lacking many connections and cell types that exist in the real olfactory system~\cite{oswald12}.  
However, we tested one alternative wiring diagram with different neurons receiving stimulus input, no E-to-I connections within OB, and no E-to-I connections within PC.  
Our predictions were robust to these changes. Second and most importantly, we tested whether our predictions held in a larger network of leaky integrate-and-fire neurons.  
This spiking network model also had more realistic network connectivity, more closely mimicking known anatomy of real olfactory systems.  

The following highlight the differences between the spiking model and the firing rate model:
\begin{itemize}
	\item Include E-to-E connections from OB to PC (lateral olfactory tract).  Also include strong E-to-I drive within PC because input from OB results in balanced 
	excitation and inhibition in PC~\cite{large16};
	\item Remove the E-to-I connections from OB to PC ($gEO$ in the firing rate model) so that the recurrent activity in PC is driven by E inputs along the lateral olfactory tract;
	\item Remove the direct sensory input to I cells in OB since granule cells do not receive direct sensory input~\cite{oswald12};
	\item Include substantial recurrent E-to-E connections within PC (see Table~\ref{table:lif_parms} for strength relative to other connections). 
\end{itemize}

The parameter $gEO$ will now refer to the strength of E-to-E connections, rather than E-to-I connections, from OB to PC. The next two sections demonstrate that our predictions hold for this LIF network model (also see \nameref{S3_file}).

\subsection*{Results are Validated in a Spiking LIF Network}

Here we show that a general leaky integrate-and-fire ({\bf LIF}) spiking neuron model of the coupled OB-PC system can satisfy all 12 data constraints.  
Rather than try to model the exact underlying physiological details of the olfactory bulb or anterior piriform cortex, 
our goal is to demonstrate that the results from the minimal firing rate model can be used as a guiding principle in a more realistic coupled 
spiking model with conductance-based synaptic input.  The LIF model does not contain all of the attributes and cell types of the olfactory system, but is a plausible model that contains: 
i) more granule than M/T cells in OB (a 4-to-1 ratio, comparable to the 3-to-1 ratio used in ~\cite{grabska17}); 
ii) E-to-E connections from OB to PC that drive the entire network within PC; iii) E-to-I (granule cell) feedback from PC to OB; iv) lack of 
sensory input to granule I cells in OB.  

We also show that the minimal firing rate model results can be applied to a generic cortical-cortical coupled population (see ~\nameref{S3_file}).

\begin{figure}[!h]
    \includegraphics[width=\columnwidth]{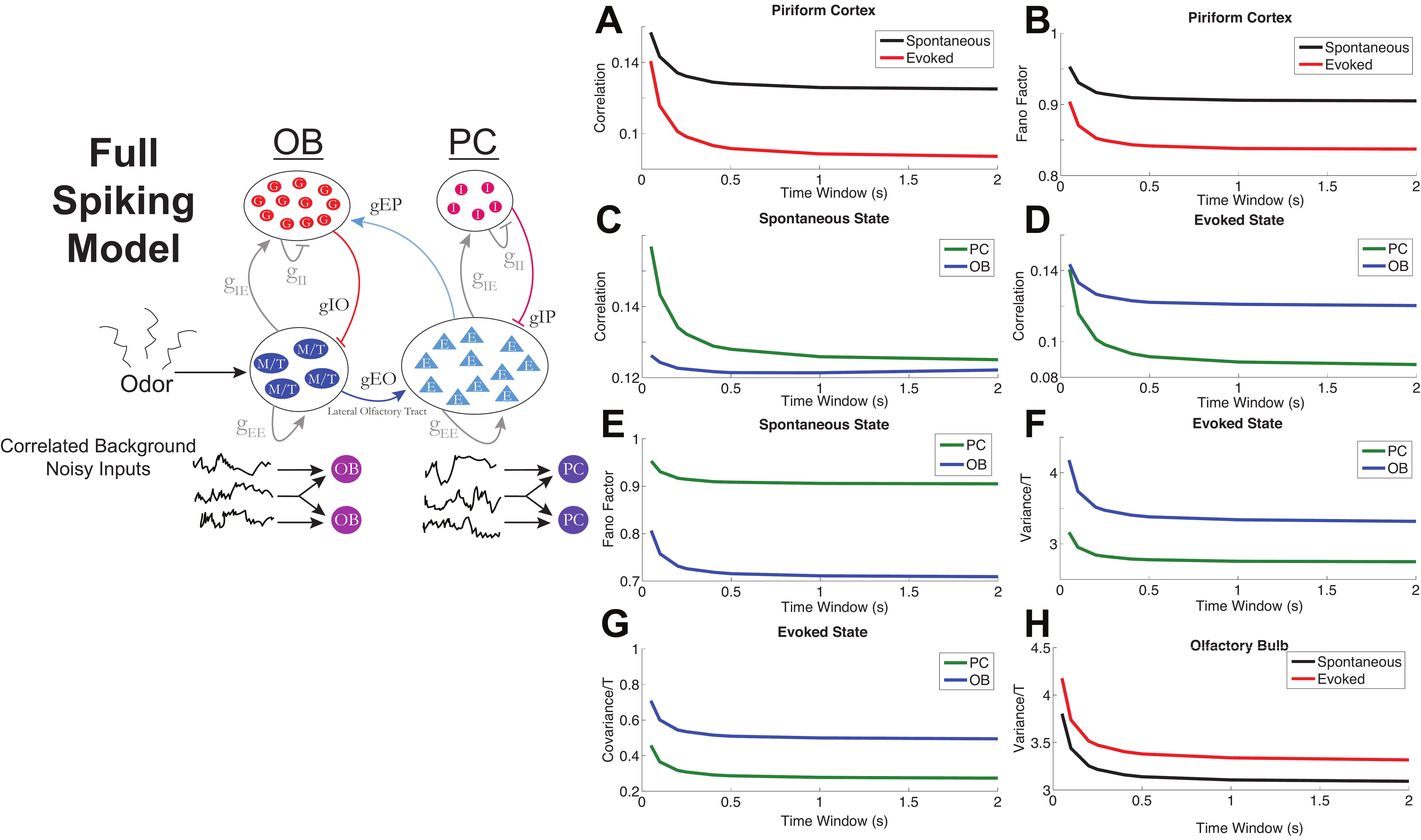}
\caption{{\bf Detailed spiking LIF model confirms the results from analytic rate model.}
Schematic of the LIF model with 2 sets of recurrently coupled E and I cells.  There are 12 types of synaptic connections.  
(A) Pairwise correlations in PC, spontaneous vs. evoked: $\rho_{PC}^{Sp}>\rho_{PC}^{Ev}$.  
(B) Variability (Fano factor) in PC, spontaneous vs evoked: $FF_{PC}^{Sp}>FF_{PC}^{Ev}$.  
(C) Correlations in the spontaneous state, PC vs. OB: $\rho_{PC}^{Sp}>\rho_{OB}^{Sp}$.
(D) Correlations in the evoked state, PC vs. OB: $\rho_{PC}^{Ev}<\rho_{OB}^{Ev}$.  
(E) Variability (Fano factor) in the spontaneous state, PC vs. OB: $FF_{PC}^{Sp}>FF_{OB}^{Sp}$.  
(F) Variability (Fano factor) in the evoked state, PC vs. OB: $\text{Var}_{PC}^{Ev}<\text{Var}_{OB}^{Ev}$ in evoked state.  
(G) Covariances in the evoked state, PC vs. OB: $\text{Cov}_{PC}^{Ev}<\text{Cov}_{OB}^{Ev}$.
(H) Variability (spike count variance) in OB, spontaneous vs. evoked: $\text{Var}_{OB}^{Sp} < \text{Var}_{OB}^{Ev}$.  
The curves show the average statistics over all $N_{OB/PC}$ cells or over a large random sample of all possible pairs.  
See {\bf Materials and Methods} for model details, and Table~\ref{table:lif_parms} and Eq~\ref{def_gs_lif} for parameter values.
}
\label{fig6}
\end{figure}

We set the four conductance strength values to:
\begin{eqnarray}
	gIO &=& 7		\nonumber \\
	gEO &=& 	10	\nonumber \\
	gIP &=& 	20	\nonumber \\
	gEP &=& 	15	; \label{def_gs_lif}
\end{eqnarray}
See Fig~\ref{fig6} or Eq~\ref{ob_lif}--\ref{pc_lif} for exact definitions of $gXY$; these conductance strength values are dimensionless scale factors.  
These values were selected to satisfy the relationships derived from the analysis of the rate model (see Fig~\ref{fig4}).  
In contrast to the minimal firing rate model, here the conductance values are all necessarily positive; an inhibitory reversal potential is used to capture the hyperpolarization that occurs upon receiving synaptic input.  

With the conductance strengths in Eq~\ref{def_gs_lif}, and other standard parameter values (see Table~\ref{table:lif_parms}) in a typical LIF model, we were able to easily satisfy all 12 constraints: see 
Table~\ref{table:frate_lif} and Fig~\ref{fig6}.  

\begin{table}[!ht]
\centering
\caption{ {\bf Population firing rate statistics from an LIF model of the OB--PC pathway. }}
\label{table:frate_lif}
\begin{tabular}{|c+c|c|}
\hline
\multicolumn{1}{|l|}{} & Mean Firing Rate (Hz) & Std. Dev. (Hz) \\ \thickhline
$\nu_{OB}^{Sp}$        & 5.5                   & 4.6            \\
$\nu_{OB}^{Ev}$        & 6.2                   & 4.8            \\ \hline
$\nu_{PC}^{Sp}$        & 2.1                   & 2.6            \\
$\nu_{PC}^{Ev}$        & 4.1                   & 5.8            \\ \hline
\end{tabular}
\begin{flushleft} See {\bf Materials and Methods} for model details, and Table~\ref{table:lif_parms} and Eq~\ref{def_gs_lif} for parameter values.  The mean and standard deviations are across the heterogeneous population.
\end{flushleft}
\end{table}

While the firing rates in the LIF network (Table~\ref{table:frate_lif}) do not \textit{quantitatively} match with the firing rates from the experimental data, a few \textit{qualitative} trends are apparent: (i) the ratio of mean 
spontaneous to evoked firing rates are similar to that observed in experimental data, for both OB and PC, (ii) the same is true of the standard deviation, (iii) the ratio of the mean OB firing rate to PC firing rate is 
similar to what is observed in the experimental data, in both spontaneous and evoked states.  Therefore, the LIF network captures the mean firing rates reasonably well.  


One difference between the LIF spiking network and the minimal firing rate model is that in the evoked state, mean background input to \textit{both} the OB and PC cells is increased, compared to the 
spontaneous state (recall that in the minimal firing rate model, only the mean input to the OB cells increased in the evoked state; this ensured that stimulus-induced changes in 
PC were due to network activity).  When the mean input to the PC cells is the same in the spontaneous and evoked states, 
10 of the 12 constraints were satisfied -- the exception was the correlation of PC in the evoked state, which decreased but is still larger than the spontaneous correlation 
(see Fig. S13 in \nameref{S2_file}). The reason is that as firing rates increase, the OB spiking is more variable and the synaptic input from OB to PC is noisier, so the input to PC activity is diffused.  

To capture the final two constraints, we allowed mean input drive to PC to increase in the evoked state. 
This has also been used in previous theoretical studies to achieve stimulus-induced decreases in spiking variability and co-variability~\cite{litwin_nn_12}.  
Churchland et al.~\cite{churchland10} used an extra source of variability in the spike generating mechanism, a doubly stochastic model, which was 
simply removed with stimulus onset.  
Thus, the mechanism we employ (increased mean input with lower input variability) is consistent with other studies that analyzed stimulus-induced changes in variability~\cite{churchland10,litwin_nn_12}.

\subsubsection*{Results of Violating Derived Relationships Between Conductance Strengths}

What happens in the full LIF spiking network when the derived relationships between the conductance strengths are violated?  Since the minimal firing rate model is different than the 
detailed spiking model in many ways, we do not expect the relationships between the conductance strengths to hold precisely.   However, 
the minimal firing rate model is still useful in providing intuition for what would otherwise be a complicated network 
with a high-dimensional parameter space.
We now demonstrate that when the relationships derived in firing rate model are violated,  
a subset of the constraints in the experimental data (Table~\ref{table:constr}) will no longer be satisfied in the large spiking network.

Because our network is heterogeneous, our ability to subsample cell pairs is limited, relative to a homogeneous network of the same size.  Also, computation for even a single 
parameter set in the spiking network require enormous computing resources.  
Thus, we cannot exhaustively explore the parameter space; indeed, the purpose of the reduction method of the firing rate model is to probe large dimensions quickly.  Instead, we perform three tests that violate the firing rate model results:
\begin{enumerate}
	\item Make $gIO > g IP$ by setting $gIO=20$ and $gIP=7$.
	\item Make $gEO > gEP$ by setting $gEO=15$ and $gEP=1$
	\item Make $gEP$ and $gIP$ relatively smaller by setting $gEP=10$ and $gIP=10$
\end{enumerate}
The original values (used in Fig~\ref{fig6}) for these parameters were given in Eq~\ref{def_gs_lif}.

The result of Test 1 is that 2 of the 12 constraints are violated (see Fig S14 in \nameref{S2_file}); most importantly stimulus-induced decorrelation of the PC cells, which is particularly important in the context of coding, was not present.  
In addition, 
the evoked PC correlation is larger than evoked OB correlation, 
violating another constraint.

The result of Test 2 is that 3 of the 12 constraints are violated (see Fig S15 in \nameref{S2_file}).  
The evoked PC correlation is larger than evoked OB correlation, and 
both the variance and covariance in PC are larger than the corresponding quantities in OB in the evoked state, which is not consistent with our data. 

The result of Test 3 is that 3 of the 12 constraints are violated: they are the same constraints that are violated in Test 2, despite quantitative differences in the statistics (see Fig S16 in \nameref{S2_file}).  
The stimulus-induced decorrelation of the PC cells does not hold for small windows, but this is also observed in our data (Fig~\ref{fig2}A), so we do not formally count this as a clear violation of data constraints.  
However, Test 1 and Test 3 show that strong PC inhibition is key for stimulus-induced decorrelation~\cite{doiron16,diesmann12,middleton12,LMD_whisker_12,litwin12,litwin11,renart10}.
 
\section*{Discussion}

As electrophysiological recording technology advances, there will be more datasets with simultaneous recordings of neurons, spanning larger regions of the nervous system.  
Such networks are inherently high-dimensional, making mechanistic analyses generally intractable without fast and reasonably accurate approximation methods. 
We have developed a computational reduction method for a multi-population firing rate model~\cite{wilsoncowan1} 
 that enables analysis of the spiking statistics.  Our work specifically enables theoretical characterizations of an important, yet hard-to-measure quantity -- synaptic connection strength -- using easy-to-measure 
 spiking statistics.  The method is computationally efficient, is validated with Monte Carlo simulations of spiking neural networks, and can provide insight into network structure.  

We applied our computational methods to 
simultaneous dual-array recordings in two distinct regions of the olfactory system: the olfactory bulb (OB) and anterior piriform cortex (PC).  
Our unique experimental dataset enables a detailed analysis of the first- and second-order spike count statistics in two activity states,
and a comparison of how these 
statistics are related between OB and PC cells.  We found twelve (12) consistent trends that held across four odors in the dataset (Table~\ref{table:constr}), and sought to identify 
what neural network attributes would account for these trends. 
We focused on four important network attributes, specifically the conductance strengths in the following connections: feedforward inhibition within OB and within PC, excitatory projections from OB to PC neurons, and finally 
excitatory projections from PC to OB.  Our reduced firing rate model predicts several relationships that are then verified with a more detailed spiking network model, specifically: 
i) inhibition within the OB has to be weaker than the inhibition in PC, ii) excitation from PC to OB is generally stronger than excitation from OB to PC, 
iii) connections that originate within PC have to relatively strong compared to connections that originate within OB.
These results make a strong prediction that to the best of our knowledge is new and might be testable with simultaneous patch-clamp recordings.  
	
In principle our theory could be used to study the structure of other network features such as background correlation, noise level, transfer function, etc..  
It is straightforward mathematically to incorporate other desired neural attributes (with the caveat of perhaps increasing the overall number of equations and terms in the approximations) without changing the basic structure of the framework.  
Here we have focused on the role of the strength of synaptic coupling; 
of course, other neural attributes can affect spike statistics (in particular, spike count correlation~\cite{cohen11,doiron16}),  some of which can conceivably change with stimuli.  
Spike count correlations can depend on intrinsic neural properties~\cite{hong12,marella08,abouzeid09,barreiro10,barreiro12,ocker14}, network architecture~\cite{rosenbaum10,litwin_nn_12,rosenbaum17} 
and synaptic inputs~\cite{renart10,diesmann12,litwin11,litwin12,middleton12,LMD_whisker_12} 
(or combinations of these~\cite{ostojic09,ly_ermentrout_09,trousdale12,BarreiroLy_RecrCorr_17}), plasticity~\cite{rosenbaum13}, as well as top-down mechanisms~\cite{mitchell09,cohen09,ruff14}.  Thus, correlation 
modulation is a rich and deep field of study, and we do not presume our result is the only plausible explanation for spike statistics modulation.

Although the minimal firing rate model did not include certain anatomical connections that are known to exist (e.g., recurrent excitation in the PC), 
the model is meant for deriving qualitative principles rather than precise quantitative modeling of the 
pathway.  We based our simplifications on insights from recent experimental work: 
recent slice physiology work 
has shown that within PC, recurrent activity is dominated by inhibition~\cite{large16};  previous 
work has also shown that inhibitory synaptic events are much more common (than excitatory synaptic events) in PC and are much easier to elicit~\cite{poo09}.  Thus, the connection from excitatory OB cells to inhibitory PC cells ($gEO$ in Fig~\ref{fig4}) 
should be thought of as the net effect of these connections along the lateral olfactory tract.  
Other theoretical analyses of effective feedforward inhibitory networks have also neglected anatomical E-to-E connections~\cite{middleton12,LMD_whisker_12}.  Furthermore, this minimal model 
was validated with a more realistic, recurrently coupled spiking network, which did include within-region excitatory connections (see Fig~\ref{fig6} and Fig S14--S16 in \nameref{S2_file}, as well as \nameref{S3_file}).

%

We have only focused on first- and second-order firing statistics, even though in principle other, higher-order statistics may be important~\cite{ohiorhenuan10,trousdale13,jovanovic16}.   
If downstream neurons use a linear decoding scheme,
 then first- and second-order spiking statistics are sufficient in quantitative measures of neural coding~\cite{kaybook,dayan2001theoretical}.  
It is currently unknown whether downstream neurons decode olfactory signals with a nonlinear decoder, but there is evidence in other sensory systems that second-order statistics are sufficient~\cite{kohn16}.  
Recent work has shown conflicting results for coding in olfactory bulb; one study found that decoding an odor in the presence of other odors might be more efficient using nonlinear decoding~\cite{grabska17}, but another has shown that linear decoding is still plausible~\cite{mathis16}.  

A second reason to neglect higher-order statistics is suggested by Fig. 5, where we show how the various data constraints narrow the scope of plausible models. Here, we saw that even with first and second- order statistics, only 1\% of the parameter sets satisfy the data constraints; including more constraints would limit the space further. In order to usefully include higher-order constraints, we would need to use a more detailed model and/or larger 
parameter spaces.
%

As a test case for our method, we used recordings from anesthetized animals.  The absence of breathing in tracheotomized rats 
in these experiments is only an approximation to olfactory processing in awake animals. 
However, there is a benefit to tracheotomized animals: the 
complex temporal firing patterns are removed, 
so that firing statistics are closer to stationarity.   
In principle, we can incorporate breathing dynamics into our framework by including an oscillatory forcing term in Eq~\ref{eqn:gen_WC_pop}; this will be the subject of future work.  
In support of this simplification, we note that there is evidence that in the anterior piriform cortex, spike count --- rather than the timing --- is most consequential for odor discrimination~\cite{miura12}. 
However, other studies have reported that timing of the stimuli in the olfactory bulb is important:~\cite{cury10,gschwend12,grabska17} 
showed decoding performance is best at the onset of odors in mammals and worsens as time 
proceeds, whereas~\cite{friedrich01} found that decoding performance improved with time in zebrafish.  These important issues are beyond the scope of this current study.

\subsection*{Relationship to Other Reduction Methods}

In computing statistics for the minimal firing rate model, we only considered equilibrium firing statistics, in which a set of stationary statistics can be solved self-consistently.  
More sophisticated methods might be used to address oscillatory firing statistics (see~\cite{nlc_15} where the adaptive quadratic integrate-and-fire model was successfully analyzed with a reduced method); capturing the firing statistics in these other regimes is a potentially interesting direction of research.  The limitation to steady-state statistics is not unique, but is shared by other approximation methods.  
Some methods are known to have issues when the system bifurcates~\cite{buice07,buice10} because truncation methods can fail~\cite{ly_tranchina_07}.  

Several authors have proposed procedures to derive population-averaged first- and second-order 
spiking statistics from the dynamics of single neurons. The microscopic dynamics in question may be given by a master equation ~\cite{buice07,bressloff09,buice10,touboul11,bressloff15}, a generalized linear model \cite{toyoizumi09,ocker2017}, or the theta model~\cite{BC_JSM_2013,BC_PLOSCB_2013}. 
(Other authors have derived rate equations at the single-neuron level, by starting with a spike response model \cite{aviel06} or  by taking the limit of slow synapses~\cite{ermentrout94}.) While we would ideally use a similar procedure to derive our rate equations, none of the approaches we note here is yet adapted to deal with our setting, a heterogeneous network of leaky integrate-and-fire neurons. Instead, we focused here on perturbing from a background state in which several populations 
(each population modeled by a single equation) 
receive correlated background input but are otherwise uncoupled. 
This allows us to narrow our focus to how spike count co-variability from common input is modulated by recurrent connections. 
%
%

We also note that other recent works have used firing rate models to explain observed patterns of correlated spiking activity in response to stimuli. 
Rosenbaum et al.~\cite{rosenbaum17} have studied the spatial structure of correlation in primate visual cortex with balanced networks~\cite{van96}; 
Keane \& Gong~\cite{keaneGong15} studied wave propagation in balanced network models. 
 



\section*{Conclusion}

Designing a spiking neural network model of two different regions that satisfies the many experimental data constraints we have outlined is a difficult problem that would often be addressed 
via undirected simulations.  We have shown that 
systematic analysis of a minimal firing rate model can yield valuable insights into the relative strength of unmeasured network connections. Furthermore, these insights are transferable to a more complex, physiologically realistic spiking model of the OB--PC pathway. 
Indeed, incorporating the relative relationships of the four conductance strengths resulted in spiking network models that satisfied the constraints.  
Strongly violating the relative relationships of these conductance strengths led to multiple violations of the data constraints.  Because our approach can be extended to other network features, we are hopeful that
the general approach we have developed -- using easy-to-measure quantities to predict hard-to-measure interactions -- will be valuable in future investigations into how whole-brain function emerges from interactions among its constituent components.

%

\section*{Materials and Methods}

\subsection*{Electrophysiological Recordings}

 

{\bf Subjects.} All procedures were carried out in accordance with the recommendations in the Guide for the Care and Use of Laboratory Animals of the National Institutes of Health and approved by 
University of Arkansas Institutional Animal Care and Use Committee (protocol \#14049). Experimental data was obtained from one adult male rat (289\,g ; \textit{Rattus Norvegicus}, Sprague-Dawley 
outbred, Harlan Laboratories, TX, USA) housed in an environment of controlled humidity (60\%) and temperature (23$^{\circ}$C) with 12\,h light-dark cycles. The experiments were performed in the light phase. 

\vspace{.1in}
\hspace{-.25in} {\bf Anesthesia.} Anesthesia was induced with isoflurane inhalation and maintained with urethane (1.5\,g/kg body weight ({\bf bw}) dissolved in saline, intraperitoneal injection ({\bf ip})). 
Dexamethasone (2\,mg/kg bw, ip) and atropine sulphate (0.4\,mg/kg bw, ip) were administered before performing surgical procedures.

\vspace{.1in}
\hspace{-.25in} {\bf Double tracheotomy surgery.} To facilitate ortho- and retronasal delivery of the odorants a double tracheotomy surgery was performed as described previously \cite{gautam12}. This 
allowed for the rat to sniff artificially while breathing naturally through the trachea bypassing the nose. A Teflon tube (OD 2.1\,mm, upper tracheotomy tube) was inserted 10\,mm into the nasopharynx through the rostral 
end of the tracheal cut.  Another Teflon tube (OD 2.3\,mm, lower tracheotomy tube) was inserted in to the caudal end of the tracheal cut to allow breathing. Both tubes were fixed and sealed to the tissues using surgical 
thread. Local anesthetic (2\% Lidocaine) was applied at all pressure points and incisions. 
Throughout the surgery and electrophysiological recordings rats' core body temperature was maintained at 37$^{\circ}$C with a thermostatically controlled heating pad.

\vspace{.1in}
\hspace{-.25in} {\bf Craniotomy surgery.} Subsequently, a craniotomy surgery was performed on the dorsal surface of the skull at two locations, one over the right Olfactory Bulb 
(2\,mm $\times$ 2\,mm, centered 8.5\,mm rostral to bregma and 1.5\,mm lateral from midline) and the other over the right anterior Pyriform Cortex (2\,mm $\times$ 2\,mm, centered 1.5\.mm caudal to bregma and 
5.5\,mm lateral from midline).

\vspace{.1in}
\hspace{-.25in} {\bf Presentation of ortho- and retronasal odorants.} The bidirectional artificial sniffing paradigm previously used for the presentation of ortho- and retronasal odorants \cite{gautam12} 
were slightly modified such that instead of a nose mask a Teflon tube was inserted into the right nostril and the left nostril was sealed by suturing.  The upper tracheotomy tube inserted into the nasopharynx was used 
to deliver odor stimuli retronasally (Fig~\ref{fig1}. We used two different odorants, Hexanal ({\bf Hexa}) and Ethyl Butyrate ({\bf EB}) by both ortho- and retronasal routes, there by constituting 4 
different odor stimuli. Each trial consisted of 10 one-second pulse presentations of an odor with 30 second interval in between two pulses, and 2-3 min in between two trials.

\vspace{.1in}
\hspace{-.25in} {\bf Electrophysiology.} Extracellular voltage was recorded simultaneously from OB and aPC using two different sets of 32-channel microelectrode arrays ({\bf MEAs}).
 (OB: A4x2tet, 4 shanks x 2 iridium tetrodes per shank, inserted 400 $\mu$m deep from dorsal surface; aPC: Buzsaki 32L, 4 shanks x 8 iridium electrode sites per shank, 
 6.5\,mm deep from dorsal surface; NeuroNexus, MI, USA). Voltages were measured with respect to an AgCl ground pellet placed in the saline-soaked gel foams covering the exposed brain surface around the inserted 
 MEAs. Voltages were digitized with 30\,kHz sample rate as described previously \cite{gautam15} using Cereplex + Cerebus, Blackrock Microsystems (UT, USA). 

Recordings were filtered between 300 and 3000\,Hz and semiautomatic spike sorting was performed using Klustakwik software, which is optimized for the types of electode arrays used here \cite{rossant16}.  
After automatic sorting, each unit was visually inspected to ensure quality of sorting. 

\subsection*{Data processing}

After the array recordings were spike sorted to identify activity from distinct cells, we further processed the data as follows:
\begin{itemize}
\item We computed average firing rate for each cell, where the average was taken over all trials and over the entire trial length (i.e., not distinguishing between spontaneous and evoked periods); units with firing rates below 0.008\,Hz  and above 49\,Hz were excluded. 
	\item When spike times from the same unit were within 0.1\,ms of each other, only the first (smaller) of the spike time was used and the subsequent spike times were discarded
\end{itemize}

We divided each 30\,s trial into two segments, representing the odor-{\bf evoked} state (first 2 seconds) and the {\bf spontaneous} state (remaining 28 seconds).  
In each state, we are interested in the random spike counts of the population in a particular window of size $T_{win}$.  For a particular time window, 
the $j^{th}$ neuron has a spike count instance $N_j$ in the time interval $[t,t+T_{win})$:
\begin{equation}\label{n_cnt}
	N_j = \sum_k \int_{t}^{t+T_{win}} \delta(t-t_k)\,dt
\end{equation}

The spike count correlation between cells $j$ and $k$ is given by:
\begin{equation}\label{rho_defn}
	\rho_{T} = \frac{ \text{Cov}(N_j,N_k) }{ \sqrt{ \text{Var}(N_j) \text{Var}(N_k) } },
\end{equation}
where the {\it covariance} of spike counts is:  
\begin{equation}\label{cov_defn}
	\text{Cov}(N_j,N_k) = \frac{1}{n-1} \sum  \left( N_j - \mu(N_j) \right) \left( N_k - \mu(N_k) \right).
\end{equation}
 Here $n$ is the total number of observations of $N_{j/k}$, and $\mu(N_j):=\frac{1}{n}\sum N_j$ is the mean spike count across $T_{win}$-windows and trials.  
 The correlation $\rho_{T}$  is a normalized measure of the the trial-to-trial variability (i.e., noise correlation), satisfying $\rho_{T}\in[-1,1]$; it is also referred to as the {\it Pearson's correlation coefficient}. 
For each cell pair, the covariance $\text{Cov}(N_j,N_k)$ and variance  $\text{Var}(N_j)$ are empirically calculated by averaging across different time windows within a trial {\it and} different trials.  

A standard measure of 
variability is the Fano Factor of spike counts, which is the variance scaled by the mean:
\begin{equation}\label{FF_defn}
	FF_k = \frac{\text{Var}(N_k)}{\mu(N_k)}.
\end{equation}

In principle, any of the statistics defined here might depend on the time $t$ as well as time window size $T_{win}$; here, we assume that $\text{Var}$, $\text{Cov}$, $FF$, and $\rho_T$ are stationary in time, and thus separate time windows based only on whether they occur in the evoked (first 2 seconds) 
or spontaneous (last 28 seconds) state.  

Each trial of experimental 
data has many time windows\footnote{an exception is when $T_{win}=2\,$s; in the evoked state, there is only 1 window per trial}; the exact number depends on the state, the value of $T_{win}$, and whether 
disjoint or overlapping windows are used.  In this paper we use overlapping windows by half the length of $T_{win}$\footnote{e.g. if the trial length is 2\,s and $T_{win}=1\,$s, then there are 3 total windows per trial: [0\,s, 1\,s], [0.5\,s, 1.5\,s], and [1\,s, 2\,s]} 
to calculate the spiking statistics.  The results are qualitatively similar for disjoint windows and importantly the relationships/constraints are the same with disjoint windows.  We limit the size of $T_{win}\leq 2\,$s 
because this is the maximum duration of the evoked state, within each trial.


The average spike count $\mu(N_j)$ of the $j^{th}$ neuron with a particular time window $T_{win}$ is related to the average firing rate $\nu_j$ of that neuron:
\begin{equation}\label{frate_defn}
	\nu_j := \frac{\mu(N_j)}{T_{win}}
\end{equation}

\subsection*{Firing Rate Model}

Recall that the activity in each representative cell is modeled by:
\begin{equation}\label{frate_firsteqn}
	\tau \frac{d x_j}{dt}  =  -x_j + \mu_j + \sigma_j \eta_j + \sum_k g_{jk} F(x_k) 
\end{equation}
where $F(x_k)$ is a transfer function mapping activity to firing rate.  Thus, the firing rate is:
\begin{equation}
	\nu_j = F(x_j).  
\end{equation}

The index of each region is denoted as follows: $j\in\{1,2,3\}$ for the 3 OB cells, and $j\in\{4,5,6\}$ for the 3 PC cells, with $j=1$ as the inhibitory granule OB cell and $j=4$ as the inhibitory PC cell (see Fig~\ref{fig4}).  In 
this paper, we set $\sigma_1=\sigma_2=\sigma_3=\sigma_{OB}$ and $\sigma_4=\sigma_5=\sigma_6=\sigma_{PC}$ (see Table~\ref{table:rateMod_parms}).  

\begin{table}[!ht]
\caption{ {\bf Parameters of the rate model (Eq~\ref{eqn:gen_WC_pop}).  The only difference between the spontaneous and evoked states, is that the mean input to OB increased in the evoked state. 
We set $\tau=1$ throughout.}}
\label{table:rateMod_parms}
\begin{tabular}{l|c|l|cc|}
\cline{2-5}
                                                       & \multicolumn{1}{l|}{\textbf{Parameter}} & \multicolumn{1}{c|}{\textbf{Definition}}       & \textbf{Spontaneous Value} & \multicolumn{1}{l|}{\textbf{Evoked Value}} \\ \hline
\multicolumn{1}{|l|}{\multirow{5}{*}{Olfactory Bulb}}  & $\mu_1$                                 & \multicolumn{1}{c|}{Mean Input}                & 13/60                      & \textbf{26/60}                                      \\
\multicolumn{1}{|l|}{}                                 & $\mu_2$                                 & \multicolumn{1}{c|}{}                          & 9/60                       & \textbf{18/60}                                      \\
\multicolumn{1}{|l|}{}                                 & $\mu_3$                                 &                                                & 7/60                       & \textbf{14/60}                                      \\
\multicolumn{1}{|l|}{}                                 & $\sigma_{OB}$                           & Background Noise Level                         & 1.4                        & 1.4                                        \\
\multicolumn{1}{|l|}{}                                 & $c_{OB}$                                & \multicolumn{1}{c|}{OB Background Correlation} & 0.3                        & 0.3                                        \\ \hline
\multicolumn{1}{|l|}{\multirow{5}{*}{Piriform Cortex}} & $\mu_4$                                 & \multicolumn{1}{c|}{Mean Input}                & 9/60                       & 9/60                                       \\
\multicolumn{1}{|l|}{}                                 & $\mu_5$                                 &                                                & 5/60                       & 5/60                                       \\
\multicolumn{1}{|l|}{}                                 & $\mu_6$                                 &                                                & 3/60                       & 3/60                                       \\
\multicolumn{1}{|l|}{}                                 & $\sigma_{PC}$                           & Background Noise Level                         & 2                          & 2                                          \\
\multicolumn{1}{|l|}{}                                 & $c_{PC}$                                & PC Background Correlation                      & 0.35                       & 0.35                                       \\ \hline
\end{tabular}
\end{table}

In the absence of coupling (i.e. $g_{jk} = 0$), any pair of activity variables, $(x_j,x_k)$, are bivariate normally distributed because the equations:
\begin{eqnarray}
	\tau \frac{d x_j}{dt} & = & -x_j + \mu_j + \sigma_j \left( \sqrt{1-c_{jk}}\xi_j(t) + \sqrt{c_{jk}} \xi_c(t) \right) \\
	\tau \frac{d x_k}{dt} & = & -x_k + \mu_k + \sigma_k \left( \sqrt{1-c_{jk}}\xi_k(t) + \sqrt{c_{jk}} \xi_c(t) \right) 
\end{eqnarray}
describe a multi-dimensional Ornstein-Uhlenbeck process~\cite{gardiner}.  Note that we have re-written $\eta_{j/k}(t)$ as sums of independent white noise processes $\xi(t)$, which is always possible for Gaussian white noise.  
Since $x_j(t) = \frac{1}{\tau}\int_0^t e^{-(t-u)/\tau} \Big[ \mu_j + \sigma_j\eta_j(u) \Big]\,du$, we calculate marginal statistics as follows: 
\begin{equation}
	\mu(j) \equiv \langle x_j  \rangle  =  \mu_j + 0  \label{eqn:mu_uncoupled}
\end{equation}

\begin{eqnarray*}
	\sigma^2(j) & \equiv & \langle (x_j - \mu(j) )^2 \rangle \\
	 & = & \left\langle   \frac{\sigma^2_j}{\tau^2} \int_0^t \int_0^t e^{-(t-u)/\tau} \eta_j(u) e^{-(t-v)/\tau} \eta_j(v) \,du\,dv  \right\rangle \\
	 & = &\frac{\sigma^2_j}{\tau^2} \lim_{t\to\infty} \int^t_0 e^{-2(t-u)/\tau} \,du=\frac{\sigma^2_j}{2\tau}
\end{eqnarray*}

A similar calculation shows that in general we have:
\begin{equation}
	\text{Cov}(j,k) = \frac{c_{jk}}{2\tau} \sigma_j \sigma_k  \label{eqn:cov_uncoupled}
\end{equation}

Thus, $(x_j,x_k)\sim \mathcal{N}\left( \left(\begin{smallmatrix}\mu_j \\ \mu_k \end{smallmatrix}\right) , \frac{1}{2\tau} \left(\begin{smallmatrix} \sigma^2_j & \sigma_j\sigma_k c_{jk} \\ \sigma_j\sigma_k c_{jk} & \sigma^2_k \end{smallmatrix}\right) \right)$.

To simplify notation, we define:
\begin{eqnarray}
	\rho_{SN}(y) &:=& \frac{1}{\sqrt{2\pi}} e^{-y^2/2}, \hbox{ the standard normal PDF} \\
	\rho_{2D}(y_1,y_2) &:=& \frac{1}{2\pi\sqrt{1-c_{jk}^2}} \exp\Big( -\frac{1}{2}\vec{y}^T  \left(\begin{smallmatrix} 1 & c_{jk} \\ c_{jk} & 1 \end{smallmatrix}\right)^{-1} \vec{y} \Big), \hbox{ bivariate standard normal} \nonumber \\
					 & &
\end{eqnarray}
With coupling, an exact expression for a joint distribution for $(x_1, x_2, x_3, x_4, x_5, x_6)$ is not explicitly known. However, we can estimate this distribution (and any derived statistics, such as means and variances) using Monte Carlo simulations.  All 
Monte Carlo simulations of the six (6) coupled SDEs were performed using a time step of 0.01 with a standard Euler-Maruyama method, for a time of 500 units 
(arbitrary, but relative to the characteristic time scale $\tau=1$) for each of the 3000 realizations.  
The activity $x_j$ was sampled at each time step after an equilibration period. 

Furthermore, we can approximate moments of the joint distribution under the assumption of weak coupling, as described in the next section.

\subsection*{Approximation of Firing Statistics in the Firing Rate Model}
We will now show how to compute approximate first and second order statistics for the firing rate model 
\textit{with coupling}; i.e., we aim to compute the mean activity $\langle x_j \rangle$, mean firing rate $\langle F(x_j) \rangle$, variance and covariances of both: $\langle  x_j x_k \rangle$ and  $\langle  F(x_j) F(x_k) \rangle$.
For a simpler exposition, we have only included twelve synaptic connections; we have excluded self (autaptic) connections and E$\to$E connections.  

An equation for each statistic can be derived by first writing Eq~\ref{frate_firsteqn} as a low-pass filter of the right-hand-side: 
\begin{eqnarray}
x_j(t) &= & \frac{1}{\tau}\int_0^t e^{-(t-u)/\tau} \Big[ \mu_j + \sigma_j\eta_j(u) + \sum_k g_{jk} F(x_k) \Big]\,du
\end{eqnarray}
We then take expectations, letting $t \rightarrow \infty$, we have:
\begin{eqnarray}
\mu(j):=\langle x_j  \rangle & = & \mu_j + \langle \sum_k g_{jk} F(x_k) \rangle  = \mu_j + \sum_{k} g_{jk} \langle F(x_k)  \rangle 
\end{eqnarray}
We assume the stochastic processes are ergodic, which is generally true for these types of stochastic differential equations, so that averaging over time is equivalent to averaging over the 
invariant measure. 

We will make several assumptions for computational efficiency. 
First, we only account for direct connections in the formulas for the first and second order statistics, assuming the terms from the indirect connections 
are either small or already accounted for in the direct connections.  We further make the following assumptions to simplify the calculations:
\begin{align}
	& \left\langle \int_0^t F(x_k(u))e^{-(t-u)/\tau}\,du \int_0^t F(x_k(v))e^{-(t-v)/\tau}\,dv \right\rangle  \approx  \frac{\tau}{2} \mathbb{E}\left[ F^2(x_k) \right]    \label{ass_Fvar}  \\
	& \hbox{where } \mathbb{E}\left[ F^2(x_k) \right] := \int F^2(\sigma(k)y+\mu(k))\,\rho_{SN}(y)\,dy    \\	
	& \left\langle \int_0^t \sigma_j\eta_j(u)e^{-(t-u)/\tau}\,du \int_0^t F(x_k(v))e^{-(t-v)/\tau}\,dv \right\rangle  \approx  \frac{\tau}{2} \mathbb{E}\left[ N_j F(x_k) \right], \hbox{ if }j\neq k \label{ass_nzFa} \\
	& \hbox{where } \mathbb{E}\left[ N_j F(x_k) \right] :=  \frac{\sigma_j}{\sqrt{2}} \iint y_1 F(\sigma(k)y_2+\mu(k))\,\rho_{2D}(y_1,y_2)\,dy_1 dy_2     \\
	& \left\langle \int_0^t \sigma_j\eta_j(u)e^{-(t-u)/\tau}\,du \int_0^t F(x_k(v))e^{-(t-v)/\tau}\,dv \right\rangle \nonumber \\
	&  \approx  \frac{\tau}{2} \frac{\sigma_k}{\sqrt{2}} \int y F(\sigma(k)y +\mu(k))\,\rho_{SN}(y)\,dy, \hbox{ if } j = k \label{ass_nzFb}  \\
	& \left\langle \int_0^t F(x_j(u))e^{-(t-u)/\tau}\,du \int_0^t F(x_k(v))e^{-(t-v)/\tau}\,dv \right\rangle  \approx  \frac{\tau}{2} \mathbb{E}\left[ F(x_j)F(x_k) \right] 		\label{ass_Fcov}  \\ 
	& \hbox{where } \mathbb{E}\left[ F(x_j) F(x_k) \right]   :=  \nonumber \\
	 & \iint F(\sigma(j)y_1+\mu(j)) F(\sigma(k)y_2+\mu(k))\,\rho_{2D}(y_1,y_2)\,dy_1 dy_2     \label{ass_end}
\end{align}
and $N_j$ denotes the random variable $\int_0^t \sigma_j\eta_j(u) e^{-(t-u)/\tau}\,du$, which is by itself normally distributed with mean 0 and variance $\sigma_j^2\tau/2$.  

The first assumption, Eq~\ref{ass_Fvar}, states that time-average of $F(x_j(t))$ multiplied by an exponential function (low-pass filter) is equal to the expected value scaled by $\tau/2$; 
the second and third, Eq~\ref{ass_nzFa} and Eq~\ref{ass_nzFb}, address $N_j$ and $F(x_k(t))$, for $j \not= k$ and $j=k$ respectively (similarly for Eq~\ref{ass_Fcov}).  

In all of the definitions for the expected values with $\rho_{2D}$, note that the underlying correlation correlation $c_{jk}$ depend on the pair of interest $(j,k)$.  
Finally, we assume that the activity variables $(x_j,x_k)$ are pairwise normally distributed with the subsequent statistics; this is sufficient to ``close" our model and solve for the statistical quantities self-consistently.  
This is implicitly a weak coupling assumption because with no coupling, $(x_j,x_k)$ are bivariate normal random variables.  

The resulting approximations for the mean activity are:
\begin{align}
	& \mu(1) = \mu_1 + \sum_{k=2,3,5,6} g_{1k}\int F(\sigma(k)y+\mu(k))\,\rho_{SN}(y)\,dy	\label{mn_x1} \\
	& \mu(2) = \mu_2 + g_{21}\int F(\sigma(1)y+\mu(1))\,\rho_{SN}(y)\,dy 		\label{mn_x2}  \\
	& \mu(3) = \mu_3 + g_{31}\int F(\sigma(1)y+\mu(1))\,\rho_{SN}(y)\,dy 		\label{mn_x3} \\ 
	& \mu(4) = \mu_4 + \sum_{k=2,3,5,6} g_{4k}\int F(\sigma(k)y+\mu(k))\,\rho_{SN}(y)\,dy \label{mn_x4} \\	
	& \mu(5) = \mu_5 + g_{54}\int F(\sigma(4)y+\mu(4))\,\rho_{SN}(y)\,dy 		\label{mn_x5} \\
	& \mu(6) = \mu_6 + g_{64}\int F(\sigma(4)y+\mu(4))\,\rho_{SN}(y)\,dy. 		\label{mn_x6}
\end{align}
The resulting approximation to the variances of the mean activity are:
\begin{eqnarray}
	\tau\sigma^2(1) &=& \frac{\sigma_1^2}{2} + \sum_{k=2,3,5,6} \frac{g^2_{1k}}{2} \text{Var}\big(F(\sigma(k)Y+\mu(k))\big)  \nonumber \\
				&  &+ \sum_{(j,k)\in\{(2,3);(5,6)\}} g_{1j}g_{1k}\text{Cov}\big(F(\sigma(j)Y_1+\mu(j)),F(\sigma(k)Y_2+\mu(k))\big)   \label{var_x1} \\
	\tau\sigma^2(2) &=& \frac{\sigma_2^2}{2} + \frac{g^2_{21}}{2} \text{Var}\big(F(\sigma(1)Y+\mu(1))\big)  \nonumber \\	
				&  & + \sigma_2 g_{21} \iint \frac{y_1}{\sqrt{2}} F(\sigma(1)y_2+\mu(1))\,\rho_{2D}(y_1,y_2)\,dy_1 dy_2 \label{var_x2}   \\
	\tau\sigma^2(3) &=& \frac{\sigma_3^2}{2} + \frac{g^2_{31}}{2} \text{Var}\big(F(\sigma(1)Y+\mu(1))\big)  \nonumber \\	
				&  & + \sigma_3 g_{31} \iint \frac{y_1}{\sqrt{2}} F(\sigma(1)y_2+\mu(1))\,\rho_{2D}(y_1,y_2)\,dy_1 dy_2 \label{var_x3}   \\
	\tau\sigma^2(4) &=& \frac{\sigma_4^2}{2} + \sum_{k=2,3,5,6} \frac{g^2_{4k}}{2} \text{Var}\big(F(\sigma(k)Y+\mu(k))\big) \nonumber \\
				&  &+ \sum_{(j,k)\in\{(2,3);(5,6)\}} g_{4j}g_{4k}\text{Cov}\big(F(\sigma(j)Y_1+\mu(j)),F(\sigma(k)Y_2+\mu(k))\big)  \\
	\tau\sigma^2(5) &=& \frac{\sigma_5^2}{2} + \frac{g^2_{54}}{2} \text{Var}\big(F(\sigma(4)Y+\mu(4))\big)  \nonumber \\	
				&  & + \sigma_5 g_{54} \iint \frac{y_1}{\sqrt{2}} F(\sigma(4)y_2+\mu(4))\,\rho_{2D}(y_1,y_2)\,dy_1 dy_2 \label{var_x5}   \\
	\tau\sigma^2(6) &=&  \frac{\sigma_6^2}{2} + \frac{g^2_{64}}{2} \text{Var}\big(F(\sigma(4)Y+\mu(4))\big)  \nonumber \\	
				&  & + \sigma_6 g_{64} \iint \frac{y_1}{\sqrt{2}} F(\sigma(4)y_2+\mu(4))\,\rho_{2D}(y_1,y_2)\,dy_1 dy_2 \label{var_x6} 				
\end{eqnarray}

In Eq~\ref{mn_x1}--\ref{var_x6}, all of the $\text{Var}$ and $\text{Cov}$ are with respect to $Y\sim \mathcal{N}(0,1)$ (for $\text{Var}$) and 
 $(Y_1,Y_2) \sim \mathcal{N}\left( \left(\begin{smallmatrix} 0 \\ 0 \end{smallmatrix}\right) , \frac{1}{2} \left(\begin{smallmatrix} 1 & c_{jk} \\ c_{jk} & 1 \end{smallmatrix}\right) \right)$ (for $\text{Cov}$);
both are easy to calculate.  The value $c_{jk}$ depends on the pairs; for example in Eq~\ref{var_x2}, the $\rho_{2D}$ has $c_{jk}=c_{OB}$, the background correlation value in the olfactory bulb but 
in Eq~\ref{var_x1}, the $\text{Cov}$ term is with respect to $\rho_{2D}$ with $c_{jk}=c_{PC}$, the background correlation value in the piriform cortex.  

Lastly, we state the formulas for the approximations to the covariances.  Although there are 15 total covariance values, we are only concerned with 6 covariance values (3 within OB and 3 within PC); we neglect all covariances \textit{between} regions.
First, our experimental data set shows that these covariance (and correlation) values are small (see Fig S9 in S2 Text).  
Second, because there is no background correlation (i.e., common input) between PC and OB in our model, 
any nonzero covariance/correlation arises strictly via direct coupling.  Thus, we cannot view OB-PC covariance from coupling as a small perturbation of the background (uncoupled) state; we do not expect our model to yield qualitatively accurate predictions for these statistics.  The formulas for the Cov of interest are:
\begin{eqnarray}
	\tau \text{Cov}(1,2) &=& \frac{1}{2}c_{OB}\sigma_1\sigma_2 +\sigma_1 \frac{g_{21}}{2} \int \frac{y}{\sqrt{2}} F(\sigma(1)y+\mu(1))\,\rho_{SN}(y)\,dy \nonumber \\
			& & +\sigma_2\frac{g_{12}}{2}\int \frac{y}{\sqrt{2}} F(\sigma(2)y+\mu(2))\,\rho_{SN}(y)\,dy \nonumber \\	 	
			& & +\sigma_2\frac{g_{13}}{2}\int \frac{y}{\sqrt{2}} F(\sigma(3)y+\mu(3))\,\rho_{SN}(y)\,dy  \nonumber \\	
			& & +\frac{1}{2}\sum_{(j,k)} g_{1j} g_{2k} \mathcal{C}(j,k) \label{cov_12}  \\
	\tau \text{Cov}(1,3) &=& \frac{1}{2}c_{OB}\sigma_1\sigma_3 +\sigma_1 \frac{g_{31}}{2} \int \frac{y}{\sqrt{2}} F(\sigma(1)y+\mu(1))\,\rho_{SN}(y)\,dy \nonumber \\
	& & +\sigma_3\frac{g_{12}}{2}\int \frac{y}{\sqrt{2}} F(\sigma(2)y+\mu(2))\,\rho_{SN}(y)\,dy \nonumber \\	 	
			& & +\sigma_3\frac{g_{13}}{2}\int \frac{y}{\sqrt{2}} F(\sigma(3)y+\mu(3))\,\rho_{SN}(y)\,dy  \nonumber \\	
			& & +\frac{1}{2}\sum_{(j,k)} g_{1j} g_{3k} \mathcal{C}(j,k) \label{cov_13} \\
	\tau \text{Cov}(2,3) &=& \frac{1}{2}c_{OB}\sigma_2\sigma_3 +\frac{g_{21}g_{31}}{2} \text{Var}\big(F(\sigma(1)Y+\mu(1))\big) 	\nonumber \\	
			     &  & + \frac{\sigma_3 g_{21}+\sigma_2 g_{31}}{2}\iint \frac{y_1}{\sqrt{2}} F(\sigma(1)y_2+\mu(1))\,\rho_{2D}(y_1,y_2)\,dy_1 dy_2   \label{cov_23} \\
	\tau \text{Cov}(4,5) &=& \frac{1}{2}c_{PC}\sigma_4\sigma_5 +\sigma_4 \frac{g_{54}}{2} \int \frac{y}{\sqrt{2}} F(\sigma(4)y+\mu(4))\,\rho_{SN}(y)\,dy \nonumber  \\
				& & +\sigma_5\frac{g_{45}}{2}\int \frac{y}{\sqrt{2}} F(\sigma(5)y+\mu(5))\,\rho_{SN}(y)\,dy \nonumber \\	 	
			& & +\sigma_5\frac{g_{46}}{2}\int \frac{y}{\sqrt{2}} F(\sigma(6)y+\mu(6))\,\rho_{SN}(y)\,dy  \nonumber \\	
			& & +\frac{1}{2}\sum_{(j,k)} g_{4j} g_{5k} \mathcal{C}(j,k) \label{cov_45}  \\
	\tau \text{Cov}(4,6) &=& \frac{1}{2}c_{PC}\sigma_4\sigma_6 +\sigma_4 \frac{g_{64}}{2} \int \frac{y}{\sqrt{2}} F(\sigma(4)y+\mu(4))\,\rho_{SN}(y)\,dy \nonumber  \\
					& & +\sigma_6\frac{g_{45}}{2}\int \frac{y}{\sqrt{2}} F(\sigma(5)y+\mu(5))\,\rho_{SN}(y)\,dy \nonumber \\	 	
			& & +\sigma_6\frac{g_{46}}{2}\int \frac{y}{\sqrt{2}} F(\sigma(6)y+\mu(6))\,\rho_{SN}(y)\,dy  \nonumber \\	
			& & +\frac{1}{2}\sum_{(j,k)} g_{4j} g_{6k} \mathcal{C}(j,k) \label{cov_46}  \\
	\tau \text{Cov}(5,6) &=& \frac{1}{2}c_{PC}\sigma_5\sigma_6 +\frac{g_{54}g_{64}}{2} \text{Var}\big(F(\sigma(4)Y+\mu(4))\big) 	\nonumber \\	
			     &  & + \frac{\sigma_6 g_{54}+\sigma_5 g_{64}}{2}\iint \frac{y_1}{\sqrt{2}} F(\sigma(4)y_2+\mu(4))\,\rho_{2D}(y_1,y_2)\,dy_1 dy_2   \label{cov_56} 
\end{eqnarray}
where 
\begin{eqnarray}
\mathcal{C}(j,k) =& \iint F(\sigma(j)y_1+\mu(j))F(\sigma(k)y_2+\mu(k))\rho_{2D}(y_1,y_2)\,dy_1dy_2  \nonumber \\
			&- \left( \int F(\sigma(j)y+\mu(j))\rho_{SN}(y)\,dy \right) \left( \int F(\sigma(k)y+\mu(k))\rho_{SN}(y)\,dy \right) \label{mathcal_defn}
\end{eqnarray}

\subsubsection*{Iteration procedure to solve for the approximate statistics self-consistently}
Based on the approximations and resulting equations described in the previous section, our objective is to solve for the statistics of $x_j$ self-consistently.  Once these are determined, the statistics of the firing rates $F(x_j)$ are approximated with the 
same pairwise normal assumption on $(x_j,x_k)$; we are {\bf not} assuming that $(F(x_j),F(x_k))$ are bivariate normal random variables.  

We use a simple iterative procedure to solve the system of coupled algebraic expression for the statistics of $x_j$. 
We first solve the system in the absence of coupling (i.e. Eq~\ref{eqn:mu_uncoupled}, \ref{eqn:cov_uncoupled}), and use these values to start the iteration;
at each step, the formulas for the means (Eq~\ref{mn_x1}--\ref{mn_x6}), variances (Eq~\ref{var_x1}--\ref{var_x6}), and covariances (Eq~\ref{cov_12}--\ref{cov_56}) are 
recalculated numerically, using the results of the previous step.  The iteration stops once \underline{all 18} statistical quantities of the activity match up to a relative tolerance of $10^{-6}$ (convergence), or after 50 total iterations (non-convergence).  The 
result with a given parameter set can either be: i) convergence, ii) non-convergence, iii) a pair of statistics with invalid covariance (non-positive definite covariance matrix), which is checked after i) and ii).  We only consider parameter sets 
where the iteration has converged and all of the covariances are valid, after which we determine whether the constraints are satisfied.
  
One subtle point is that we did not use any of the numerically calculated $\text{Cov}$ values in the bivariate normal distributions $\rho_{2D}$; rather, the correlation value is always $c_{jk}$ which is either 0, $c_{OB}$, 
or $c_{PC}$ depending on the pair.  In principle, one can use a fully iterative procedure where the formulas for the $\text{Cov}$ (Eq~\ref{cov_12}--\ref{cov_56}) are used in $\rho_{2D}$; however, we found that the resulting covariance matrices (for  
$\rho_{2D}$) can fail to be positive semi-definite. 
Handling this case requires additional code in the program and slower calculations for each parameter set, which 
detracts from 
the purpose of our method.  We checked some parameter sets comparing the results of the two procedures, 
and the results are quantitatively similar.

The standard normal $\rho_{SN}$ and bivariate $\rho_{2D}$ PDFs have state variable(s) $y_{1,2}$ discretized from -3 to 3 with a mesh size of 0.01; integrals in Eq~\ref{mn_x1}--\ref{cov_56} are computed using the trapezoidal rule.

\subsubsection*{Simplified network with four coupling parameters}

To further simplify the network, we:
\begin{itemize}
	\item set $\tau=1$,
	\item assume feedforward inhibitory connections within a region have the same strength: $g_{21} = g_{31} =: gIO$ and $g_{54} = g_{64} =: gIP$,
	\item assume cross-region excitatory connections are equal from the presynaptic cell, i.e., $g_{15} = g_{16} =: gEP$ and $g_{42} = g_{43} =: gEO$.
	\item assume $\sigma_1=\sigma_2=\sigma_3=:\sigma_{OB}$ and $\sigma_4=\sigma_5=\sigma_6=:\sigma_{PC}$
	\item assume $g_{12}=g_{13}=g_{45}=g_{46}=:g_\epsilon=0.1$
\end{itemize}
Now there are only 4 variable coupling parameters: $gIO$, $gEO$, $gIP$, $gEP$.

The above formulas for the statistics of $x_j$ reduce to:
\begin{eqnarray}
	\mu(1) &=& \mu_1 + gEP  \int \Big( F(\sigma(5)y+\mu(5)) + F(\sigma(6)y+\mu(6)) \Big)\,\rho_{SN}(y)\,dy	\nonumber \\
		 & & 	     + g_\epsilon \int \Big( F(\sigma(2)y+\mu(2)) + F(\sigma(3)y+\mu(3)) \Big)\,\rho_{SN}(y)\,dy \label{mnX1} \\
	\mu(2) &=& \mu_2 + gIO\int F(\sigma(1)y+\mu(1))\,\rho_{SN}(y)\,dy 		\label{mnX2}  \\
	\mu(3) &=& \mu_3 + gIO\int F(\sigma(1)y+\mu(1))\,\rho_{SN}(y)\,dy 		\label{mnX3} \\ 
	\mu(4) &=& \mu_4 + gEO \int \Big( F(\sigma(2)y+\mu(2)) + F(\sigma(3)y+\mu(3)) \Big)\,\rho_{SN}(y)\,dy 	 \nonumber \\
		 & &  + g_\epsilon \int \Big( F(\sigma(5)y+\mu(5)) + F(\sigma(6)y+\mu(6)) \Big)\,\rho_{SN}(y)\,dy	\label{mnX4} \\	
	\mu(5) &=& \mu_5 + gIP\int F(\sigma(4)y+\mu(4))\,\rho_{SN}(y)\,dy 		\label{mnX5} \\
	\mu(6) &=& \mu_6 + gIP \int F(\sigma(4)y+\mu(4))\,\rho_{SN}(y)\,dy; 		\label{mnX6}
\end{eqnarray}
the variances are:  
\begin{eqnarray}
	\sigma^2(1) &=& \frac{\sigma^2_{OB}}{2} + \frac{(gEP)^2}{2} \text{Var}\Big(F(\sigma(5)Y_1+\mu(5)) + F(\sigma(6)Y_2+\mu(6))\Big) \nonumber \\	
			& & +\frac{g^2_\epsilon}{2} \text{Var}\Big(F(\sigma(2)Y_1+\mu(2)) + F(\sigma(3)Y_2+\mu(3))\Big) \label{varX1} \\
	\sigma^2(2) &=& \frac{\sigma^2_{OB}}{2} + \frac{(gIO)^2}{2} \text{Var}\big(F(\sigma(1)Y+\mu(1))\big)  \nonumber \\	
				&  & + \sigma_{OB} gIO \iint \frac{y_1}{\sqrt{2}} F(\sigma(1)y_2+\mu(1))\,\rho_{2D}(y_1,y_2)\,dy_1 dy_2 \label{varX2}   \\
	\sigma^2(3) &=& \sigma^2(2)	 \label{varX3}   \\
	\sigma^2(4) &=& \frac{\sigma^2_{PC}}{2} + \frac{(gEO)^2}{2} \text{Var}\Big(F(\sigma(2)Y_1+\mu(2)) + F(\sigma(3)Y_2+\mu(3))\Big)  \nonumber \\
		& & +\frac{g^2_\epsilon}{2} \text{Var}\Big(F(\sigma(5)Y_1+\mu(5)) + F(\sigma(6)Y_2+\mu(6))\Big) \label{varX4}	\\
	\sigma^2(5) &=& \frac{\sigma^2_{PC}}{2} + \frac{(gIP)^2}{2} \text{Var}\big(F(\sigma(4)Y+\mu(4))\big)  \nonumber \\	
				&  & + \sigma_{PC} gIP \iint \frac{y_1}{\sqrt{2}} F(\sigma(4)y_2+\mu(4))\,\rho_{2D}(y_1,y_2)\,dy_1 dy_2 \label{varX5}   \\
	\sigma^2(6) &=&  \sigma^2(5); \label{varX6} 				
\end{eqnarray}
the covariances are:
\begin{eqnarray}
	\text{Cov}(1,2) &=& \frac{1}{2}c_{OB}\sigma^2_{OB} +\sigma_{OB} \frac{gIO}{2} \int \frac{y}{\sqrt{2}} F(\sigma(1)y+\mu(1))\,\rho_{SN}(y)\,dy \nonumber \\
				& &  \sigma_{OB} \frac{g_\epsilon}{2} \int \frac{y}{\sqrt{2}} F(\sigma(2)y+\mu(2))\,\rho_{SN}(y)\,dy  \nonumber \\
				& & g_\epsilon gIO * \mathcal{C}(1,2) \label{covX12}  \\
	\text{Cov}(1,3) &=& \text{Cov}(1,2) \label{covX13}  \\
	\text{Cov}(2,3) &=& \frac{1}{2}c_{OB}\sigma^2_{OB}+\frac{g^2_{IO}}{2} \text{Var}\big(F(\sigma(1)Y+\mu(1))\big) 	\nonumber \\	
			     &  & + \sigma_{OB} gIO \iint \frac{y_1}{\sqrt{2}} F(\sigma(1)y_2+\mu(1))\,\rho_{2D}(y_1,y_2)\,dy_1 dy_2   \label{covX23} \\
	\text{Cov}(4,5) &=& \frac{1}{2}c_{PC}\sigma^2_{PC} +\sigma_{PC} \frac{gIP}{2} \int \frac{y}{\sqrt{2}} F(\sigma(4)y+\mu(4))\,\rho_{SN}(y)\,dy \nonumber \\
			& &  \sigma_{PC} \frac{g_\epsilon}{2} \int \frac{y}{\sqrt{2}} F(\sigma(5)y+\mu(5))\,\rho_{SN}(y)\,dy  \nonumber \\
				& & g_\epsilon gIP * \mathcal{C}(4,5)	\label{covX45}  \\
	\text{Cov}(4,6) &=& \text{Cov}(4,5)   \label{covX46}  \\
	\text{Cov}(5,6) &=& \frac{1}{2}c_{PC}\sigma^2_{PC} +\frac{g^2_{IP}}{2} \text{Var}\big(F(\sigma(4)Y+\mu(4))\big) 	\nonumber \\	
			     &  & + \sigma_{PC} gIP \iint \frac{y_1}{\sqrt{2}} F(\sigma(4)y_2+\mu(4))\,\rho_{2D}(y_1,y_2)\,dy_1 dy_2   \label{covX56} 
\end{eqnarray}
See Eq~\ref{mathcal_defn} for the definition of $\mathcal{C}$.

\subsection*{Leaky Integrate-and-Fire Model of the OB--PC Circuit}

We use a generic spiking neural network model of leaky integrate-and-fire neurons to test the results of the theory.  
There were $N_{OB}=100$ total OB cells, of which we set 80\% (80) to be granule (I-)cells and 20\% (20) to be mitral/tufted ({\bf M/T}) E-cells.  There are known to be many 
more granule cells than M/T cells in the OB; this ratio of 4-to-1 is similar to other models of OB (see~\cite{grabska17} who used 3-to-1).  
The equations for the OB cells are, indexed by $k\in\{1,2,\dots,N_{OB}\}$:
\begin{eqnarray}\label{ob_lif}
	\tau_m \frac{d v_k}{dt} & = & \mu_{OB}-v_k-g_{k, XI}(t)(v_k-\mathcal{E}_I)-g_{k, XE}(t)(v_k-\mathcal{E}_E) \nonumber \\	
	     & & - g_{k,XPC}(t - \tau_{\Delta,PC})(v_k-\mathcal{E}_I) +\sigma_{OB}\left(\sqrt{1-\tilde{c}_{OB}}\eta_k(t) + \sqrt{\tilde{c}_{OB}}\xi_o(t) \right) \nonumber \\
	v_k(t^*) & \geq & \theta_k  \Rightarrow v_k(t^*+\tau_{ref})=0 \nonumber \\
	g_{k,XE}(t) &=& \frac{\gamma_{XE}}{p_{XE} \left(0.2 N_{OB} \right) }\sum_{k'\in\{\hbox{ presyn OB E-cells}\}  } G_{k'}(t) \nonumber \\
	g_{k,XI}(t) &=& \frac{\gamma_{XI}}{p_{XI} \left(0.8 N_{OB} \right)}\sum_{k'\in\{\hbox{presyn OB I-cells}\}} G_{k'}(t) \nonumber \\
	g_{k,XPC}(t) &=& \frac{\gamma_{X,PC}}{p_{X,PC} \left(0.8 N_{PC} \right)} \sum_{j'\in\{\hbox{presyn PC E-cells}\}} G_{j'}(t) \nonumber \\
	\tau_{d,X}\frac{d G_k}{dt} &=& -G_k + A_k  \nonumber \\
	\tau_{r,X} \frac{d A_k}{dt} &=& -A_k + \tau_{r,X} \alpha_X \sum_{l} \delta(t-t_{k,l}).  \label{eqn:OB_LIF_all}
\end{eqnarray}
The conductance values in the first equation $g_{k,XI}$, $g_{k,XE}$, and $g_{k,XPC}$ depend on the type of neuron $v_k$ ($X\in\{ E, I\}$).  The last conductance,  
$g_{X,PC}(t - \tau_{\Delta,PC})(v_k-\mathcal{E}_E)$, models the excitatory presynaptic input (feedback) from the PC cells with a time delay of $\tau_{\Delta,PC}$.  The conductance variables $g_{k,XY}(t)$ are dimensionless because this model was 
derived from scaling the original (raw) conductance variables by the leak conductance with the same dimension.  
The leak, inhibitory and excitatory reversal potentials are 0, $\mathcal{E}_I$, and $\mathcal{E}_E$, respectively with $\mathcal{E}_I<0<\mathcal{E}_E$ 
(the voltage is scaled to be dimensionless, see Table~\ref{table:lif_parms}).  
$\xi_k(t)$ are uncorrelated white noise processes and $\xi_o(t)$ is the common noise term to all $N_{OB}$ cells.

The second equation describes the refractory period at spike time $t^*$: when the neuron's voltage crosses 
threshold $\theta_j$ (see below for distribution of thresholds), 
the neuron goes into a refractory period for $\tau_{ref}$, after which we set the neuron's voltage to 0.  

The parameter $\gamma_{XY}$ gives the relative weight of a connection from neuron type $Y$ to neuron type $X$; the parameter $p_{XY}$ is the probability that any such connection exists ($X,Y\in\{E,I\}$). $G_k$ is the synaptic variable associated with each cell, and dependent only on that cell's spike times; its dynamics are given by the final two equations in Eq~\ref{eqn:OB_LIF_all} and depend on whether $k \in  \{E,I\}$.

Finally, two of the parameters above can be equated with coupling parameters in the reduced model:
\begin{equation}
gEP =  \gamma_{E,PC}; \quad gIO = \gamma_{EI}
\end{equation}
which are dimensionless scale factors for the synaptic conductances.

\begin{table}[!ht]
\centering
\caption{{\bf Fixed parameters for the LIF OB--PC model, see Eqs~\ref{ob_lif}--\ref{pc_lif}. }}
\label{table:lif_parms}
\begin{tabular}{|lcccccccccc|}
\hline
\multicolumn{11}{|c|}{\textbf{Same for both OB and PC}}                                                                                                                                               \\ \hline
\textbf{Parameter}                & $\tau_m$ & $\tau_{ref}$ & $\mathcal{E}_I$ & $\mathcal{E}_E$ & $\tau_{d,I}$ & $\tau_{r,I}$  & $\tau_{d,E}$  & $\tau_{r,E}$  & $\alpha_I$               & $\alpha_E$               \\ \hline
                   & 20\,ms   & 2\,ms        & -2.5            & 6.5             & 10\,ms       & 2\,ms         & 5\,ms         & 1\,ms         								& 2\,Hz                       & 1\,Hz                        \\ \thickhline
\textbf{Parameter} & $N$    	 & Spont. $\mu$    & Evoked $\mu$ &   $\sigma$     & $\tilde{c}$  &   $\gamma_{EE}$ & $\gamma_{IE}$ & $\gamma_{II}$ &  $\tau_{\Delta,PC/OB}$ &       \\ \hline
\textbf{OB}        & 100     		 & 0.6             & 0.9$^*$      				& 0.05         & 0.5                  & 2             & 4             & 2            		        &  	10\,ms  & $ $  \\
\textbf{PC}        & 100      		 & 0               & 0.4  					& 0.1          & 0.8                    & 5             & 8             & 6             		        &      	5\,ms     & $ $  \\ \hline
\end{tabular}
\begin{flushleft} All 12 probabilities of connections are set to $p_{XY}=0.30$; otherwise connections were chosen randomly and independently (Erd\H{o}s-R\'enyi graphs).  
The synaptic time delay from OB to PC is $\tau_{\Delta,OB}=10\,$ms, and from PC to OB is $\tau_{\Delta,PC}=5\,$ms.  
The scaled voltages from mV is: (V+Vreset)/(Vth+Vreset), corresponding for 
example to Vreset=Vleak=-65\,mV, Vth=-55\,mV (on average), excitatory reversal potential of 0\,mV and inhibitory reversal potential of -90\,mV.  {\bf *}Note: in the evoked state, 
only the {\bf M/T} (E-cells) in OB receive a larger $\mu$ input from 0.6 to 0.9; the granule cells in OB have $\mu=0.6$ even in the evoked state.
\end{flushleft}
\end{table}

The PC cells had similar functional form but with different parameters (see Table~\ref{table:lif_parms} for parameter values).  We modeled $N_{PC}=100$ total PC cells, of which 80\% were excitatory and 20\% inhibitory.  
The equations, indexed by $j\in\{1,2,\dots,N_{PC}\}$ are:
%
\begin{eqnarray}\label{pc_lif}
	\tau_m \frac{d v_j}{dt} & = & \mu_{PC}-v_j-g_{j,XI}(t)(v_j-\mathcal{E}_I)-g_{j,XE}(t)(v_j-\mathcal{E}_E) \nonumber \\	
	     & & - g_{j,XOB}(t - \tau_{\Delta,OB})(v_j-\mathcal{E}_E) +\sigma_{PC}\left(\sqrt{1-\tilde{c}_{PC}}\eta_j(t) + \sqrt{\tilde{c}_{PC}}\xi_p(t) \right) \nonumber \\
	v_j(t^*) & \geq & \theta_j  \Rightarrow v_j(t^*+\tau_{ref})=0 \nonumber \\
	g_{j,XE}(t) &=& \frac{\gamma_{XE}}{p_{XE} \left(0.8 N_{PC} \right)}\sum_{j'\in\{\hbox{presyn PC E-cells}\}} G_{j'}(t) \nonumber \\
	g_{j,XI}(t) &=& \frac{\gamma_{XI}}{p_{XI} \left(0.2 N_{PC} \right)}\sum_{j'\in\{\hbox{presyn PC I-cells}\}} G_{j'}(t) \nonumber \\
		g_{j,XOB}(t) &=& \frac{\gamma_{X,OB}}{p_{X,OB} \left(0.2 N_{OB} \right)} \sum_{k'\in\{\hbox{presyn OB E-cells}\}} G_{k'}(t) \nonumber \\
	\tau_{d,X}\frac{d G_j}{dt} &=& -G_j + A_j  \nonumber \\
	\tau_{r,X} \frac{d A_j}{dt} &=& -A_j + \tau_{r,X} \alpha_X \sum_{l} \delta(t-t_{j,l}).
\end{eqnarray}
Excitatory synaptic input from the OB cells along the lateral olfactory tract is modeled by: $g_{X,OB}(t - \tau_{\Delta,OB})(v_j-\mathcal{E}_E)$.  The common noise term for the 
PC cells $\xi_p(t)$ is independent of the common noise term for the OB cells $\xi_o(t)$.  
Two of the parameters above can be equated with coupling parameters in the reduced model:
\begin{equation}
gEO =  \gamma_{E,OB}; \quad gIP = \gamma_{EI}
\end{equation}

The values of the parameters 
that were not stated in Table~\ref{table:lif_parms} were varied in the paper: 
$$ gIO, \hspace{.5in} gEO, \hspace{.5in} gIP, \hspace{.5in} gEP. $$

To model two activity states, we allowed mean inputs to vary (see Table~\ref{table:lif_parms}). In contrast to the reduced model, we increased both inputs to PC cells (from $\mu_{PC}=0$ in the spontaneous state to 
$\mu_{PC}=0.4$ in the evoked state) as well as to OB cells; $\mu_{OB}=0.6$ in the spontaneous state to $\mu_{OB}=0.9$ in the evoked state only for {\bf M/T} cells (OB granule cell 
input is the same for spontaneous and evoked). 

Finally, we model heterogeneity by setting the threshold values $\theta_j$ in the following way.  Both OB and PC cells had the following distributions for $\theta_j$:
\begin{eqnarray}\label{thres_distr}
	\theta_j &\sim& e^{\mathcal{N}} 
\end{eqnarray}
where $\mathcal{N}$ is normal distribution with mean $-\sigma^2_\theta/2$ and standard deviation $\sigma_\theta$, so that $\{\theta_j\}$ has a 
log-normal distribution with mean 1 and variance: $e^{\sigma_\theta^2}-1$.  We set $\sigma_\theta=0.1$, which results in firing rates ranges seen in the experimental data.  
Since the number of cells are modest with regards to sampling ($N_{OB}=100$, $N_{PC}=100$), we evenly sampled the log-normal distribution from the 5$^{th}$ to 95$^{th}$ percentiles (inclusive).  

We remark that the synaptic delays of $\tau_{\Delta,PC}$ and $\tau_{\Delta,OB}$ were set to modest values to capture the appreciable distances between OB and PC.  This is a reasonable choice 
based on evidence that stimulation in PC elicit a response in OB 5-10\,ms later~\cite{neville03}.

In all Monte Carlo simulations of the coupled LIF network, we used a time step of 0.1\,ms, with 2\,s of biology time for each of the 50,000 realizations (i.e., over 27.7 hours of biology time), enough simulated statistics to effectively have convergence.

\section*{Supporting Information}

\paragraph*{S1 Text.}
\label{S1_file}
{\bf Experimental Data Statistics by Odor.} This file shows the trial-averaged spiking statistics of the experimental data dissected by a specific odor.
Contains Figs. S1-S8, and Tables S1-S2.

\paragraph*{S2 Text.}
\label{S2_file}
{\bf Supplementary Figures for the Main Modeling.} This file contains supplemental figures from modeling and analysis. Contains Figs. S9-S16.

\paragraph*{S3 Text.}
\label{S3_file}
{\bf Supplementary Material: Cortical-Cortical Network.} This file contains supplemental modeling results on a generic cortical-cortical coupled network.  Contains Figs. S17-S21, and Table S3.

\paragraph*{S1 Table.} 
{\bf Average population firing rate by odor and activity state.}
\paragraph*{S2 Table.} 
{\bf Standard deviation of population firing rate by odor and activity state.}
\paragraph*{S3 Table.} 
{\bf Fixed parameters for the LIF Cortical-cortical model.}

\paragraph*{S1 Figure.} 
{\bf Experimental statistics by odor and activity state: Fano Factor.}
\paragraph*{S2 Figure.}
{\bf Experimental statistics by odor and activity state: spike count variance.}
\paragraph*{S3 Figure.}
{\bf Experimental statistics by odor and region: Fano Factor.}
\paragraph*{S4 Figure.}
{\bf Experimental statistics by odor and region: spike count variance.}
\paragraph*{S5 Figure.}
{\bf Experimental statistics by odor and activity state: spike count correlation.}
\paragraph*{S6 Figure.}
{\bf Experimental statistics by odor and activity state: spike count covariance.}
\paragraph*{S7 Figure.}
{\bf Experimental statistics by odor and region: spike count correlation.}
\paragraph*{S8 Figure.}
{\bf Experimental statistics by odor and region: spike count covariance.}
\paragraph*{S9 Figure.}
{\bf Cross-region correlations are smaller than within-region correlations.}
\paragraph*{S10 Figure.}
{\bf Fast analytic approximation accurately captures statistics of a multi-population firing rate model.}
\paragraph*{S11 Figure.}
{\bf Experimental observations constrain conductance parameters in analytic model.}
\paragraph*{S12 Figure.}
{\bf Analytic approximation results are robust to choice of transfer function.}
\paragraph*{S13 Figure.}
{\bf Mean input to PC must increase in the evoked state.}
\paragraph*{S14 Figure.}
{\bf Violating derived relationship $gIO < gIP$ results in statistics that are inconsistent with experimental observations.}
\paragraph*{S15 Figure.}
{\bf Violating derived relationship $gEP > gEO$ results in statistics that are inconsistent with experimental observations.}
\paragraph*{S16 Figure.}
{\bf Violating derived relationship $gEP,  gIP \gg gEO, gIO$ results in statistics that are inconsistent with experimental observations.}
\paragraph*{S17 Figure.}
{\bf Minimal firing rate model to analyze synaptic conductance strengths.}
\paragraph*{S18 Figure.}
{\bf Detailed spiking LIF model confirms the results from analytic rate model.}
\paragraph*{S19 Figure.}
{\bf Violating derived relationship $\vert gI1\vert < \vert gI2\vert$ results in statistics that are inconsistent with experimental observations.}
\paragraph*{S20 Figure.}
{\bf Violating derived relationship $gE2,  gI2 \gg gE1, gI1$ results in statistics that are inconsistent with experimental observations.}
\paragraph*{S21 Figure.}
{\bf iolating derived relationship $gE2 > gE1$ results in statistics that are inconsistent with experimental observations.}


\section*{Author Contributions}
Conceived and designed research: AKB CL.  Derived expressions in theoretical methods: CL.  Analyzed the data: AKB SHG WLS CL.  Conceived and designed electrophysiological experiments: SHG WLS.  
Wrote the paper: AKB SHG WLS CL.  

\nolinenumbers

%
%
%

\begin{thebibliography}{10}

\bibitem{prevedel14}
Prevedel R, Yoon YG, Hoffmann M, Pak N, Wetzstein G, Kato S, et~al.
\newblock Simultaneous whole-animal 3D imaging of neuronal activity using
  light-field microscopy.
\newblock Nature methods. 2014;11(7):727--730.

\bibitem{ahrens13}
Ahrens MB, Orger MB, Robson DN, Li JM, Keller PJ.
\newblock Whole-brain functional imaging at cellular resolution using
  light-sheet microscopy.
\newblock Nature methods. 2013;10(5):413--420.

\bibitem{lemon15}
Lemon WC, Pulver SR, H{\"o}ckendorf B, McDole K, Branson K, Freeman J, et~al.
\newblock Whole-central nervous system functional imaging in larval Drosophila.
\newblock Nature communications. 2015;6.

\bibitem{brainInit_13}
Kandel ER, Markram H, Matthews PM, Yuste R, Koch C.
\newblock Neuroscience thinks big (and collaboratively).
\newblock Nature Reviews Neuroscience. 2013;14(9):659--664.

\bibitem{brunel}
Brunel N.
\newblock Dynamics of Sparsely Connected Networks of Excitatory and Inhibitory
  Spiking Neurons.
\newblock Journal of Computational Neuroscience. 2000;8:183--208.

\bibitem{brunelhakim}
Brunel N, Hakim V.
\newblock Fast global oscillations in networks of integrate-and-fire neurons
  with low firing rates.
\newblock Neural Computation. 1999;11:1621--1671.

\bibitem{doiron16}
Doiron B, Litwin-Kumar A, Rosenbaum R, Ocker G, Josi{\'c} K.
\newblock The mechanics of state-dependent neural correlations.
\newblock Nature Neuroscience. 2016;19(3):383--393.

\bibitem{BarreiroLy_RecrCorr_17}
Barreiro AK, Ly C.
\newblock When do correlations increase with firing rates in recurrent
  networks?
\newblock PLoS Computational Biology. 2017;13(4):e1005506.

\bibitem{zohary94}
Zohary E, Shadlen M, Newsome W.
\newblock Correlated neuronal discharge rate and its implications for
  psychophysical performance.
\newblock Nature. 1994;370(6485):140--143.

\bibitem{bair01}
Bair W, Zohary E, Newsome WT.
\newblock Correlated firing in macaque visual area MT: time scales and
  relationship to behavior.
\newblock The Journal of Neuroscience. 2001;21(5):1676--1697.

\bibitem{ecker11}
Ecker A, Berens P, Tolias A, Bethge M.
\newblock The effect of noise correlations in populations of diversely tuned
  neurons.
\newblock The Journal of Neuroscience. 2011;31(40):14272--14283.

\bibitem{moreno14}
Moreno-Bote R, Beck J, Kanitscheider I, Pitkow X, Latham P, Pouget A.
\newblock Information-limiting correlations.
\newblock Nature neuroscience. 2014;17(10):1410--1417.

\bibitem{kohn16}
Kohn A, Coen-Cagli R, Kanitscheider I, Pouget A.
\newblock Correlations and Neuronal Population Information.
\newblock Annual review of neuroscience. 2016;39(0).

\bibitem{wilsoncowan1}
Wilson HR, Cowan JD.
\newblock Excitatory and Inhibitory Interactions in Localized Populations of
  Model Neurons.
\newblock Biophysical Journal. 1972;12:1--24.

\bibitem{churchland10}
Churchland MM, Yu BM, et~al.
\newblock Stimulus onset quenches neural variability: a widespread cortical
  phenomenon.
\newblock Nature Neuroscience. 2010;13:369--378.

\bibitem{miura12}
Miura K, Mainen Z, Uchida N.
\newblock Odor representations in olfactory cortex: distributed rate coding and
  decorrelated population activity.
\newblock Neuron. 2012;74:1087--1098.

\bibitem{LMD_whisker_12}
Ly C, Middleton JW, Doiron B.
\newblock Cellular and circuit mechanisms maintain low spike co-variability and
  enhance population coding in somatosensory cortex.
\newblock Frontiers in Computational Neuroscience. 2012;6:1--26.
\newblock doi:{10.3389/fncom.2012.00007}.

\bibitem{fellous03}
Fellous JM, Rudolph M, Destexhe A, Sejnowski TJ.
\newblock Synaptic background noise controls the input/output characteristics
  of single cells in an in vitro model of in vivo activity.
\newblock Neuroscience. 2003;122(3):811--829.

\bibitem{prescott03}
Prescott SA, De~Koninck Y.
\newblock Gain control of firing rate by shunting inhibition: roles of synaptic
  noise and dendritic saturation.
\newblock Proceedings of the National Academy of Sciences.
  2003;100(4):2076--2081.

\bibitem{cardin08}
Cardin JA, Palmer LA, Contreras D.
\newblock Cellular mechanisms underlying stimulus-dependent gain modulation in
  primary visual cortex neurons in vivo.
\newblock Neuron. 2008;59(1):150--160.

\bibitem{murakami05}
Murakami M, Kashiwadani H, Kirino Y, Mori K.
\newblock State-dependent sensory gating in olfactory cortex.
\newblock Neuron. 2005;46(2):285--296.

\bibitem{renart10}
Renart A, de~la Rocha J, Bartho P, Hollender L, Parga N, Reyes A, et~al.
\newblock The asynchronous state in cortical circuits.
\newblock Science. 2010;327:587--590.

\bibitem{oswald12}
Oswald AM, Urban NN.
\newblock There and back again: the corticobulbar loop.
\newblock Neuron. 2012;76(6):1045--1047.

\bibitem{boyd12}
Boyd AM, Sturgill JF, Poo C, Isaacson JS.
\newblock Cortical feedback control of olfactory bulb circuits.
\newblock Neuron. 2012;76(6):1161--1174.

\bibitem{markopoulos12}
Markopoulos F, Rokni D, Gire DH, Murthy VN.
\newblock Functional properties of cortical feedback projections to the
  olfactory bulb.
\newblock Neuron. 2012;76(6):1175--1188.

\bibitem{large16}
Large A, Vogler N, Mielo S, Oswald AMM.
\newblock Balanced feedforward inhibition and dominant recurrent inhibition in
  olfactory cortex.
\newblock Proceedings of the National Academy of Sciences.
  2016;113(8):2276--2281.

\bibitem{poo09}
Poo C, Isaacson J.
\newblock Odor representations in olfactory cortex: ``sparse" coding, global
  inhibition, and oscillations.
\newblock Neuron. 2009;62:850--861.

\bibitem{burton15}
Burton SD, Urban NN.
\newblock Rapid Feedforward Inhibition and Asynchronous Excitation Regulate
  Granule Cell Activity in the Mammalian Main Olfactory Bulb.
\newblock The Journal of Neuroscience. 2015;35(42):14103--14122.

\bibitem{grabska17}
Grabska-Barwi{\'n}ska A, Barthelm{\'e} S, Beck J, Mainen ZF, Pouget A, Latham
  PE.
\newblock A probabilistic approach to demixing odors.
\newblock Nature Neuroscience. 2017;20:98--106.

\bibitem{litwin_nn_12}
Litwin-Kumar A, Doiron B.
\newblock Slow dynamics and high variability in balanced cortical networks with
  clustered connections.
\newblock Nature Neuroscience. 2012;15(11):1498--1505.

\bibitem{diesmann12}
Tetzlaff T, Helias M, Einevoll GT, Diesmann M.
\newblock Decorrelation of neural-network activity by inhibitory feedback.
\newblock PLoS Computational Biology. 2012;8:e1002596.

\bibitem{middleton12}
Middleton JW, Omar C, Simons BDDJ.
\newblock Neural correlation is stimulus modulated by feedforward inhibitory
  circuitry.
\newblock The Journal of Neuroscience. 2012;32:506--518.

\bibitem{litwin12}
Litwin-Kumar A, Chacron M, Doiron B.
\newblock The spatial structure of stimuli shapes the timescale of correlations
  in population spiking activity.
\newblock PLoS Computational Biology. 2012;8(9):e1002667.

\bibitem{litwin11}
Litwin-Kumar A, Oswald AM, Urban NN, Doiron B.
\newblock Balanced synaptic input shapes the correlation between neural spike
  trains.
\newblock PLoS Computational Biology. 2011;7:e1002305.

\bibitem{cohen11}
Cohen MR, Kohn A.
\newblock Measuring and interpreting neuronal correlations.
\newblock Nature Neuroscience. 2011;14:811--819.

\bibitem{hong12}
Hong S, Ratt{\'e} S, Prescott SA, De~Schutter E.
\newblock Single neuron firing properties impact correlation-based population
  coding.
\newblock The Journal of Neuroscience. 2012;32(4):1413--1428.

\bibitem{marella08}
Marella S, Ermentrout B.
\newblock Class-II neurons display a higher degree of stochastic
  synchronization than class-I neurons.
\newblock Physical Review E. 2008;77:041918.

\bibitem{abouzeid09}
Abouzeid A, Ermentrout B.
\newblock Type-II phase resetting curve is optimal for stochastic synchrony.
\newblock Physical Review E. 2009;80(1):011911.

\bibitem{barreiro10}
Barreiro AK, Shea-Brown E, Thilo EL.
\newblock Time scales of spike-train correlation for neural oscillators with
  common drive.
\newblock Physical Review E. 2010;81:011916.

\bibitem{barreiro12}
Barreiro AK, Thilo EL, Shea-Brown E.
\newblock A-current and type I/type II transition determine collective spiking
  from common input.
\newblock Journal of Neurophysiology. 2012;108(6):1631.

\bibitem{ocker14}
Ocker GK, Doiron B.
\newblock Kv7 channels regulate pairwise spiking covariability in health and
  disease.
\newblock Journal of neurophysiology. 2014;112(2):340--352.

\bibitem{rosenbaum10}
Rosenbaum R, Trousdale J, Josi{\'c} K.
\newblock Pooling and correlated neural activity.
\newblock Frontiers in Computational Neuroscience. 2010;4.

\bibitem{rosenbaum17}
Rosenbaum R, Smith MA, Kohn A, Rubin JE, Doiron B.
\newblock The spatial structure of correlated neuronal variability.
\newblock Nature Neuroscience. 2017;20(1):107--114.

\bibitem{ostojic09}
Ostojic S, Brunel N, Hakim V.
\newblock How connectivity, background activity, and synaptic properties shape
  the cross-correlation between spike trains.
\newblock The Journal of Neuroscience. 2009;29:10234--10253.

\bibitem{ly_ermentrout_09}
Ly C, Ermentrout B.
\newblock Synchronization Dynamics of Two Coupled Neural Oscillators Receiving
  Shared and Unshared Noisy Stimuli.
\newblock Journal of Computational Neuroscience. 2009;26:425--443.
\newblock doi:{10.1007/s10827-008-0120-8}.

\bibitem{trousdale12}
Trousdale J, Hu Y, Shea-Brown E, Josi{\'c} K.
\newblock Impact of network structure and cellular response on spike time
  correlations.
\newblock PLoS Comput Biol. 2012;8(3):e1002408.

\bibitem{rosenbaum13}
Rosenbaum R, Rubin JE, Doiron B.
\newblock Short-term synaptic depression and stochastic vesicle dynamics reduce
  and shape neuronal correlations.
\newblock Journal of neurophysiology. 2013;109(2):475--484.

\bibitem{mitchell09}
Mitchell JF, Sundberg KA, Reynolds JH.
\newblock Spatial attention decorrelates intrinsic activity fluctuations in
  macaque area V4.
\newblock Neuron. 2009;63(6):879--888.

\bibitem{cohen09}
Cohen MR, Maunsell JH.
\newblock Attention improves performance primarily by reducing interneuronal
  correlations.
\newblock Nature Neuroscience. 2009;12:1594--1600.

\bibitem{ruff14}
Ruff DA, Cohen MR.
\newblock Attention can either increase or decrease spike count correlations in
  visual cortex.
\newblock Nature neuroscience. 2014;17(11):1591--1597.

\bibitem{ohiorhenuan10}
Ohiorhenuan IE, Mechler F, Purpura KP, Schmid AM, Hu Q, Victor JD.
\newblock Sparse coding and high-order correlations in fine-scale cortical
  networks.
\newblock Nature. 2010;466(7306):617--621.

\bibitem{trousdale13}
Trousdale J, Hu Y, Shea-Brown E, Josi{\'c} K.
\newblock A generative spike train model with time-structured higher order
  correlations.
\newblock Frontiers in computational neuroscience. 2013;7.

\bibitem{jovanovic16}
Jovanovi{\'c} S, Rotter S.
\newblock Interplay between Graph Topology and Correlations of Third Order in
  Spiking Neuronal Networks.
\newblock PLOS Computational Biology. 2016;12(6):e1004963.

\bibitem{kaybook}
Kay SM.
\newblock Fundamentals of Statistical Signal Processing, Volume 1: Estimation
  Theory.
\newblock Prentice Hall PTR; 1993.

\bibitem{dayan2001theoretical}
Dayan P, Abbott LF.
\newblock Theoretical neuroscience: Computational and mathematical modeling of
  neural systems.
\newblock Taylor \& Francis; 2001.

\bibitem{mathis16}
Mathis A, Rokni D, Kapoor V, Bethge M, Murthy VN.
\newblock Reading out olfactory receptors: feedforward circuits detect odors in
  mixtures without demixing.
\newblock Neuron. 2016;91:1110--1123.

\bibitem{cury10}
Cury KM, Uchida N.
\newblock Robust odor coding via inhalation-coupled transient activity in the
  mammalian olfactory bulb.
\newblock Neuron. 2010;68(3):570--585.

\bibitem{gschwend12}
Gschwend O, Beroud J, Carleton A.
\newblock Encoding odorant identity by spiking packets of rate-invariant
  neurons in awake mice.
\newblock PloS one. 2012;7(1):e30155.

\bibitem{friedrich01}
Friedrich RW, Laurent G.
\newblock Dynamic optimization of odor representations by slow temporal
  patterning of mitral cell activity.
\newblock Science. 2001;291(5505):889--894.

\bibitem{nlc_15}
Nicola W, Ly C, Campbell SA.
\newblock One-Dimensional Population Density Approaches to Recurrently Coupled
  Networks of Neurons with Noise.
\newblock SIAM Journal on Applied Mathematics. 2015;75:2333--2360.

\bibitem{buice07}
Buice MA, Cowan JD.
\newblock Field-theoretic approach to fluctuation effects in neural networks.
\newblock Physical Review E. 2007;75(5):051919.

\bibitem{buice10}
Buice MA, Cowan JD, Chow CC.
\newblock Systematic fluctuation expansion for neural network activity
  equations.
\newblock Neural Computation. 2010;22:377--426.

\bibitem{ly_tranchina_07}
Ly C, Tranchina D.
\newblock Critical {A}nalysis of {D}imension {R}eduction by a {M}oment
  {C}losure {M}ethod in a {P}opulation {D}ensity {A}pproach to {N}eural
  {N}etwork {M}odeling.
\newblock Neural Computation. 2007;19:2032--2092.

\bibitem{bressloff09}
Bressloff PC.
\newblock Stochastic neural field theory and the system-size expansion.
\newblock SIAM Journal on Applied Mathematics. 2009;70(5):1488--1521.

\bibitem{touboul11}
Touboul J, Ermentrout B.
\newblock Finite-size and correlation-induced effects in mean-field dynamics.
\newblock Journal of Computational Neuroscience. 2011;31:453--484.

\bibitem{bressloff15}
Bressloff PC.
\newblock Path-Integral Methods for Analyzing the Effects of Fluctuations in
  Stochastic Hybrid Neural Networks.
\newblock Journal of Mathematical Neuroscience. 2015;5(4).

\bibitem{toyoizumi09}
Toyoizumi T, Rad KR, Paninski L.
\newblock Mean-field approximations for coupled populations of generalized
  linear model spiking neurons with Markov refractoriness.
\newblock Neural computation. 2009;21(5):1203--1243.

\bibitem{ocker2017}
Ocker GK, Josic K, Shea-Brown E, Buice MA.
\newblock Linking structure and activity in nonlinear spiking networks; 2017.

\bibitem{BC_JSM_2013}
Buice MA, Chow CC.
\newblock Beyond mean field theory: statistical field theory for neural
  networks.
\newblock Journal of Statistical Mechanics: Theory and Experiment. 2013; p.
  P03003.
\newblock doi:{10.1088/1742-5468/2013/03/P03003}.

\bibitem{BC_PLOSCB_2013}
Buice M, Chow CC.
\newblock Dynamic finite size effects in spiking neural networks.
\newblock PLoS Comput Biol. 2013;9:e1002872.

\bibitem{van96}
van Vreeswijk C, Sompolinsky H.
\newblock Chaos in neuronal networks with balanced excitatory and inhibitory
  activity.
\newblock Science. 1996;274:1724--1726.

\bibitem{keaneGong15}
Keane A, Gong P.
\newblock Propagating waves can explain irregular neural dynamics.
\newblock Journal of Neuroscience. 2015;35(4):1591--1605.

\bibitem{ermentrout94}
Ermentrout B.
\newblock Reduction of conductance-based models with slow synapses to neural
  nets.
\newblock Neural Computation. 1994;6(4):679--695.

\bibitem{aviel06}
Aviel Y, Gerstner W.
\newblock From spiking neurons to rate models: A cascade model as an
  approximation to spiking neuron models with refractoriness.
\newblock Physical Review E. 2006;73(5):051908.

\bibitem{truccolo05}
Truccolo W, Eden UT, Fellows MR, Donoghue JP, Brown EN.
\newblock A point process framework for relating neural spiking activity to
  spiking history, neural ensemble, and extrinsic covariate effects.
\newblock Journal of neurophysiology. 2005;93(2):1074--1089.

\bibitem{gautam12}
Gautam SH, Verhagen JV.
\newblock Retronasal odor representations in the dorsal olfactory bulb of rats.
\newblock The Journal of Neuroscience. 2012;32(23):7949--7959.

\bibitem{gautam15}
Gautam SH, Hoang TT, McClanahan K, Grady SK, Shew WL.
\newblock Maximizing sensory dynamic range by tuning the cortical state to
  criticality.
\newblock PLoS Computational Biology. 2015;11(12):e1004576.

\bibitem{rossant16}
Rossant C, Kadir S, Goodman D, Schulman J, Hunter M, Saleem A, et~al.
\newblock Spike sorting for large, dense electrode arrays.
\newblock Nature neuroscience. 2016;19(4):634--641.

\bibitem{gardiner}
Gardiner CW.
\newblock Handbook of stochastic methods.
\newblock Springer-Verlag; 1985.

\bibitem{neville03}
Neville KR, Haberly LB.
\newblock Beta and gamma oscillations in the olfactory system of the
  urethane-anesthetized rat.
\newblock Journal of Neurophysiology. 2003;90(6):3921--3930.

\end{thebibliography}

\setcounter{page}{1}

\newpage

\begin{center}
{\Large
\textbf{Experimental Data Statistics by Odor}
}


\vspace{0.5cm}

{\large

Andrea K. Barreiro$^{\#}$ ; Shree Hari Gautam$^{\ddag}$ ; Woodrow L. Shew$^{\ddag}$ ; Cheng Ly$^{\dag}$ 
}
\vspace{0.5cm}

{\small

{\it \# Department of Mathematics, Southern Methodist University, Dallas, TX 75275  U.S.A.}
\\
{\it \ddag Department of Physics, University of Arkansas, Fayetteville, AR  72701  U.S.A.}
\\
{\it \dag Department of Statistical Sciences and Operations Research, Virginia Commonwealth University, Richmond, VA 23284  U.S.A.}
\\
$\ast$ E-mail: abarreiro@smu.edu ; shgautam@uark.edu; woodrowshew@gmail.com ; CLy@vcu.edu
}
\end{center}

\renewcommand\thefigure{S\arabic{figure}}    
\setcounter{figure}{0}   
 
\setcounter{table}{0}
\renewcommand{\thetable}{S\arabic{table}}


\begin{table}[!ht]
\centering
\caption{ {\bf Average population firing rate (Hz) by odor and activity state.}}
\label{table:mn_frate}
\begin{tabular}{|c|lllll|}
\hline
\multicolumn{1}{|l|}{Mean across population} & All Odors & Odor 1 & Odor 2 & Odor 3 & Odor 4 \\ \hline
$\nu_{OB}^{Sp}$                              &      1.97     &    1.44    &  1.8       &     2.62   &  2.24       \\
$\nu_{OB}^{Ev}$                              &        4.66  &       4.91 &  3.28      &      5.41  &  5.34      \\ \hline
$\nu_{PC}^{Sp}$                              &         0.75 &        0.56 &  0.91      &      0.74  &  0.79      \\
$\nu_{PC}^{Ev}$                              &         1.45  &        1.6 &   1.26      &      1.7  &   1.23     \\ \hline
\end{tabular}
\end{table}

\begin{table}[!ht]
\centering
\caption{ {\bf Standard deviation of firing rate across the population (Hz) by odor and activity state.}}
\label{table:std_frate}
\begin{tabular}{|c|lllll|}
\hline
\multicolumn{1}{|l|}{Std. Dev. across population} & All Odors & Odor 1 & Odor 2 & Odor 3 & Odor 4 \\ \hline
$\nu_{OB}^{Sp}$                              &      3.28     &     2.34   &    3.07     &      4.32  & 3.58       \\
$\nu_{OB}^{Ev}$                              &       7.14    &      7.55 &    5.55       &      8  &   8.04     \\ \hline
$\nu_{PC}^{Sp}$                              &        0.93   &      0.83  &  1.08       &     0.96   &  0.95      \\
$\nu_{PC}^{Ev}$                              &        1.58   &       2.09 &   1.45      &        1.93 &  1.18      \\ \hline
\end{tabular}
\end{table}

\begin{figure}
\centering
 \includegraphics[width=\textwidth]{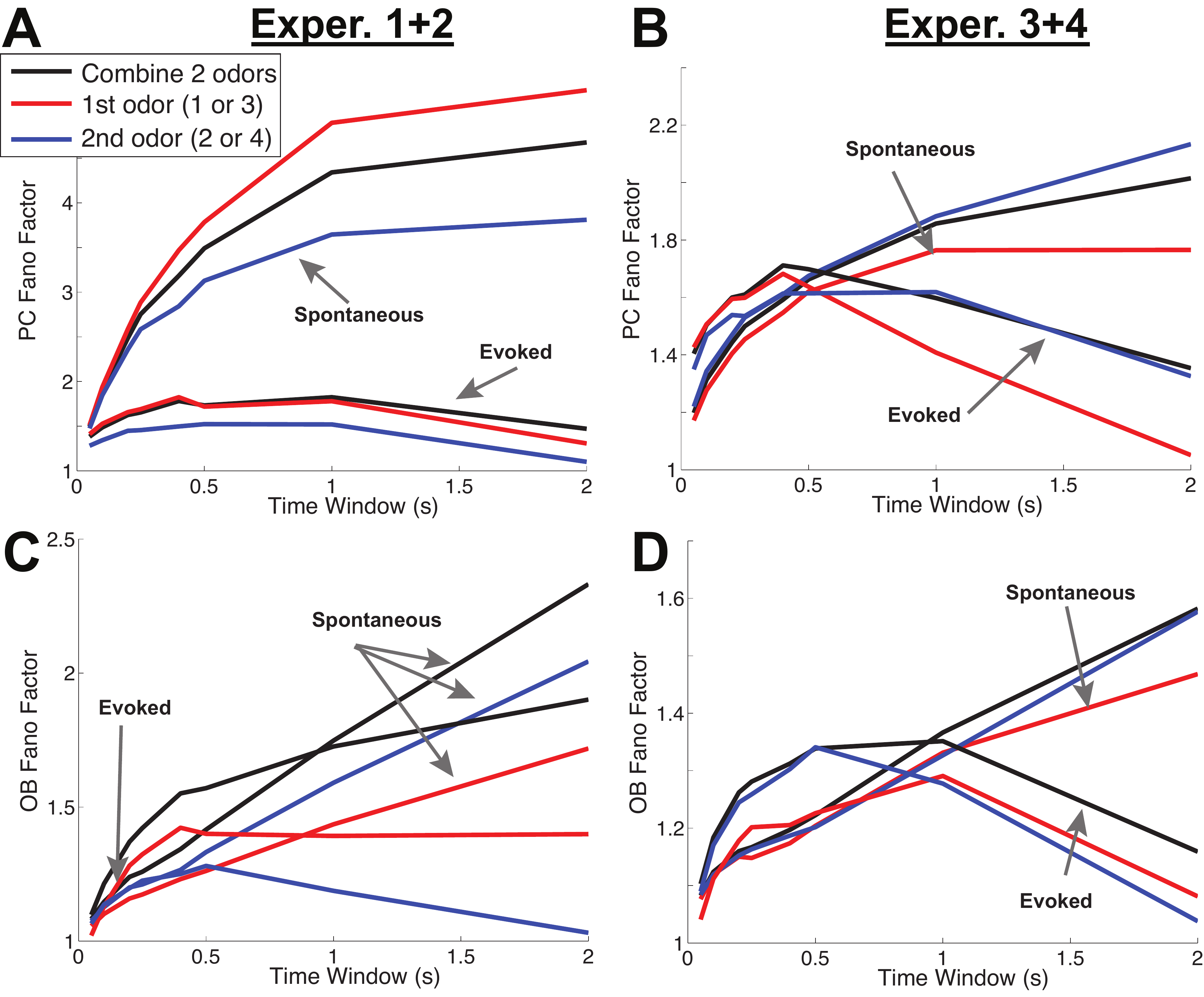}
\caption{\label{fig:FF_spEvk} {\bf Experimental statistics by odor and activity state: Fano Factor.} Comparing the {\bf mean} Fano Factor across all simultaneously cells with: 
i) pairs from the 2 stimuli (black), ii) from the first odor (red), iii) from the second odor (blue).  A and B are the PC cells, C and D are the OB cells.  
The left column A), C) is from \texttt{data040515\_exp1+2.mat}, and the right column B), D) is from \texttt{data040515\_exp3+4.mat}
The spontaneous and evoked states in groups of 3 curves are denoted by the gray arrows.
} 
\end{figure}

\begin{figure}
\centering
 \includegraphics[width=\textwidth]{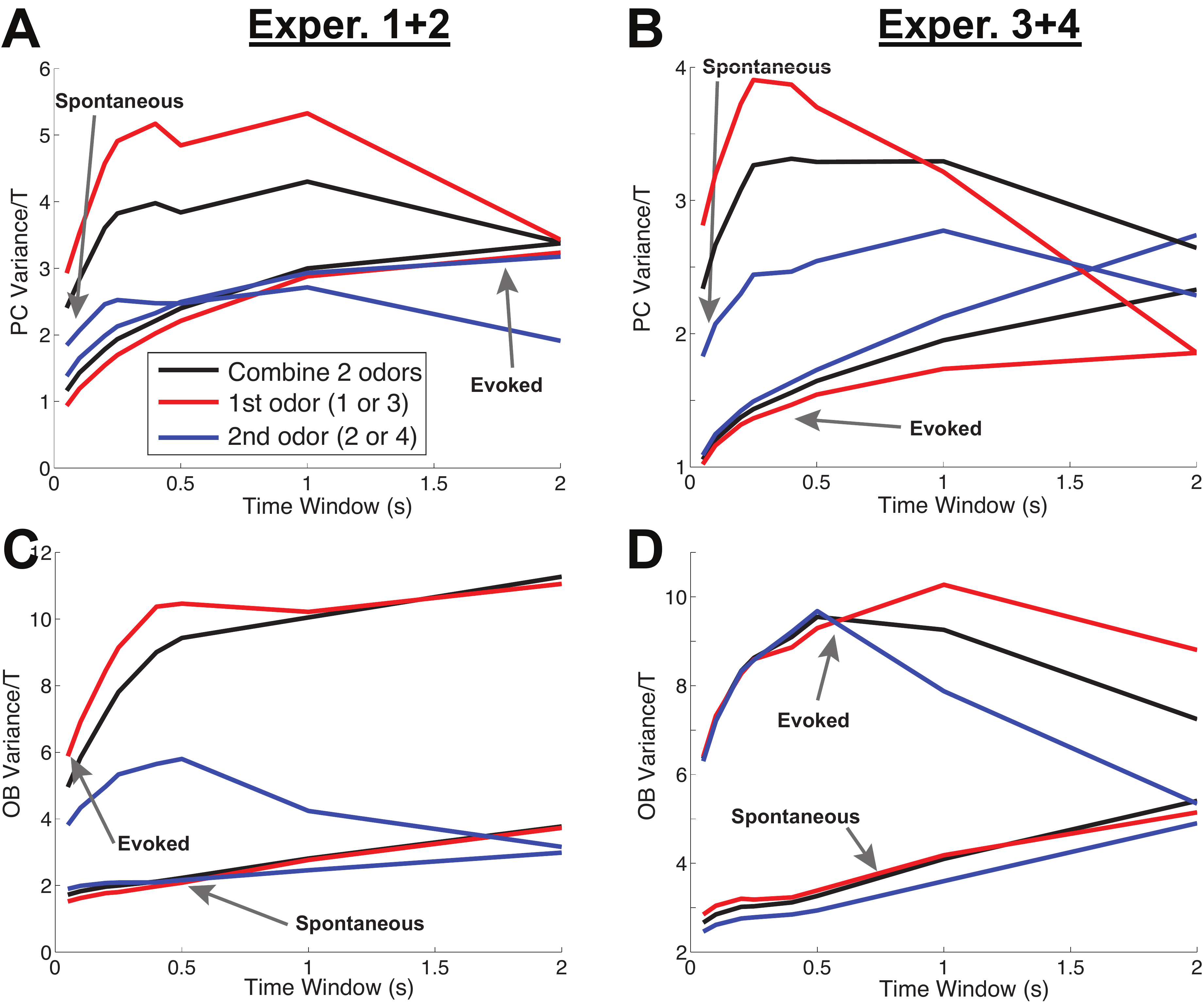}
\caption{\label{fig:VR_spEvk} {\bf Experimental statistics by odor and activity state: spike count variance.} Similar to Fig. \ref{fig:FF_spEvk} but comparing the {\bf mean} variance divided by time window across all simultaneously cells with: 
i) pairs from the 2 stimuli (black), ii) from the first odor (red), iii) from the second odor (blue).    A and B are the PC cells, C and D are the OB cells.  
The left column A), C) is from \texttt{data040515\_exp1+2.mat}, and the right column B), D) is from \texttt{data040515\_exp3+4.mat}
The spontaneous and evoked states in groups of 3 curves are denoted by the gray arrows.
} 
\end{figure}

\begin{figure}
\centering
 \includegraphics[width=\textwidth]{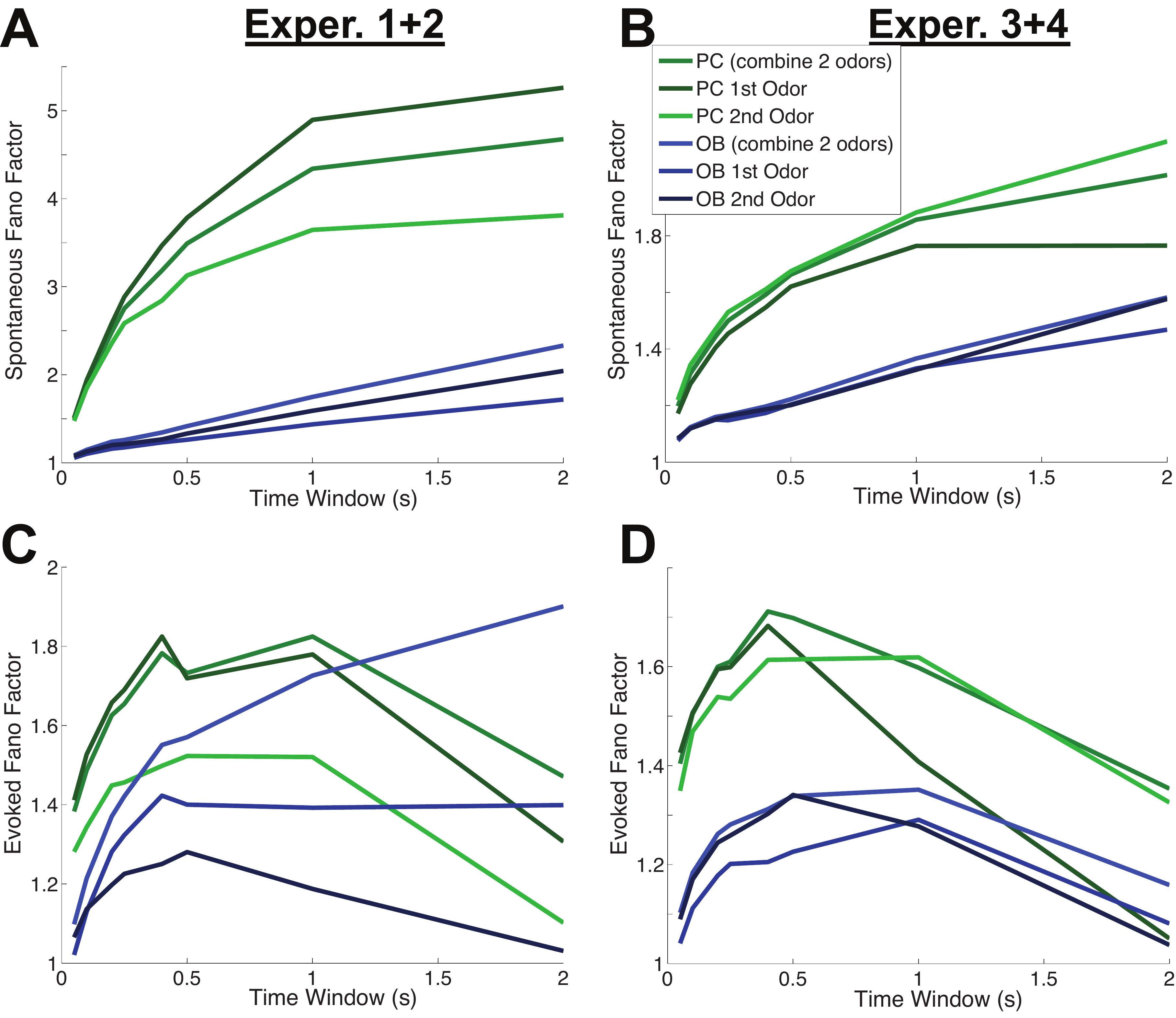}
\caption{\label{fig:FF_pcOB} {\bf Experimental statistics by odor and region: Fano Factor.} Comparing the {\bf mean} Fano Factor between recorded PC (3 green curve) and OB (3 blue curves) cells, with: 
i) pairs from the 2 stimuli , ii) from the first odor, iii) from the second odor (see figure legend for color convention).  A and B is for the spontaneous state, C and D is for the evoked state.  
The left column A), C) is from \texttt{data040515\_exp1+2.mat}, and the right column B), D) is from \texttt{data040515\_exp3+4.mat}. 
} 
\end{figure}

\begin{figure}
\centering
 \includegraphics[width=\textwidth]{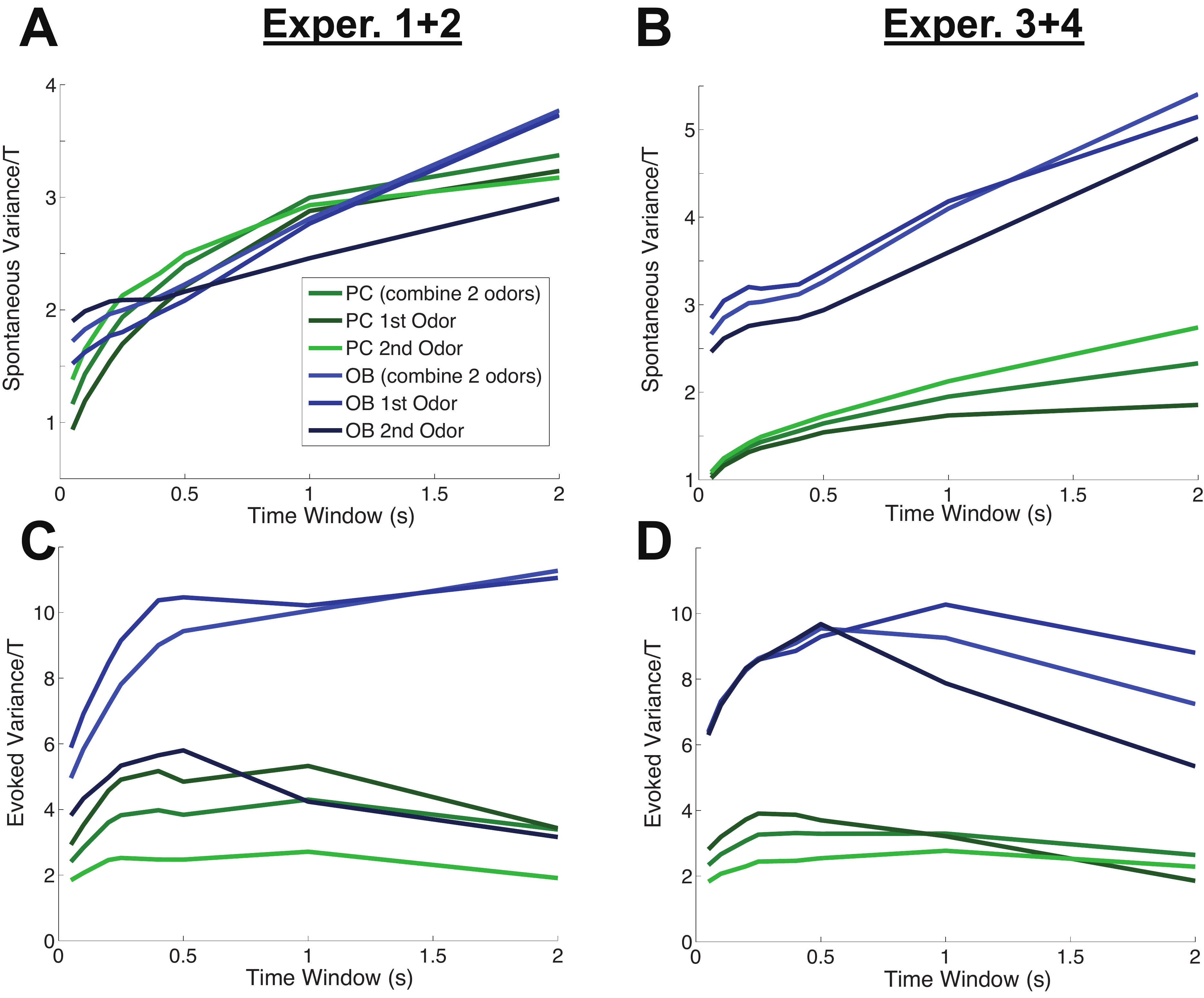}
\caption{\label{fig:VR_pcOB}  {\bf Experimental statistics by odor and region: spike count variance.} Comparing the {\bf mean} variance divided by time window between recorded PC (3 green curve) and OB (3 blue curves) cells, with: 
i) pairs from the 2 stimuli , ii) from the first odor, iii) from the second odor (see figure legend for color convention).  A and B is for the spontaneous state, C and D is for the evoked state.  
The left column A), C) is from \texttt{data040515\_exp1+2.mat}, and the right column B), D) is from \texttt{data040515\_exp3+4.mat}. 
} 
\end{figure}


\begin{figure}
\centering
 \includegraphics[width=0.8\textwidth]{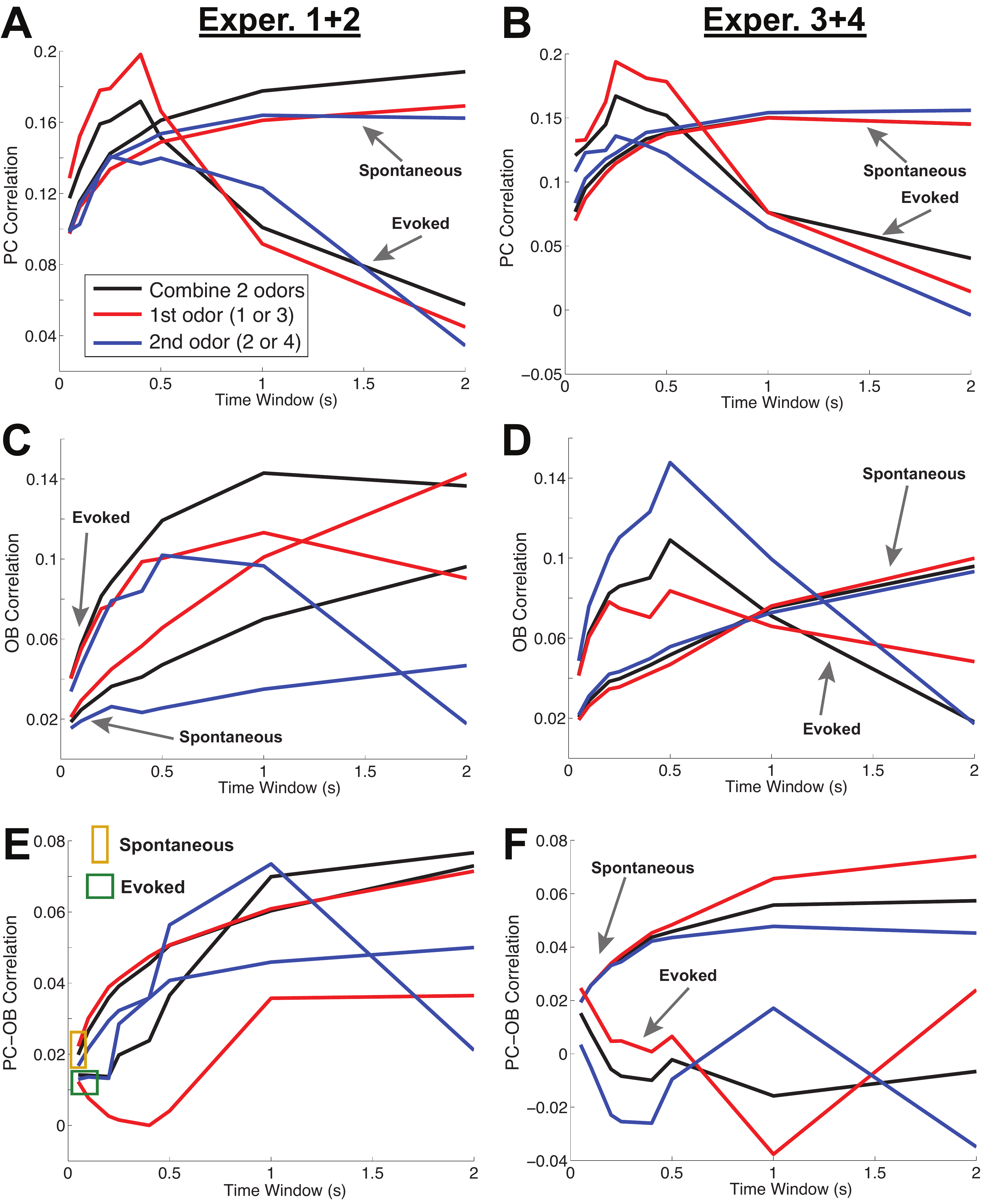}
\caption{\label{fig:corr_spEvk}  {\bf Experimental statistics by odor and activity state: spike count correlation.} Comparing the {\bf mean} spike count correlation across all simultaneously recorded pairs with: 
i) pairs from the 2 stimuli (black), ii) from the first odor (red), iii) from the second odor (blue).  
The left column A), C), E) is from \texttt{data040515\_exp1+2.mat}, and the right column B), D), F) is from \texttt{data040515\_exp3+4.mat}
The spontaneous and evoked states in groups of 3 curves are denoted by the gray arrows.
} 
\end{figure}

\begin{figure}
\centering
 \includegraphics[width=0.8\textwidth]{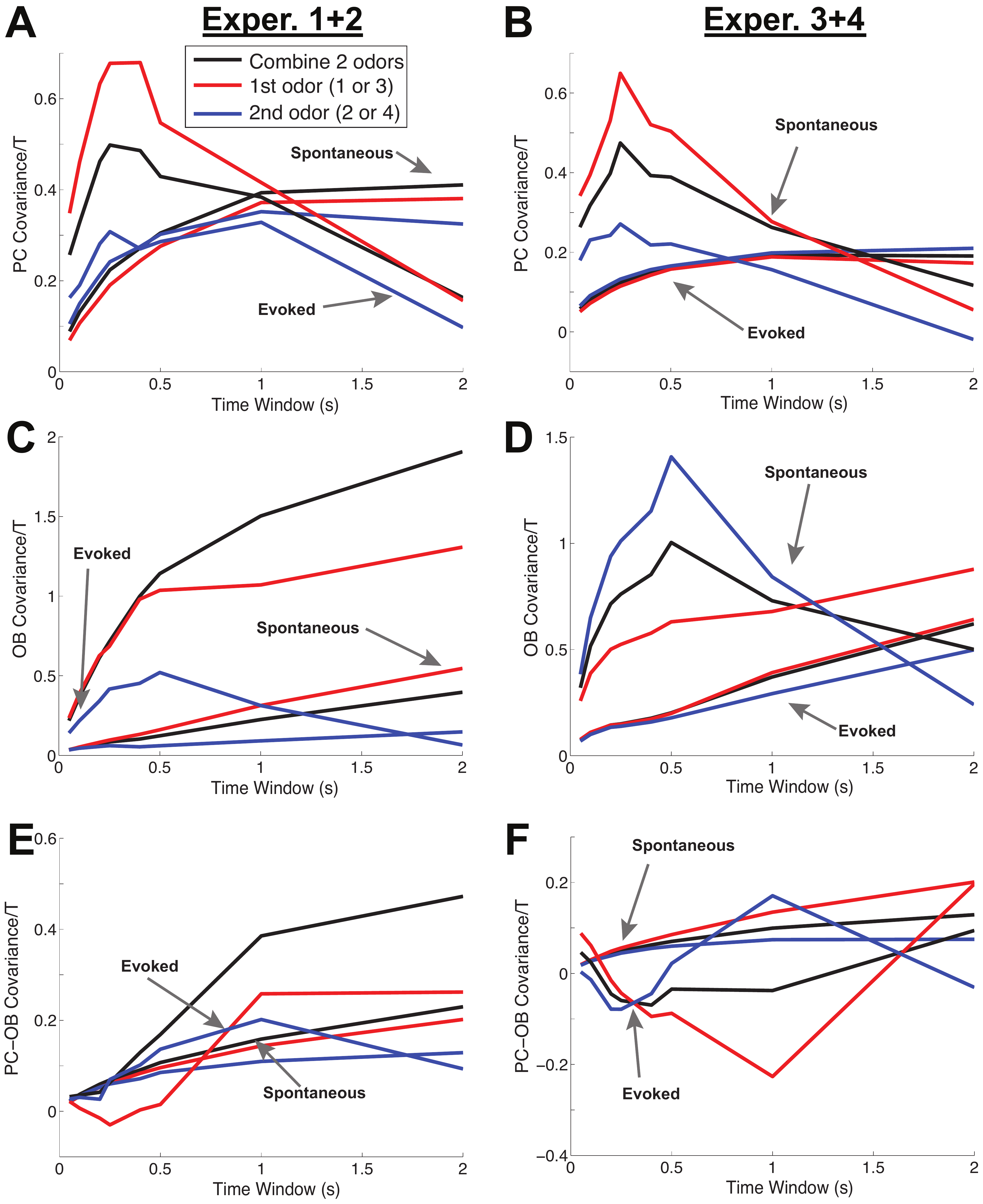}
\caption{\label{fig:cov_spEvk}  {\bf Experimental statistics by odor and activity state: spike count covariance.} Similar to Fig. \ref{fig:corr_spEvk} but comparing the {\bf mean} spike count covariance divided by time window across all simultaneously recorded pairs with: 
i) pairs from the 2 stimuli (black), ii) from the first odor (red), iii) from the second odor (blue).  
The left column A), C), E) is from \texttt{data040515\_exp1+2.mat}, and the right column B), D), F) is from \texttt{data040515\_exp3+4.mat}
The spontaneous and evoked states in groups of 3 curves are denoted by the gray arrows.
} 
\end{figure}

\begin{figure}
\centering
 \includegraphics[width=\textwidth]{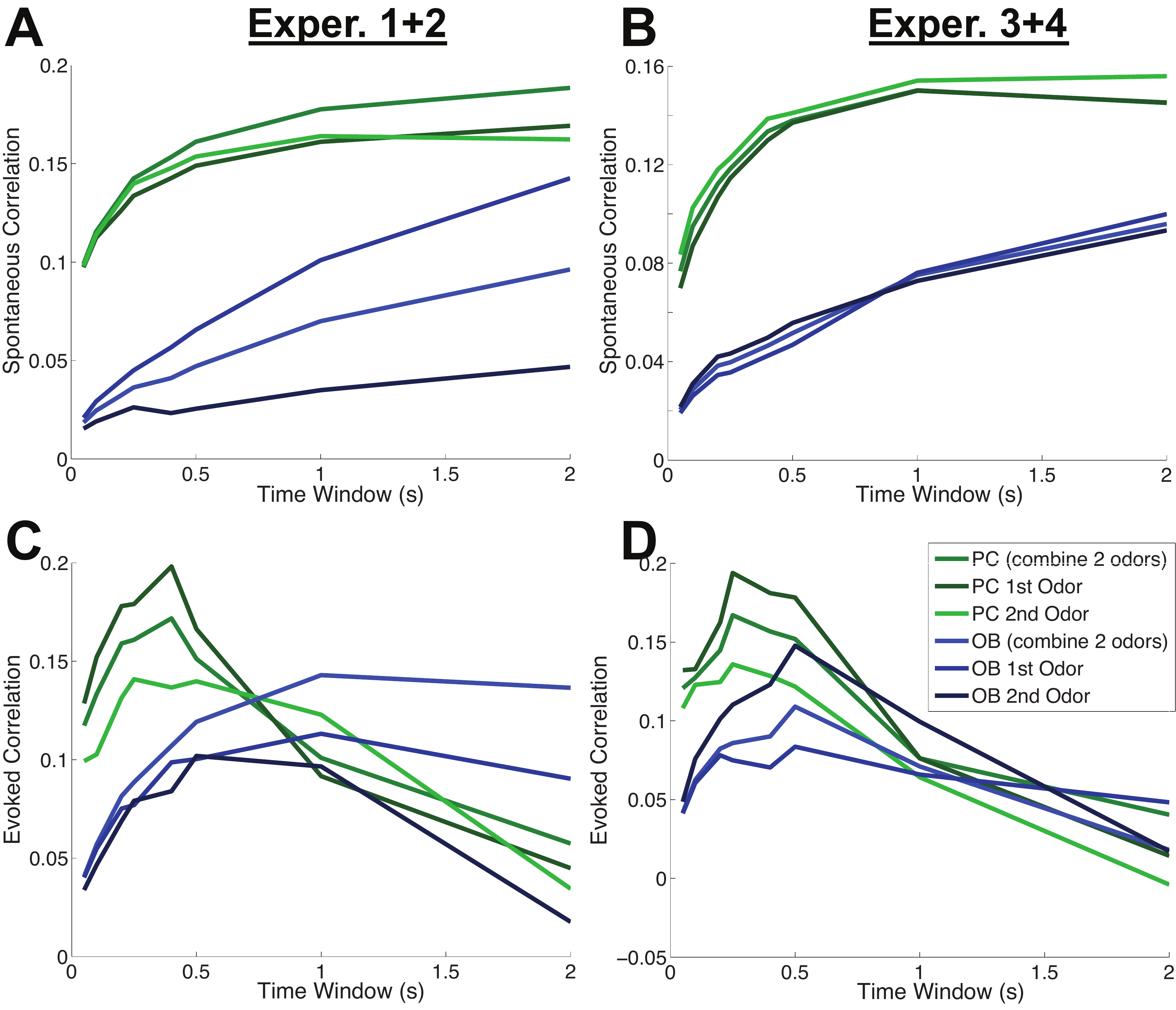}
\caption{\label{fig:corr_pcOB}  {\bf Experimental statistics by odor and region: spike count correlation.} Comparing the {\bf mean} spike count correlation between all pairs of PC (3 green curve) and OB (3 blue curves) cells, with: 
i) pairs from the 2 stimuli , ii) from the first odor, iii) from the second odor (see figure legend for color convention).  This data constraint was chosen because for larger time windows, it held by odor and experiments.  
The left column A), C) is from \texttt{data040515\_exp1+2.mat}, and the right column B), D) is from \texttt{data040515\_exp3+4.mat}. 
} 
\end{figure}

\begin{figure}
\centering
 \includegraphics[width=\textwidth]{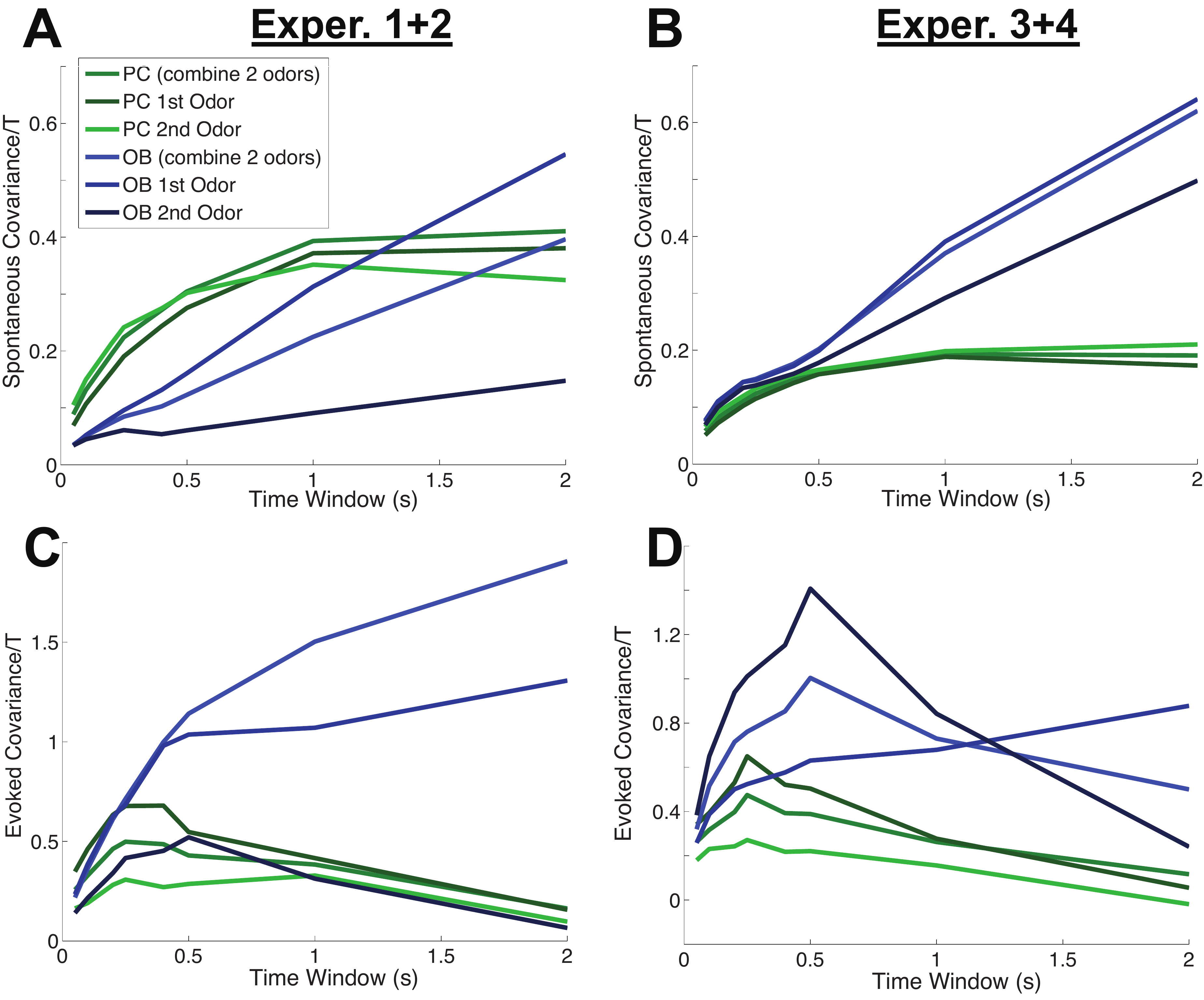}
\caption{\label{fig:cov_pcOB}  {\bf Experimental statistics by odor and region: spike count covariance.} Comparing the {\bf mean} spike count covariance divided by time window between all pairs of PC (3 green curve) and OB (3 blue curves) cells, with: 
i) pairs from the 2 stimuli , ii) from the first odor, iii) from the second odor (see figure legend for color convention).  
The left column A), C), E) is from \texttt{data040515\_exp1+2.mat}, and the right column B), D), F) is from \texttt{data040515\_exp3+4.mat}. 
} 
\end{figure}

\clearpage

\setcounter{page}{1}

\begin{center}
{\Large
\textbf{Supplementary Figures for the Main Modeling}
}


\vspace{0.5cm}

{\large

Andrea K. Barreiro$^{\#}$ ; Shree Hari Gautam$^{\ddag}$ ; Woodrow L. Shew$^{\ddag}$ ; Cheng Ly$^{\dag}$ 
}
\vspace{0.5cm}

{\small

{\it \# Department of Mathematics, Southern Methodist University, Dallas, TX 75275  U.S.A.}
\\
{\it \ddag Department of Physics, University of Arkansas, Fayetteville, AR  72701  U.S.A.}
\\
{\it \dag Department of Statistical Sciences and Operations Research, Virginia Commonwealth University, Richmond, VA 23284  U.S.A.}
\\
$\ast$ E-mail: abarreiro@smu.edu ; shgautam@uark.edu; woodrowshew@gmail.com ; CLy@vcu.edu
}
\end{center}

\renewcommand\thefigure{S\arabic{figure}}    
\setcounter{figure}{8}   
 
\setcounter{table}{2}
\renewcommand{\thetable}{S\arabic{table}}

\begin{figure}
\centering
 \includegraphics[width=\textwidth]{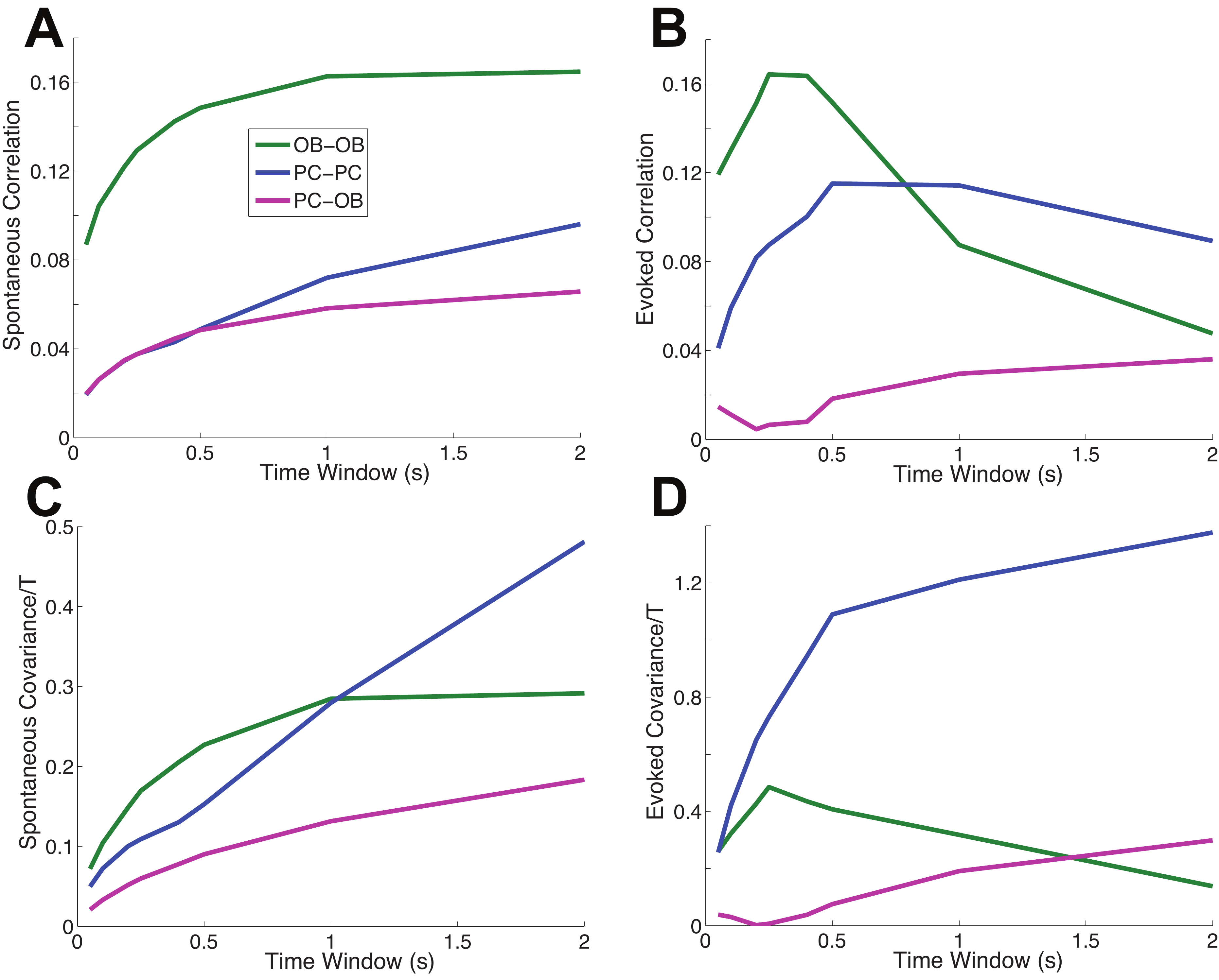}
\caption{\label{fig:lowPCOB} {\bf Cross-region correlations are smaller than within-region correlations.} The experimental data shows that the PC-OB correlation and covariance is small (on average) compared to both OB and PC.  
A: In the spontaneous state, the (average) Fano Factor of the PC cells is larger than the OB cells.  
B: In the evoked state, the (average) variance of spike counts of OB cells is larger than the PC cells; here, we have divided by the time window for illustration purposes 
(which obviously does not change the relationship).  In both A and B, there are 73 PC cells and 41 OB cells.  
C: In the evoked state, the (average) OB covariance is larger than the PC covariance.  D: The evoked variance among OB cells is larger than the spontaneous OB variance.  
In C and D, the covariances were scaled by the time window for illustration purposes, and there were 1298 pairs of PC cells and 406 pairs of OB cells.
} 
\end{figure}

\begin{figure}
\centering
 \includegraphics[width=\textwidth]{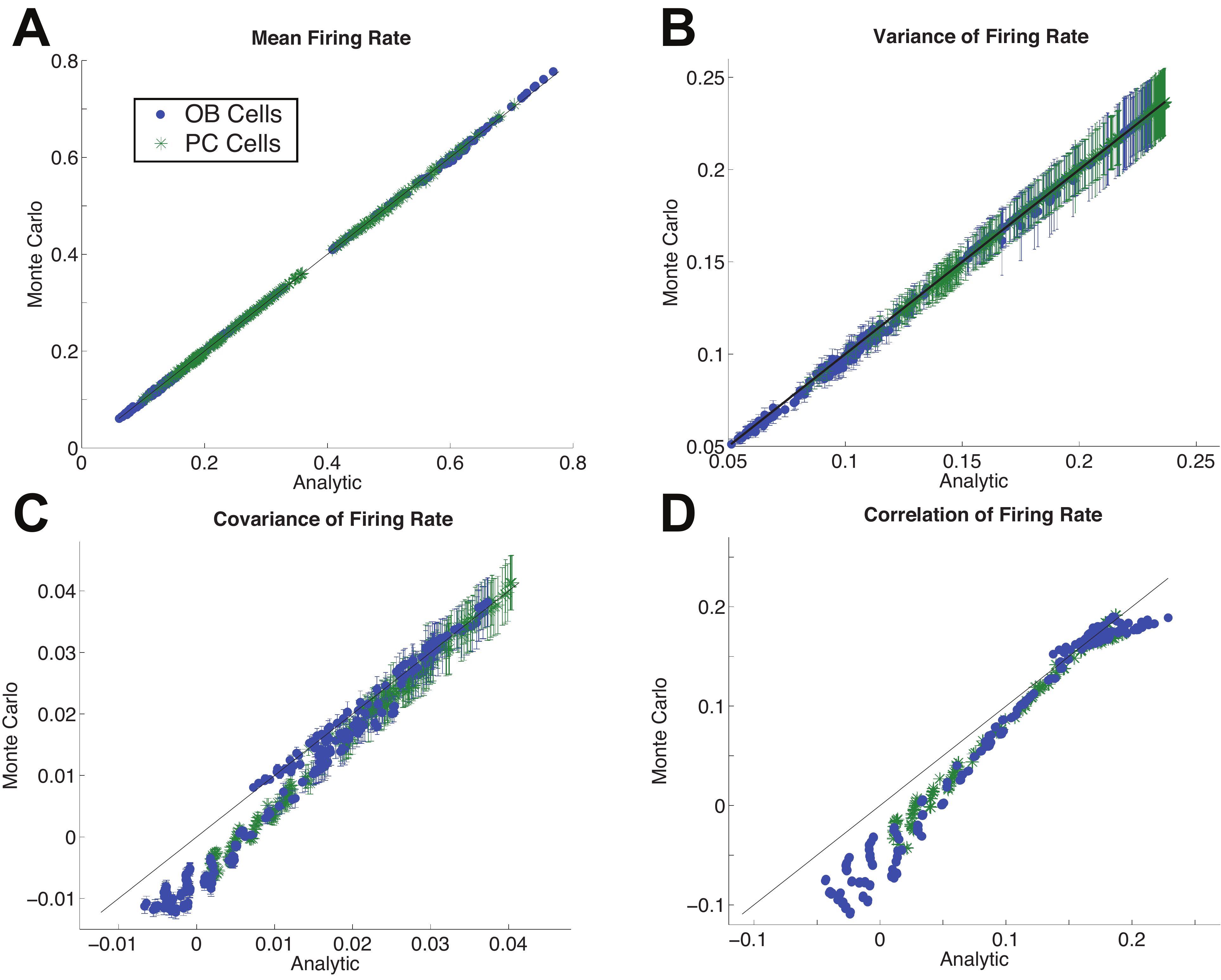}
\caption{\label{fig:mcAn} {\bf Fast analytic approximation accurately captures statistics of a multi-population firing rate model.} Comparing the results of the fast analytic approximation to Monte Carlo simulations from 100 randomly selected parameters in the 6 equation rate model: $-2\leq gIO <0$, $-2\leq gIP<0$, $0<gEO\leq2$, $0<gEP\leq2$.  
Comparing 4 important firing rate statistics on a cell by cell basis (i.e., not the average across the population); the statistics for the activity $X_j$ are just as accurate (not shown).  
A: The mean firing rate $F(X_j)$.  B: The variance of the firing rate $Var(F(X_j))$.  
C: The covariance of the firing rate between OB pairs and PC pairs (we do not focus on OB--PC pairs): $Cov(F(X_j),F(X_k))$.  
D: The correlation of the firing rate between OB and PC pairs: $\rho=Cov(F(X_j),F(X_k))/\sqrt{Var(F(X_j))Var(F(X_k))}$.  The fast analytic approximation is accurate (dots lie on the diagonal line).  Error bars 
are shown in B and C, representing 95\% confidence intervals assuming a normal distribution for finite number of realizations, or 1.96 standard deviations above and below the mean.
} 
\end{figure}

\begin{figure}
\centering
 \includegraphics[width=0.9\textwidth]{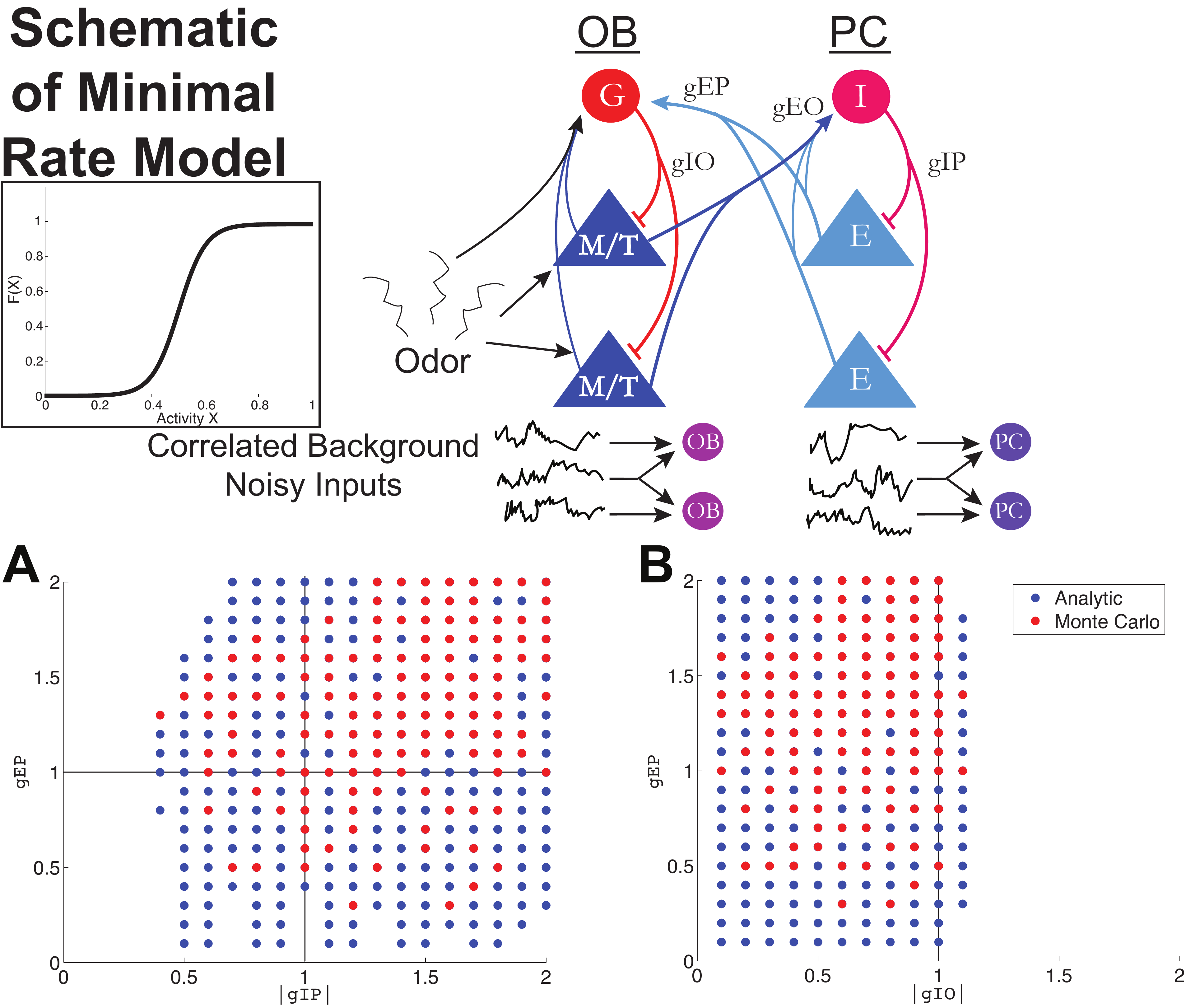}
\caption{\label{fig:wcOther2t} {\bf Experimental observations constrain conductance parameters in analytic model.} The final 2 relationships between the 4 conductance parameters from the fast analytic theory for the rate model not shown in the main text with 
$F(X)=\frac{1}{2}\left(1+\tanh((X-0.5)/0.1) \right)$.  
A: Both $gEP$ and $| gIP |$ are relatively large.  B: $| gIO |$ is relatively small.  
} 
\end{figure}

\begin{figure}
\centering
 \includegraphics[width=\textwidth]{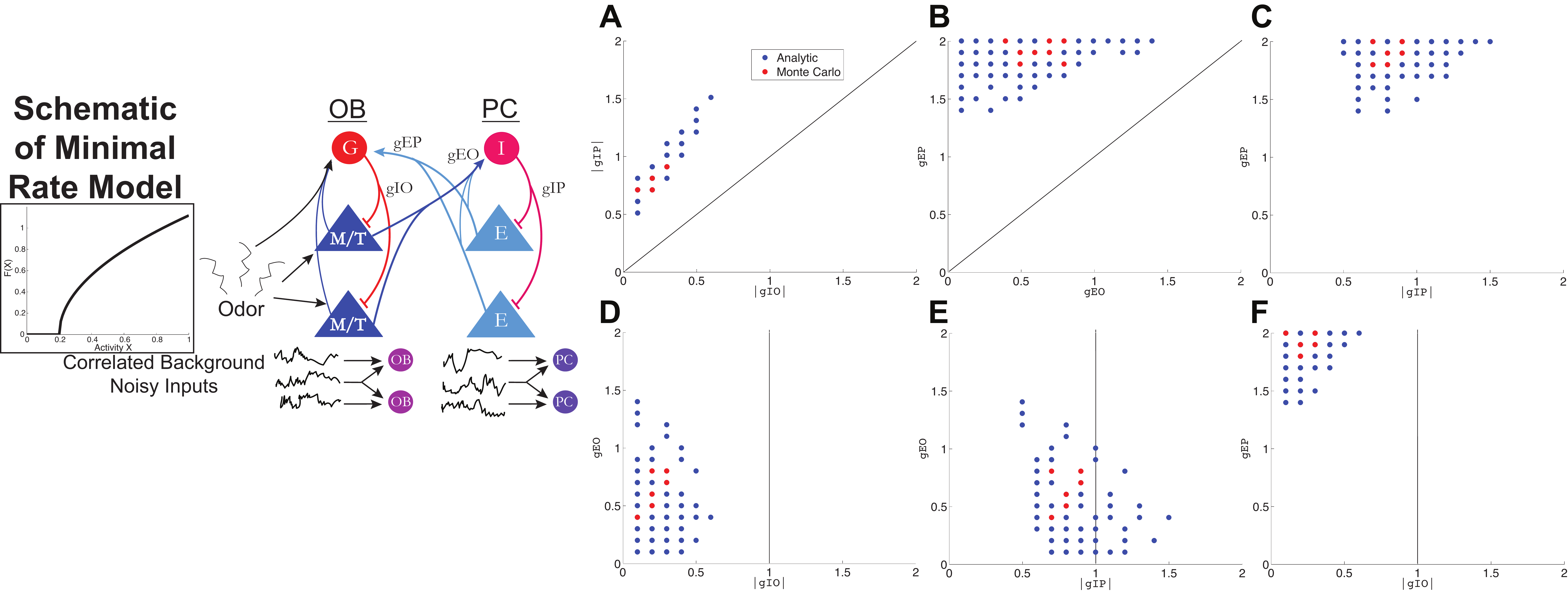}
\caption{\label{fig:wcSqrt} {\bf Analytic approximation results are robust to choice of transfer function.} The results of the fast analytic theory for the rate model using a truncated square root transfer function 
$F(X)=1.25\sqrt{X-0.2}H(X-0.2)$ are qualitatively similar to the results with the more common sigmoidal function in the main text.  Here we have omitted the E to I connections within OB and PC because 
it does not qualitatively change the results.  
A: The inhibitory conductance within the PC population $| gIP |$ is larger than in the OB population $gOP$.  
B: The excitatory conductance from PC to OB $gEP$ is generally larger than OB to PC $gEO$.  C: Both $gEP$ and $| gIP |$ are relatively large.  
D: $| gIO |$ is relatively small.  E: $| gIP |$ is relatively large.  F: Again, $| gIO |$ is relatively small.
} 
\end{figure}

\begin{figure}
\centering
 \includegraphics[width=\textwidth]{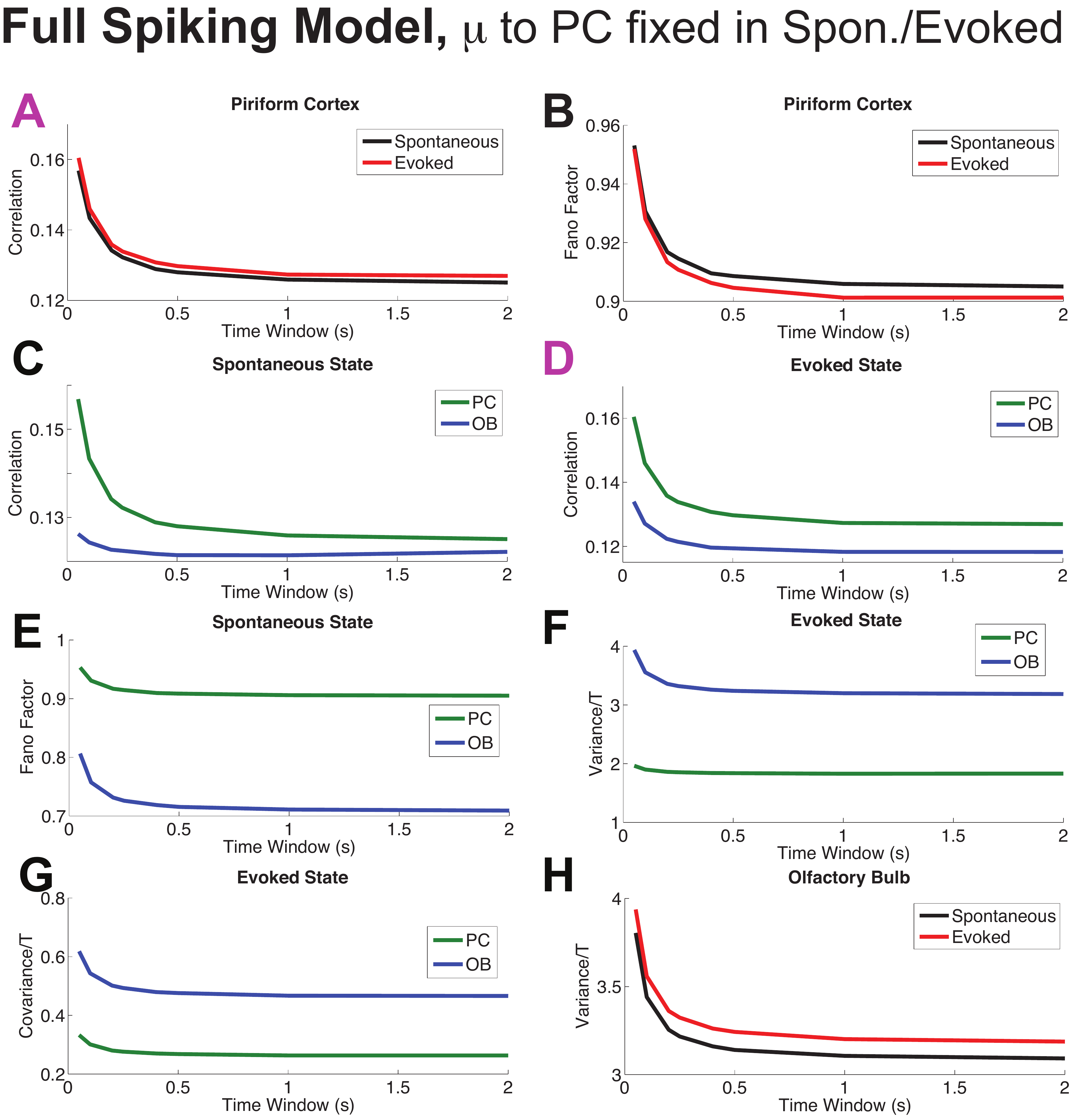}
\caption{\label{fig:fixMuPC} {\bf Mean input to PC must increase in the evoked state.}  Showing the results of the full LIF spiking model when the mean input to PC is the same in spontaneous and evoked states: $\mu_{PC}=0$. 
The rest of the parameters are the same as in Figure 6 (see main text).  
The firing rates are: $\nu_{OB}^{Sp}=5.5\pm4.6$, $\nu_{OB}^{Ev}=5.7\pm4.6$, $\nu_{PC}^{Sp}=2.096\pm2.6$, and $\nu_{PC}^{Ev}=2.13\pm2.6$, 
which barely satisfies the constraint from the experimental data that $\nu_{PC}^{Ev}>\nu_{PC}^{Sp}$.  
The 8 panels show the constraints on the 2nd order spiking statistics in the same format as in Figure 6 of the main text.  
The evoked PC correlations decrease but not enough; panels A and D with magenta coloring show the 2 constraints that are violated.
} 
\end{figure}



\begin{figure}
\centering
 \includegraphics[width=\textwidth]{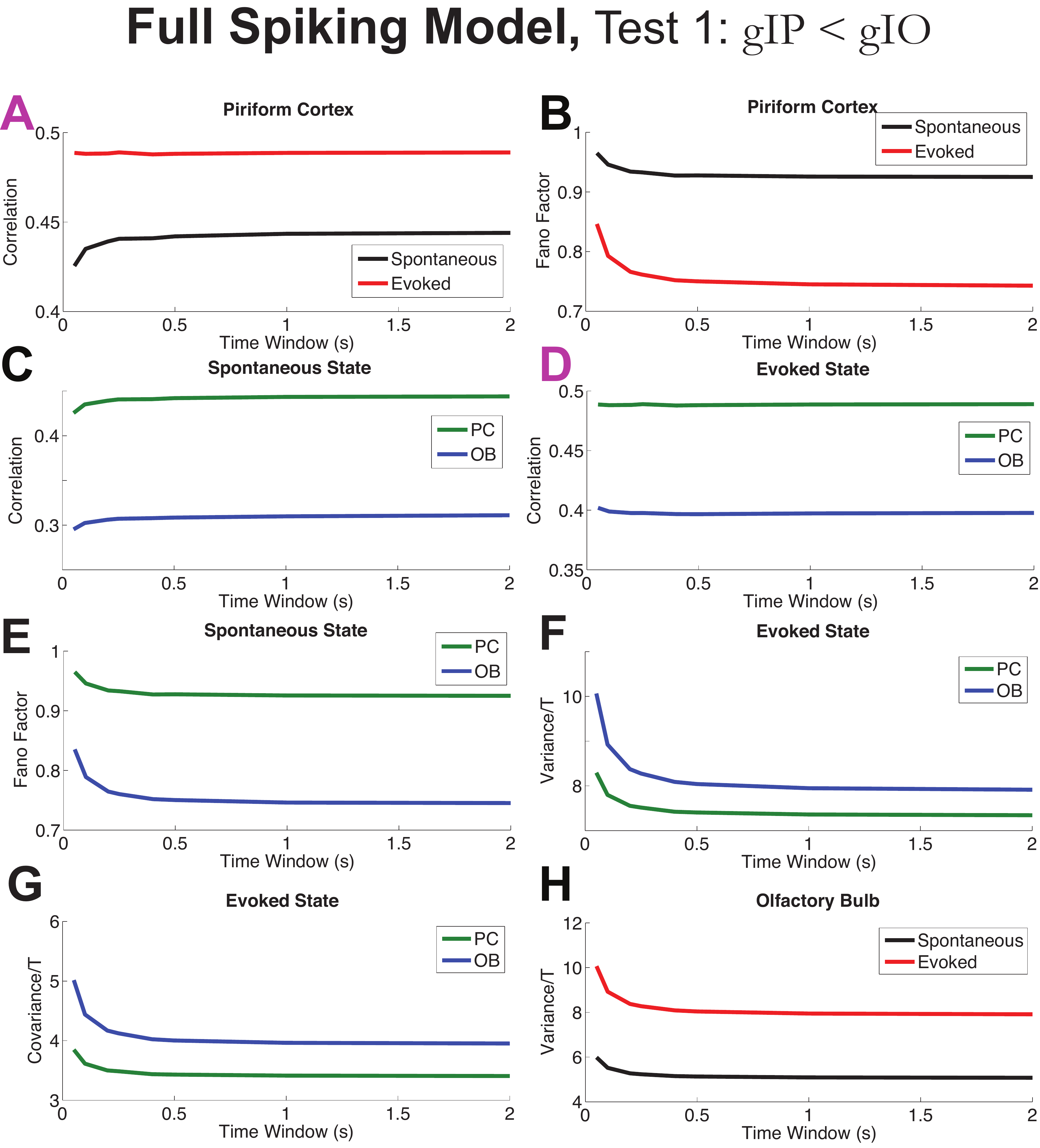}
\caption{\label{fig:test1}  {\bf Violating derived relationship $| gIO | < |gIP |$ results in statistics that are inconsistent with experimental observations.} Showing the results of the full LIF spiking model when $gIP<gIO$; specifically, we set $gIP=7$ and $gIO=20$ and set the values of the rest of the parameters to those used in 
Figure xxx (see main text).  
The firing rates are: $\nu_{OB}^{Sp}=7.82\pm5.64$, $\nu_{OB}^{Ev}=13.42\pm8.36$, $\nu_{PC}^{Sp}=3.8\pm2.82$, and $\nu_{PC}^{Ev}=9.67\pm6.36$.  
The 8 panels show the constraints on the 2nd order spiking statistics in the same format as in Figure xxx of the main text.  Two constraints are violated; 
the panels with magenta letters (i.e., A, D) are constraints that are violated.
} 
\end{figure}

\begin{figure}
\centering
 \includegraphics[width=\textwidth]{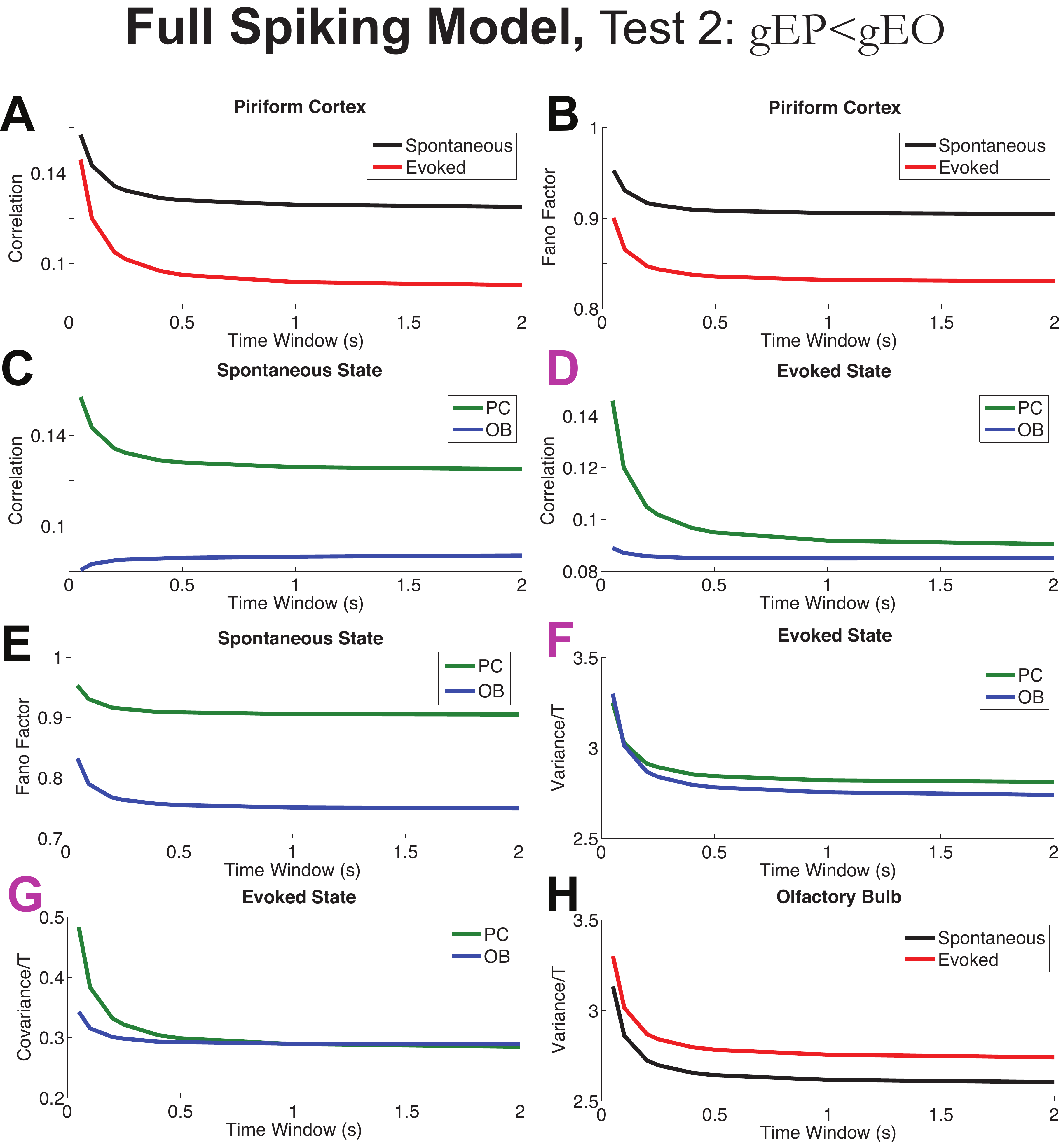}
\caption{\label{fig:test2} {\bf Violating derived relationship $gEP > gEO$ results in statistics that are inconsistent with experimental observations.} 
Showing the results of the full LIF spiking model when $gEP<gEO$; specifically, we set $gEP=1$ and $gEO=15$; 
we set the values of the rest of the parameters to those used in Figure 6 (see main text).  
The firing rates are: $\nu_{OB}^{Sp}=4.42\pm 4.09$, $\nu_{OB}^{Ev}=4.63\pm 4.01$, $\nu_{PC}^{Sp}=2.1\pm 2.64$, and $\nu_{PC}^{Ev}=4.17\pm 5.81$.  
The 8 panels show the constraints on the 2nd order spiking statistics.  Three constraints are violated (D, F, G in magenta); note that in G the constraints are violated for small time windows and 
almost indistinguishable for large time windows. } 
\end{figure}

\begin{figure}
\centering
 \includegraphics[width=\textwidth]{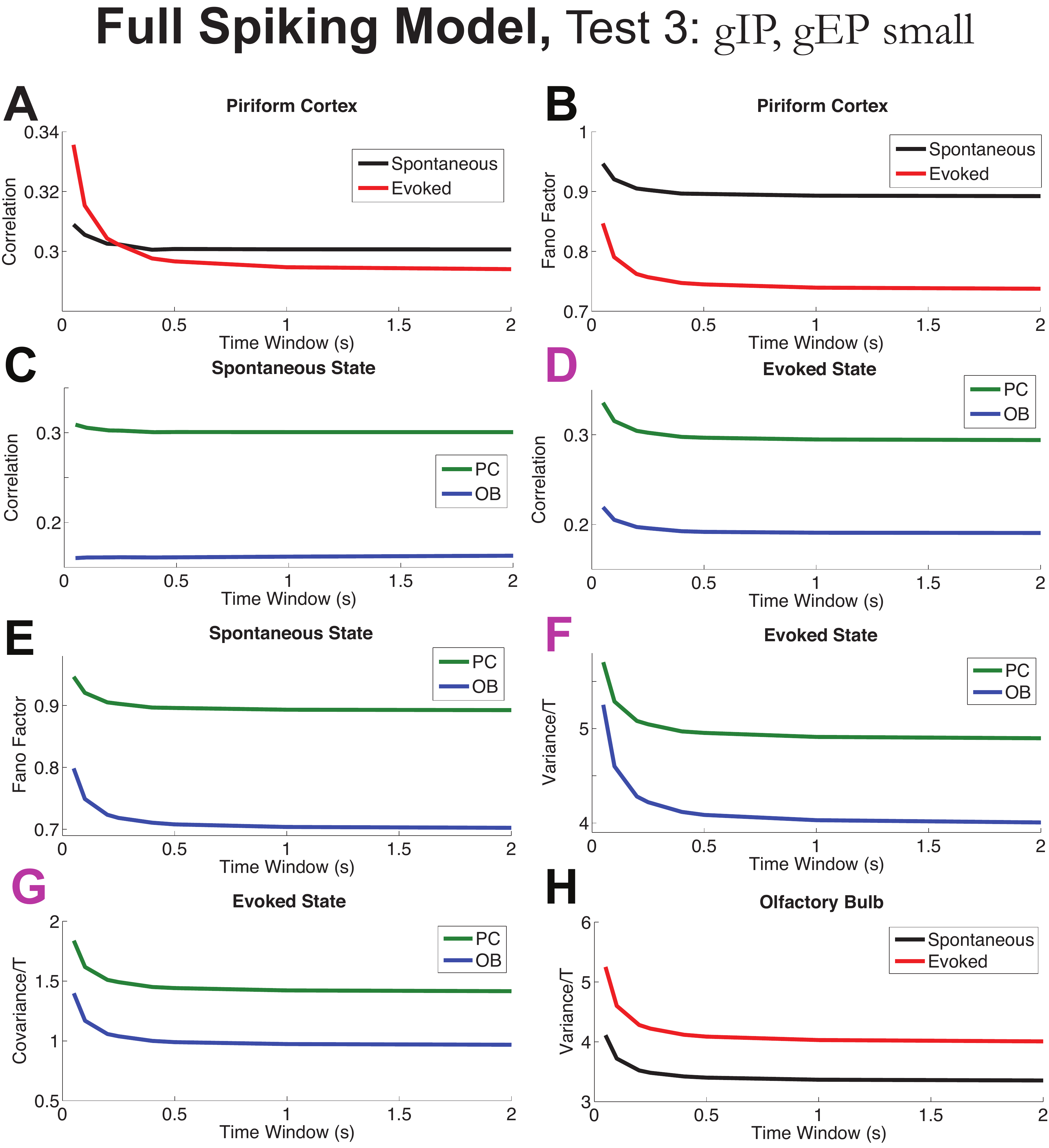}
\caption{\label{fig:test3}  {\bf Violating derived relationship $gEP,  gIP \gg gEO, gIO$ results in statistics that are inconsistent with experimental observations.}Showing the results of the full LIF spiking model when $gEP$ and $gIP$ are both relatively small; specifically, we set $gEP=10$ and $gIP=10$ and 
set the values of the rest of the parameters to those used in Figure 6 (see main text).  
The firing rates are: $\nu_{OB}^{Sp}=5.98\pm4.85$, $\nu_{OB}^{Ev}=8.17\pm5.84$, $\nu_{PC}^{Sp}=3.03\pm2.74$, and $\nu_{PC}^{Ev}=6.95\pm6.1$.  
The 8 panels show the constraints on the 2nd order spiking statistics.  The panels with magenta letters (i.e., D, F, G) are constraints that are violated.
} 
\end{figure}



\clearpage

\setcounter{page}{1}

\begin{center}
{\Large
\textbf{Supplementary Material: Cortical-Cortical Network}
}


\vspace{0.5cm}

{\large

Andrea K. Barreiro$^{\#}$ ; Shree Hari Gautam$^{\ddag}$ ; Woodrow L. Shew$^{\ddag}$ ; Cheng Ly$^{\dag}$ 
}
\vspace{0.5cm}

{\small

{\it \# Department of Mathematics, Southern Methodist University, Dallas, TX 75275  U.S.A.}
\\
{\it \ddag Department of Physics, University of Arkansas, Fayetteville, AR  72701  U.S.A.}
\\
{\it \dag Department of Statistical Sciences and Operations Research, Virginia Commonwealth University, Richmond, VA 23284  U.S.A.}
\\
$\ast$ E-mail: abarreiro@smu.edu ; shgautam@uark.edu; woodrowshew@gmail.com ; CLy@vcu.edu
}
\end{center}

\renewcommand\thefigure{S\arabic{figure}}    
\setcounter{figure}{16}   
 
\setcounter{table}{2}
\renewcommand{\thetable}{S\arabic{table}}

This text shows that the theoretical framework can be applied to a canonical cortical-cortical strongly coupled region.  We use the same experimental data constraints from our 
simultaneous dual-array recordings in the olfactory bulb and piriform cortex, but note that the anatomical connections are {\bf not} the ones described here.

\section*{Minimal Firing Rate Model}

\begin{figure}
\centering
 \includegraphics[width=\columnwidth]{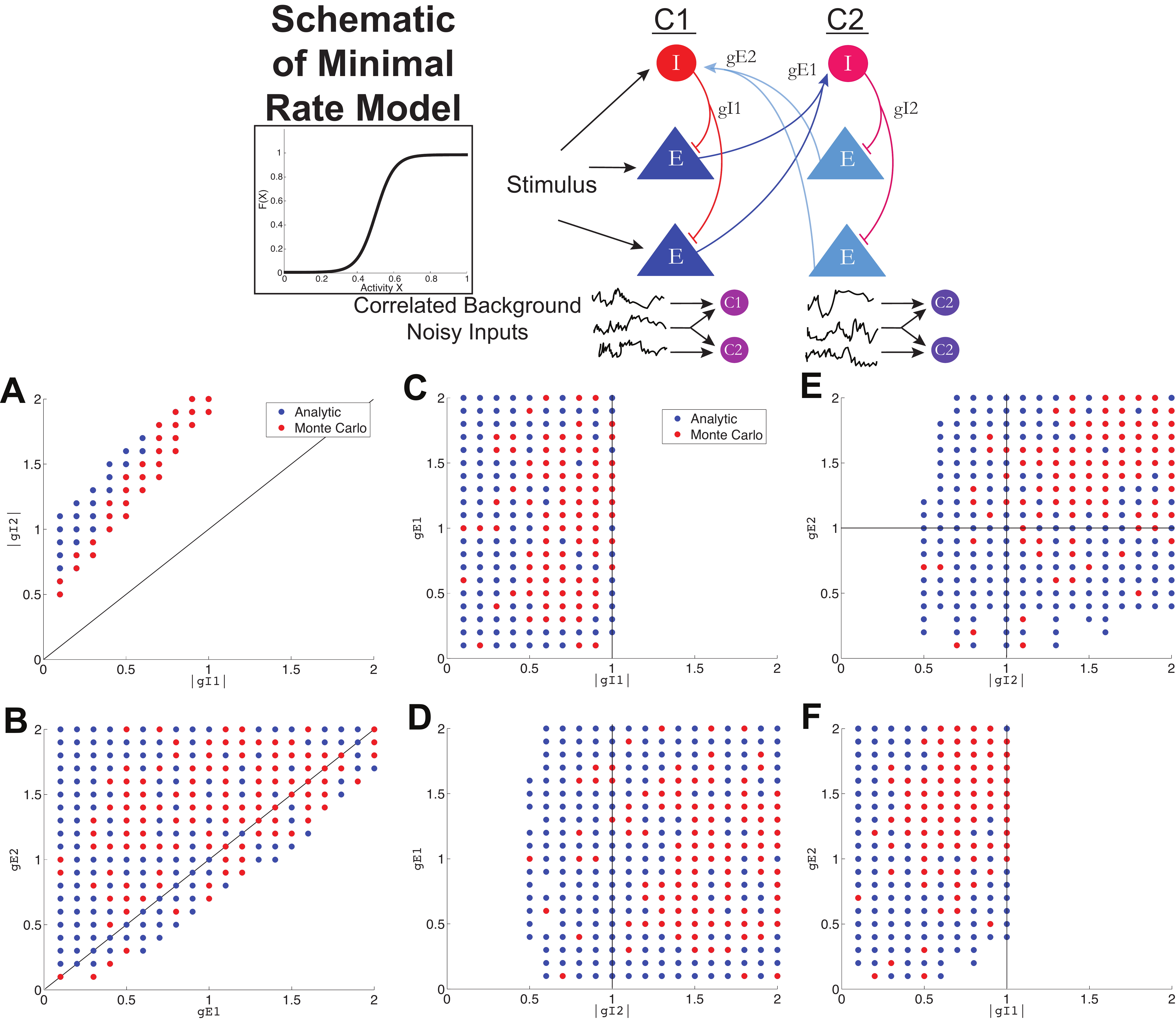}
\caption{{\bf Minimal firing rate model to analyze synaptic conductance strengths.}
This firing rate model only incorporates a subset of the conductances.  
Each plot shows parameter sets that satisfy all 12 data constraints in Table 1 (main text, substitute OB with C1, PC with C2), projected into a two-dimensional plane in parameter space. 
The blue dots show the result of the fast analytic method that satisfy all constraints; the red dots show the Monte Carlo simulations that satisfy 
all 12 constraints.  For computational purposes, we only tested the Monte Carlo on parameter sets that first satisfied the constraints in the fast analytic method.  
(A) The magnitude of the inhibition within C2 ($|gI2|$) is greater than the magnitude of the inhibition within C1 ($|gI1|$);  all dots are above the diagonal line.  
(B) The excitation from C2 to C1 ($gE2$) is generally (but not always) larger than the excitation from C1 to C2 ($gE1$).  
(C) The inhibition within C1 is generally weak; dots are to the left of the vertical line.  
(D) The inhibition within C2 is generally strong; dots are to the right of the vertical line.  (E) Shows again that excitation from C2 and inhibition within C2 are both strong.  
(F) Shows again that excitation from C1 to C2 is relatively small. See Table 3 (main text) for parameter values.  
}
\label{figS17}
\end{figure}

The minimal firing rate model results in Fig~\ref{figS17} have the same parameters and configuration as in the main text except the E to I connections within a region are omitted.  
The derived relationships are qualitatively the same as in the main text:

$$ |gI1| < gE1 < gE2 \lesssim |gI2|. $$

\section*{Leaky Integrate-and-Fire Model of the generic Cortical--Cortical Circuit}

We use a generic spiking neural network model of leaky integrate-and-fire neurons to test the results of the theory again.  The following model is very much like the LIF model in the 
main text, with the main differences being in the network connection strengths and the size of C1 (60 here instead of 100 for OB in the main text).  
There were $N_{C1}=60$ total C1 cells, of which we set 80\% (48) to be excitatory and 20\% (12) inhibitory.  The equations for the C1 cells are, indexed by $k\in\{1,2,\dots,N_{C1}\}$:
\begin{eqnarray}\label{ob_lif}
	\tau_m \frac{d v_k}{dt} & = & \mu_{C1}-v_k-g_{k, XI}(t)(v_k-\mathcal{E}_I)-g_{k, XE}(t)(v_k-\mathcal{E}_E) \nonumber \\	
	     & & - g_{k,XC2}(t - \tau_{\Delta,C2})(v_k-\mathcal{E}_E) +\sigma_{C1}\left(\sqrt{1-\tilde{c}_{C1}}\eta_k(t) + \sqrt{\tilde{c}_{C1}}\xi_o(t) \right) \nonumber \\
	v_k(t^*) & \geq & \theta_k  \Rightarrow v_k(t^*+\tau_{ref})=0 \nonumber \\
	g_{k,XE}(t) &=& \frac{\gamma_{XE}}{p_{XE} \left(0.8 N_{C1} \right) }\sum_{k'\in\{\hbox{ presyn C1 E-cells}\}  } G_{k'}(t) \nonumber \\
	g_{k,XI}(t) &=& \frac{\gamma_{XI}}{p_{XI} \left(0.2 N_{C1} \right)}\sum_{k'\in\{\hbox{presyn C1 I-cells}\}} G_{k'}(t) \nonumber \\
	g_{k,XC2}(t) &=& \frac{\gamma_{X,C2}}{p_{X,C2} \left(0.8 N_{C2} \right)} \sum_{j'\in\{\hbox{presyn C2 E-cells}\}} G_{j'}(t) \nonumber \\
	\tau_{d,X}\frac{d G_k}{dt} &=& -G_k + A_k  \nonumber \\
	\tau_{r,X} \frac{d A_k}{dt} &=& -A_k + \tau_{r,X} \alpha_X \sum_{l} \delta(t-t_{k,l}).  \label{eqn:OB_LIF_all}
\end{eqnarray}
The conductance values in the first equation $g_{k,XI}$, $g_{k,XE}$, and $g_{k,XC2}$ depend on the type of neuron $v_k$ ($X\in\{ E, I\}$).  The last conductance,  
$g_{X,C2}(t - \tau_{\Delta,C2})(v_k-\mathcal{E}_E)$, models the excitatory presynaptic input (feedback) from the C2 cells with a time delay of $\tau_{\Delta,C2}$.  The conductance variables $g_{k,XY}(t)$ are dimensionless because this model was 
derived from scaling the original (raw) conductance variables by the leak conductance with the same dimension.  
The leak, inhibitory and excitatory reversal potentials are 0, $\mathcal{E}_I$, and $\mathcal{E}_E$, respectively with $\mathcal{E}_I<0<\mathcal{E}_E$ 
(the voltage is scaled to be dimensionless, see Table~\ref{tableS:lif_parms}).  
$\xi_k(t)$ are uncorrelated white noise processes and $\xi_o(t)$ is the common noise term to all $N_{C1}$ cells.

The second equation describes the refractory period at spike time $t^*$: when the neuron's voltage crosses 
threshold $\theta_j$ (see below for distribution of thresholds), 
the neuron goes into a refractory period for $\tau_{ref}$, after which we set the neuron's voltage to 0.  

The parameter $\gamma_{XY}$ gives the relative weight of a connection from neuron type $Y$ to neuron type $X$; the parameter $p_{XY}$ is probability that any such connection exists ($X,Y\in\{E,I\}$). $G_k$ is the synaptic variable associated with each cell, and dependent only on that cell's spike times; its dynamics are given by the final two equations in Eq~\ref{eqn:OB_LIF_all} and depend on whether $k \in  \{E,I\}$.

Finally, two of the parameters above can be equated with coupling parameters in the reduced model:
\begin{equation}
gE2 =  \gamma_{I,C2}; \quad gI1 = \gamma_{EI}
\end{equation}
which are dimensionless scale factors for the synaptic conductances.

\begin{table}[!ht]
\centering
\caption{{\bf Fixed parameters for the LIF Cortical--Cortical model. }}
\label{tableS:lif_parms}
\begin{tabular}{|lcccccccccc|}
\hline
\multicolumn{11}{|c|}{\textbf{Same for both C1 and C2}}                                                                                                                                               \\ \hline
\textbf{Parameter}                & $\tau_m$ & $\tau_{ref}$ & $\mathcal{E}_I$ & $\mathcal{E}_E$ & $\tau_{d,I}$ & $\tau_{r,I}$  & $\tau_{d,E}$  & $\tau_{r,E}$  & $\alpha_I$               & $\alpha_E$               \\ \hline
                   & 20\,ms   & 2\,ms        & -2.5            & 6.5             & 10\,ms       & 2\,ms         & 5\,ms         & 1\,ms         								& 2\,Hz                       & 1\,Hz                        \\ \hline
\textbf{Parameter} & $N$    	 & Spont. $\mu$    & Evoked $\mu$ &   $\sigma$     & $\tilde{c}$  &   $\gamma_{EE}$ & $\gamma_{IE}$ & $\gamma_{II}$ & $\gamma_{E,C2/C1}$          & $\tau_{\Delta,C2/C1}$       \\ \hline
\textbf{C1}        & 60     		 & 0.6             & 0.9      				& 0.05         & 0.5                  & 2             & 4             & 6            		 		& 1                        &  	10\,ms \\
\textbf{C2}        & 100      		 & 0               & 0.4  					& 0.1          & 0.8                    & 2             & 4             & 6             			& 1 &                         	5\,ms \\ \hline
\end{tabular}
\begin{flushleft} 
See Eqs~\ref{ob_lif}--\ref{pc_lif}.  All 12 probabilities of connections are set to $p_{XY}=0.30$ and were randomly chosen (Erd\H{o}s-R\'enyi graphs).  
The synaptic time delay from C1 to C2 is $\tau_{\Delta,C1}=10\,$ms, and from C2 to C1 is $\tau_{\Delta,C2}=5\,$ms.  
The scaled voltages from mV is: (V+Vreset)/(Vth+Vreset), corresponding for 
example to Vreset=Vleak=-65\,mV, Vth=-55\,mV (on average), excitatory reversal potential of 0\,mV and inhibitory reversal potential of -90\,mV.
\end{flushleft}
\end{table}

\begin{figure}
\centering
 \includegraphics[width=\textwidth]{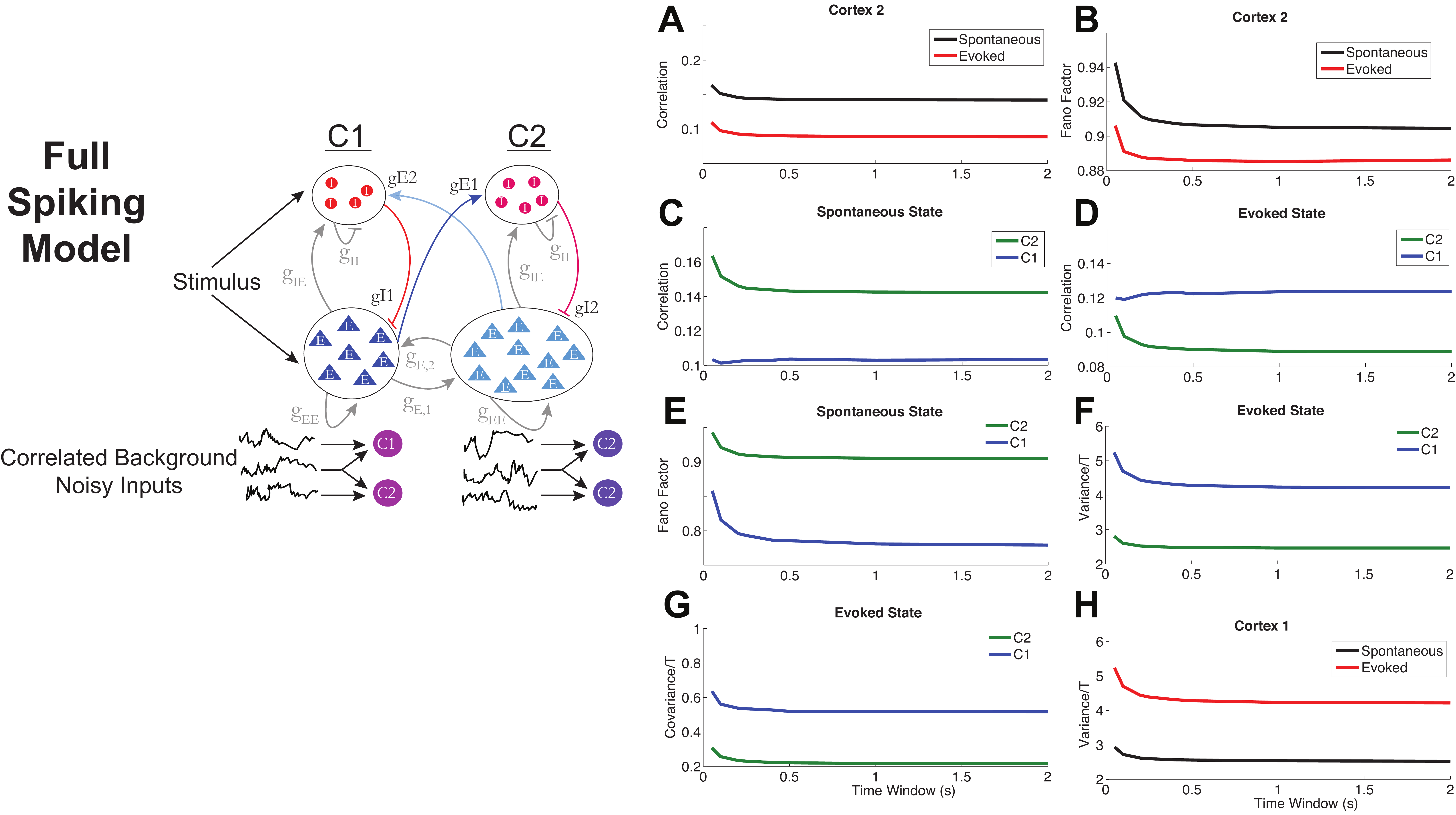}
\caption{{\bf Detailed spiking LIF model confirms the results from analytic rate model.}
Schematic of the LIF model with 2 sets of recurrently coupled E and I cells.  There are 12 types of synaptic connections.  
(A) Pairwise correlations in C2, spontaneous vs. evoked: $\rho_{C2}^{Sp}>\rho_{C2}^{Ev}$.  
(B) Variability (Fano factor) in C2, spontaneous vs evoked: $FF_{C2}^{Sp}>FF_{C2}^{Ev}$.  
(C) Correlations in the spontaneous state, C2 vs. C1: $\rho_{C2}^{Sp}>\rho_{C1}^{Sp}$.
(D) Correlations in the evoked state, C2 vs. C1: $\rho_{C2}^{Ev}<\rho_{C1}^{Ev}$.  
(E) Variability (Fano factor) in the spontaneous state, C2 vs. C1: $FF_{C2}^{Sp}>FF_{C1}^{Sp}$.  
(F) Variability (Fano factor) in the evoked state, C2 vs. C1: $Var_{C2}^{Ev}<Var_{C1}^{Ev}$ in evoked state.  
(G) Covariances in the evoked state, C2 vs. C1: $Cov_{C2}^{Ev}<Cov_{C1}^{Ev}$.
(H) Variability (spike count variance) in C1, spontaneous vs. evoked: $Var_{C1}^{Sp} < Var_{C1}^{Ev}$.  
The curves show the average statistics over all $N_{C1/C2}$ cells or over all possible pairs $N_{C1/C2}(N_{C1/C2}-1)/2$.  
We set $gI1=7$, $gE1=10$, $gI2=20$, $gE2=15$.  See text for model details, and Table~\ref{tableS:lif_parms} for parameter values.
}
\label{figS18}
\end{figure}

The C2 cells had similar functional form but with different parameters (see Table~\ref{tableS:lif_parms} for parameter values).  We modeled $N_{C2}=100$ total C2 cells, of which 80\% were excitatory and 20\% inhibitory.  
The equations, indexed by $j\in\{1,2,\dots,N_{C2}\}$ are:
\begin{eqnarray}\label{pc_lif}
	\tau_m \frac{d v_j}{dt} & = & \mu_{C2}-v_j-g_{j,XI}(t)(v_j-\mathcal{E}_I)-g_{j,XE}(t)(v_j-\mathcal{E}_E) \nonumber \\	
	     & & - g_{j,XC1}(t - \tau_{\Delta,C1})(v_j-\mathcal{E}_E) +\sigma_{C2}\left(\sqrt{1-\tilde{c}_{C2}}\eta_j(t) + \sqrt{\tilde{c}_{C2}}\xi_p(t) \right) \nonumber \\
	v_j(t^*) & \geq & \theta_j  \Rightarrow v_j(t^*+\tau_{ref})=0 \nonumber \\
	g_{j,XE}(t) &=& \frac{\gamma_{XE}}{p_{XE} \left(0.8 N_{C2} \right)}\sum_{j'\in\{\hbox{presyn C2 E-cells}\}} G_{j'}(t) \nonumber \\
	g_{j,XI}(t) &=& \frac{\gamma_{XI}}{p_{XI} \left(0.2 N_{C2} \right)}\sum_{j'\in\{\hbox{presyn C2 I-cells}\}} G_{j'}(t) \nonumber \\
		g_{j,XC1}(t) &=& \frac{\gamma_{X,C1}}{p_{X,C1} \left(0.8 N_{C1} \right)} \sum_{k'\in\{\hbox{presyn C1 E-cells}\}} G_{k'}(t) \nonumber \\
	\tau_{d,X}\frac{d G_j}{dt} &=& -G_j + A_j  \nonumber \\
	\tau_{r,X} \frac{d A_j}{dt} &=& -A_j + \tau_{r,X} \alpha_X \sum_{l} \delta(t-t_{j,l}).
\end{eqnarray}
Excitatory synaptic input from the C1 cells along the lateral olfactory tract is modeled by: $g_{X,C1}(t - \tau_{\Delta,C1})(v_j-\mathcal{E}_E)$.  The common noise term for the 
C2 cells $\xi_p(t)$ is independent of the common noise term for the C1 cells $\xi_o(t)$.  
Two of the parameters above can be equated with coupling parameters in the reduced model:
\begin{equation}
gE1 =  \gamma_{I,C1}; \quad gI2 = \gamma_{EI}
\end{equation}

The values of the parameters 
that were not stated in Table~\ref{tableS:lif_parms} were varied: 
$$ gI1, \hspace{.5in} gE1, \hspace{.5in} gI2, \hspace{.5in} gE2. $$

To model two activity states, we allowed mean inputs to vary (see Table~\ref{tableS:lif_parms}). In contrast to the reduced model, we increased both inputs to C2 cells (from $\mu_{C2}=0$ in the spontaneous state to 
$\mu_{C2}=0.4$ in the evoked state) as well as to C1 cells (from $\mu_{C1}=0.6$ in the spontaneous state to $\mu_{C1}=0.9$ in the evoked state).

Finally, we model heterogeneity by setting the threshold values $\theta_j$ in the following way.  Both C1 and C2 cells had the following distributions for $\theta_j$:
\begin{eqnarray}\label{thres_distr}
	\theta_j &\sim& e^{\mathcal{N}} 
\end{eqnarray}
where $\mathcal{N}$ is normal distribution with mean $-\sigma^2_\theta/2$ and standard deviation $\sigma_\theta$, so that $\{\theta_j\}$ has a 
log-normal distribution with mean 1 and variance: $e^{\sigma_\theta^2}-1$.  We set $\sigma_\theta=0.1$, which results in firing rates ranges seen in the experimental data.  
Since the number of cells are modest with regards to sampling ($N_{C1}=60$, $N_{C2}=100$), we evenly sampled the log-normal distribution from the 5$^{th}$ to 95$^{th}$ percentiles (inclusive).

\subsection*{Violating Derived Relationships Between Conductance Strengths}

Similar to the main text, we demonstrate here that violating the relationships derived in the main text results in 
a subset of the 12 constraints in the experimental data no longer being satisfied in the full spiking network.

Due to the large amount of computing resources required, 
we cannot exhaustively explore the parameter space; 
recall that the purpose of the method we developed in the minimal firing rate model is for faster computation.  Instead, we distill results into three tests that are exactly the same as in the main text:
\begin{enumerate}
	\item Make $gI1 > g I2$ by setting $gI1=20$ and $gI2=7$.
	\item Make $gE1 > gE2$ by setting $gE1=15$ and $gE2=1$
	\item Make $gE2$ and $gI2$ relatively smaller by setting $gE2=10$ and $gI2=10$
\end{enumerate}

The result of Test 1 is that 7 of the 12 constraints are violated (see Fig~\ref{fig:test1}); most importantly stimulus-induced decorrelation of the C2 cells, 
which is particularly important in the context of coding, was not present.  
In addition, the C2 firing rates are larger than the C1 firing rates in both states, the evoked C2 correlation is larger than evoked C1 correlation, the spontaneous C2 Fano Factor is larger than spontaneous C1 Fano Factor, and both the variance and covariance 
of C2 is larger than C1 in the evoked state (all of which violate the constraints from our data).

\begin{figure}
\centering
 \includegraphics[width=\textwidth]{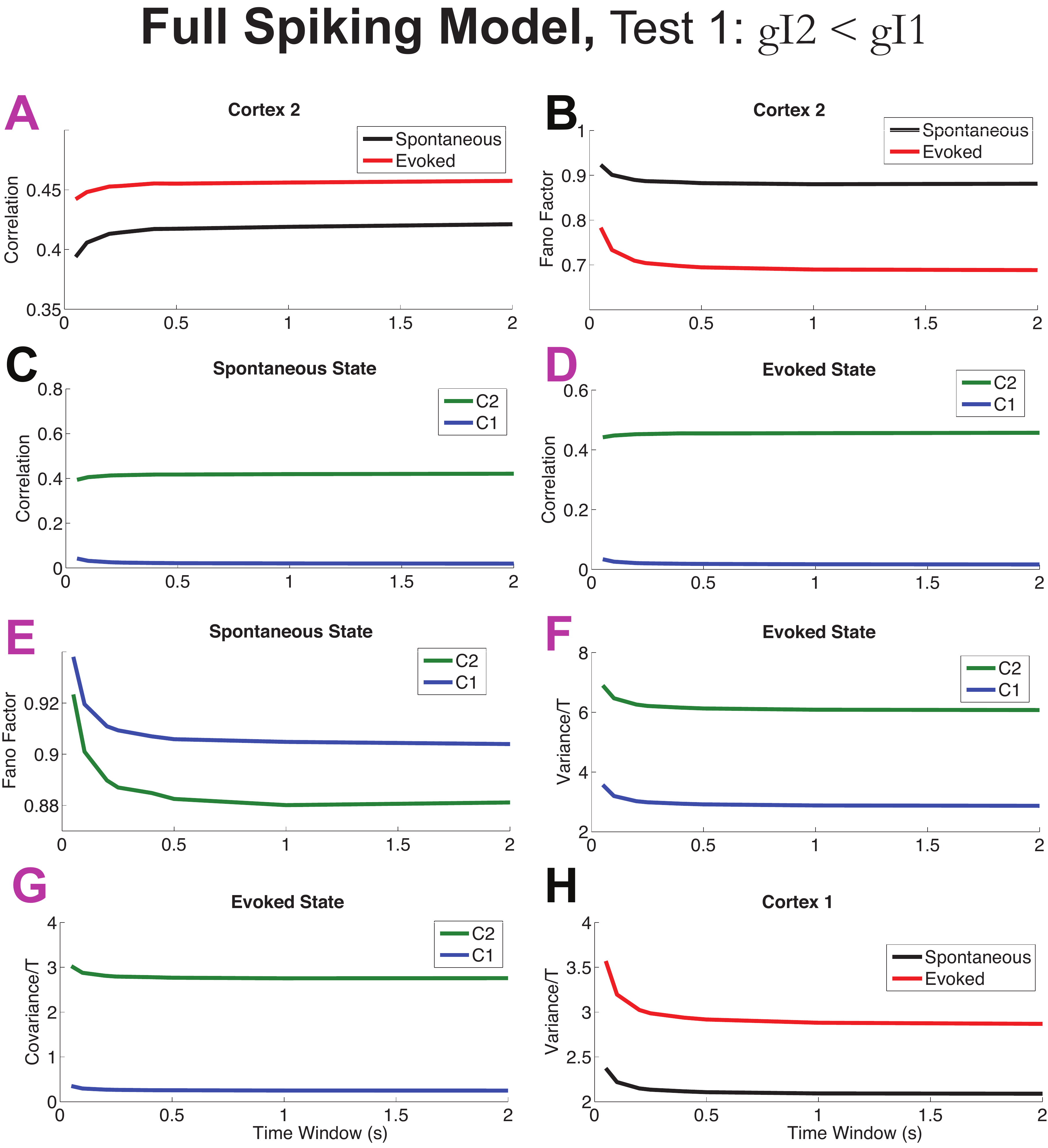}
\caption{\label{fig:test1}  {\bf Violating derived relationship $| gI1 | < |gI2 |$ results in statistics that are inconsistent with experimental observations.} Showing the results of the full LIF spiking model when $gI2<gI1$; specifically, we set $gI2=7$ and $gI1=20$ and set the values of the rest of the parameters to those used previously.  
The firing rates are: $\nu_{C1}^{Sp}=2.96\pm5$, $\nu_{C1}^{Ev}=5.94\pm11.67$, $\nu_{C2}^{Sp}=3.43\pm1.59$, and $\nu_{C2}^{Ev}=8.85\pm3.38$, which violates the constraint from the experimental data that $\nu_{C1}>\nu_{C2}$ in both states.  
The 8 panels show the constraints on the 2nd order spiking statistics in the same format as before.  The panels with magenta letters (i.e., A, D, E, F, G) are constraints that are violated.
} 
\end{figure}

The result of Test 3 is that 4 of the 12 constraints are violated (see Fig~\ref{fig:test3}), including again stimulus-induced decorrelation of the C2 cells.  The evoked C2 correlation is larger than evoked C1 correlation, and 
both the variance and covariance of C2 are larger than the corresponding quantities in C1 in the evoked state.
Both these two tests (1 and 3) indicate that these two qualitative relationships 
(stronger effective inhibition within C2 and stronger effective presynaptic inputs from C2) are robust with respect to both the detailed LIF spiking model and the minimal firing rate model.

\begin{figure}
\centering
 \includegraphics[width=\textwidth]{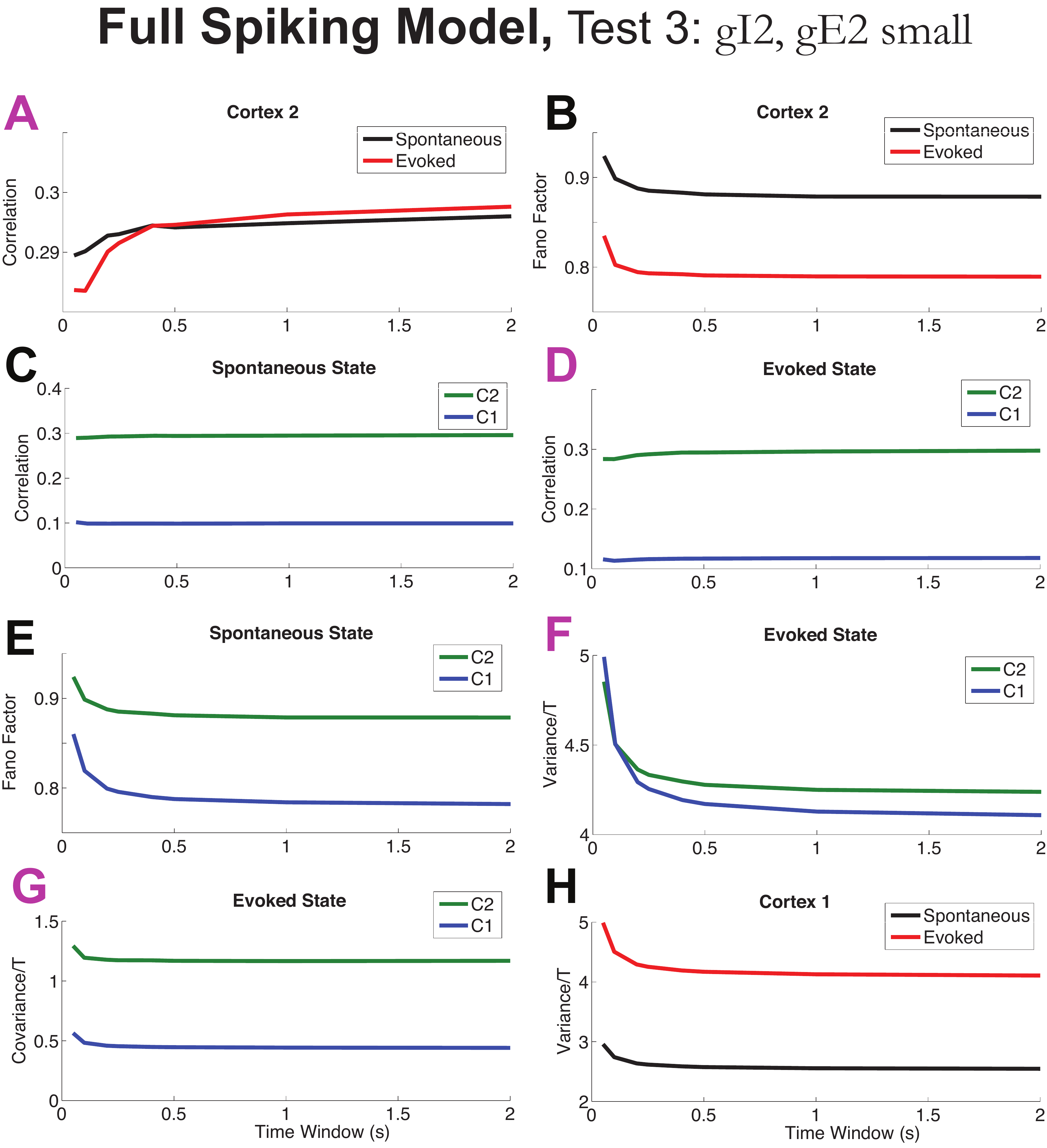}
\caption{\label{fig:test3}  {\bf Violating derived relationship $gE2,  gI2 \gg gE1, gI1$ results in statistics that are inconsistent with experimental observations.}  Showing the results of the full LIF spiking model when $gE2$ and $gI2$ are both relatively small; specifically, we set $gE2=10$ and $gI2=10$ and 
set the values of the rest of the parameters to those used in Figure 5 (see main text).  
The firing rates are: $\nu_{C1}^{Sp}=3.85\pm3.56$, $\nu_{C1}^{Ev}=8.2\pm7.08$, $\nu_{C2}^{Sp}=2.92\pm2.31$, and $\nu_{C2}^{Ev}=6.45\pm6.17$, which violates the constraint from the experimental data that $\nu_{C1}>\nu_{C2}$ in both states.  
The 8 panels show the constraints on the 2nd order spiking statistics.  The panels with magenta letters (i.e., A, D, F, G) are constraints that are violated.
} 
\end{figure}

The result of Test 2 is not as straightforward as the others.  We did not exhaustively search 
parameter space due to the vast computational resources this would require, but in several parameter sets with $gE1>gE2$, we found the resulting network statistics could still satisfy all of the constraints 
(e.g., with $gE1=15$ and $gE2=1$, as well as with $gE1=20$ and $gE2=1$).  
The reason for this may be that in the two coupled recurrent networks we chose very different $gI1$ and $gI2$ values to begin with (7 and 20, respectively), and would thus require $gE1$ and $gE2$ to be significantly different to 
counter-balance this.  Also, notice in the minimal firing rate model results in Fig 4B that there are a significant number of red dots below the diagonal, indicating that the relationship $gE2>gE1$ does not have to strictly hold.  However, 
we did find a condition where this test demonstrates the value of the minimal firing rate model; we changed $\tilde{c}_{C1}$ from $0.5$ to $0.6$ (recall $\tilde{c}_{C2}=0.8$).  
(Note that in the minimal firing model that $c_{C1}=0.3$ and $c_{C2}=0.35$, relatively close in value.)  
The result of Test 2 ($gE1=15$ and $gE2=1$) with $\tilde{c}_{C1}=0.6$ is that one constraint is violated: $\rho^{Sp}_{C2}$ is no longer 
less than $\rho^{Sp}_{C1}$ (see Fig~\ref{fig:test2}).  
This suggests that the 
relationship that $gE1 > gE2$ is not as robust as the others and can be violated.

\begin{figure}
\centering
 \includegraphics[width=\textwidth]{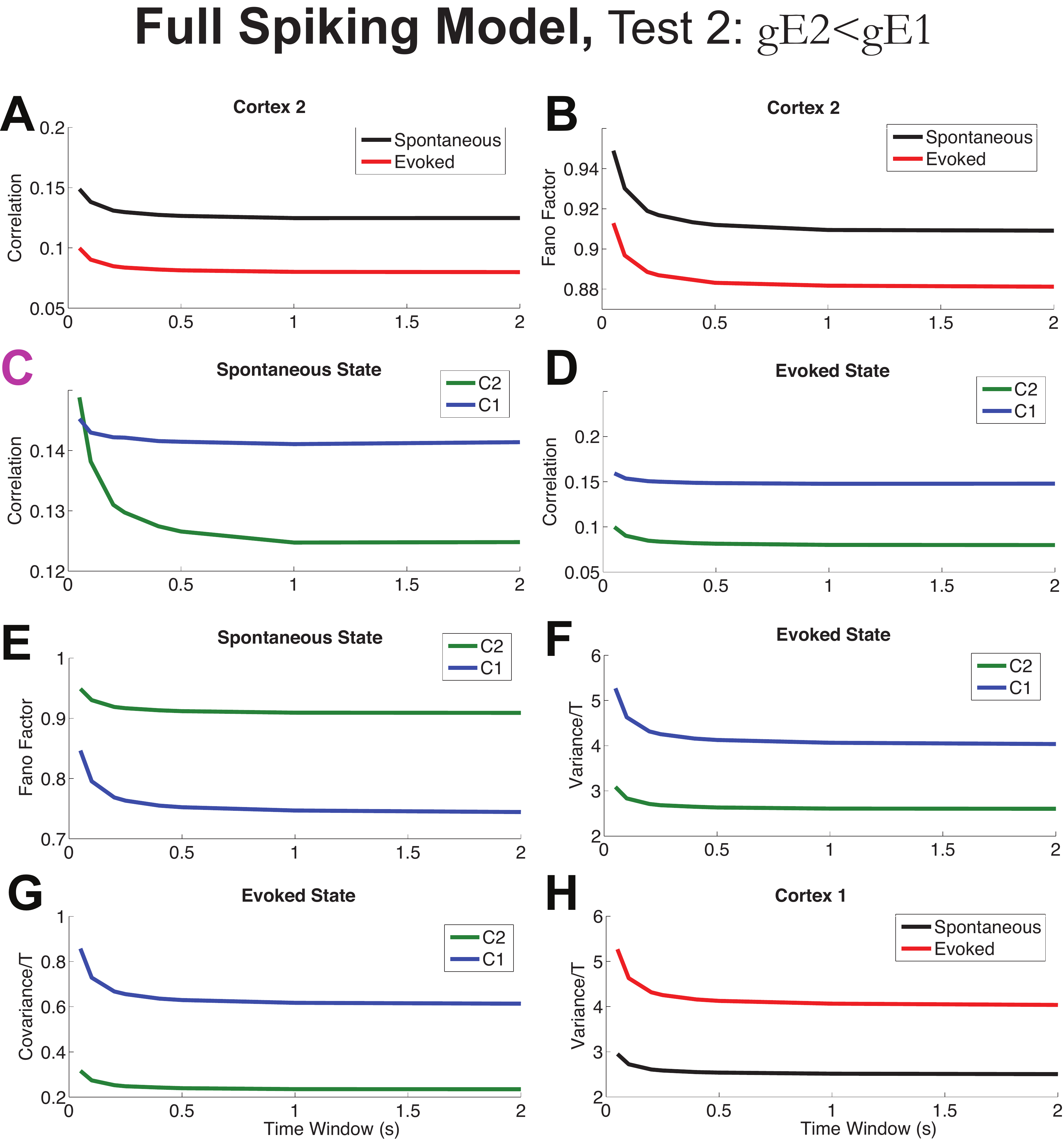}
\caption{\label{fig:test2} {\bf Violating derived relationship $gE2 > gE1$ results in statistics that are inconsistent with experimental observations.} Showing the results of the full LIF spiking model when $gE2<gE1$; specifically, we set $gE2=1$ and $gE1=15$, \underline{\textit{and with} $\tilde{c}_{C1}=0.6$ instead of $0.5$}; 
we set the values of the rest of the parameters to those used in Figure 5 (see main text).  
The firing rates are: $\nu_{C1}^{Sp}=3.75\pm 2.61$, $\nu_{C1}^{Ev}=8.73\pm 5.12$, $\nu_{C2}^{Sp}=2.28\pm 3.32$, and $\nu_{C2}^{Ev}=4.87\pm 9.2$.  
The 8 panels show the constraints on the 2nd order spiking statistics.  Only 1 constraint is violated, panel C in magenta. } 
\end{figure}

\end{document}